\RequirePackage{rotating}

\documentclass[fleqn,usenatbib]{mnras}

\usepackage{newtxtext,newtxmath}

\usepackage[T1]{fontenc}
\usepackage{pifont}
\DeclareRobustCommand{\VAN}[3]{#2}
\let\VANthebibliography\thebibliography
\def\thebibliography{\DeclareRobustCommand{\VAN}[3]{##3}\VANthebibliography}

\usepackage{graphicx}	
\usepackage{amsmath}	
\usepackage{pdflscape}
\usepackage{rotating}
\usepackage{hyperref}
\usepackage{longtable}
\usepackage{soul}

\title[Extreme emission line galaxies detected in JADES JWST/NIRSpec]{{Extreme emission line galaxies detected in JADES JWST/NIRSpec I: inferred galaxy properties }}

\author[K. Boyett et al.]{Kit Boyett,$^{1,2,3}$\thanks{E-mail: kit.boyett@unimelb.edu.au}
Andrew J. Bunker,$^{3}$
Emma Curtis-Lake,$^{4}$
Jacopo Chevallard,$^{3}$
Alex J. Cameron,$^{3}$ \newauthor
Gareth C. Jones,$^{3}$
Aayush Saxena,$^{3}$
Stéphane Charlot,$^{5}$
Mirko Curti,$^{6,7,8}$
Imaan E.B.  Wallace,$^{3}$
Santiago Arribas,$^{9}$ \newauthor
Stefano Carniani,$^{13}$
Chris Willott,$^{18}$
Stacey Alberts,$^{12}$
Daniel J. Eisenstein,$^{19}$
Kevin Hainline,$^{12}$
Ryan Hausen,$^{20}$ \newauthor
Benjamin D. Johnson,$^{19}$
Marcia Rieke,$^{12}$
Brant Robertson,$^{21}$
Daniel P. Stark,$^{12}$
Sandro Tacchella,$^{7}$ \newauthor
Christina C. Williams,$^{22}$
Zuyi Chen,$^{12}$
Eiichi Egami,$^{12}$ 
Ryan Endsley,$^{23}$
Nimisha Kumari,$^{24}$
Isaac Laseter,$^{25}$\newauthor
Tobias J. Looser,$^{7}$
Michael V. Maseda,$^{25}$
Jan Scholtz,$^{7,8}$
Irene Shivaei,$^{9}$
Charlotte Simmonds,$^{7,8}$ 
Renske Smit,$^{26}$\newauthor
Hannah \"Ubler,$^{7,8}$
Joris Witstok$^{7,8}$
\\
$^1$School of Physics, University of Melbourne, Parkville 3010, VIC, Australia\\
$^2$ARC Centre of Excellence for All Sky Astrophysics in 3 Dimensions (ASTRO 3D), Australia\\
$^3$Department of Physics, University of Oxford, Denys Wilkinson Building, Keble Road, Oxford OX1 3RH, UK\\
$^4$Centre for Astrophysics Research, Department of Physics, Astronomy and Mathematics, University of Hertfordshire, Hatfield AL10 9AB, UK\\
$^5$Sorbonne Universit\'e, CNRS, UMR 7095, Institut d'Astrophysique de Paris, 98 bis bd Arago, 75014 Paris, France\\
$^{6}$European Southern Observatory, Karl-Schwarzschild-Strasse 2, 85748 Garching, Germany \\
$^{7}$Kavli Institute for Cosmology, University of Cambridge, Madingley Road, Cambridge, CB3 0HA, UK.\\ 
$^{8}$Cavendish Laboratory - Astrophysics Group, University of Cambridge, 19 JJ Thomson Avenue, Cambridge, CB3 0HE, UK. \\
$^9$Centro de Astrobiolog\'ia (CAB), CSIC–INTA, Cra. de Ajalvir Km.~4, 28850- Torrej\'on de Ardoz, Madrid, Spain\\
$^{10}$Cosmic Dawn Center (DAWN), Copenhagen, Denmark \\
$^{11}$Niels Bohr Institute, University of Copenhagen, Jagtvej 128, DK-2200, Copenhagen, Denmark \\
$^{12}$Steward Observatory University of Arizona 933 N. Cherry Avenue Tucson AZ 85721, USA\\
$^{13}$Scuola Normale Superiore, Piazza dei Cavalieri 7, I-56126 Pisa, Italy\\
$^{14}$European Space Agency, European Space Astronomy Centre, Camino Bajo del Castillo s/n, 28692 Villafranca del Castillo, Madrid, Spain\\
$^{15}$ATG Europe for the European Space Agency, ESTEC, Noordwijk, The Netherlands\\
$^{16}$Department of Physics and Astronomy, University College London, Gower Street, London WC1E 6BT, UK\\
$^{17}$Max-Planck-Institut f\"ur Astronomie, K\"onigstuhl 17, D-69117, Heidelberg, Germany\\
$^{18}$NRC Herzberg, 5071 West Saanich Rd, Victoria, BC V9E 2E7, Canada\\
$^{19}$Center for Astrophysics $|$ Harvard \& Smithsonian, 60 Garden St., Cambridge MA 02138 USA\\
$^{20}$Department of Physics and Astronomy, The Johns Hopkins University, 3400 N. Charles St., Baltimore, MD 21218\\
$^{21}$Department of Astronomy and Astrophysics University of California, Santa Cruz, 1156 High Street, Santa Cruz CA 96054, USA \\
$^{22}$NSF’s National Optical-Infrared Astronomy Research Laboratory, 950 North Cherry Avenue, Tucson, AZ 85719, USA\\
$^{23}$Department of Astronomy, University of Texas, Austin, TX 78712, USA\\
$^{24}$AURA for European Space Agency, Space Telescope Science Institute, 3700 San Martin Drive. Baltimore, MD, 21210\\
$^{25}$Department of Astronomy, University of Wisconsin-Madison, 475 N. Charter St., Madison, WI 53706 USA\\
$^{26}$Astrophysics Research Institute, Liverpool John Moores University, 146 Brownlow Hill, Liverpool L3 5RF, UK\\
}

\date{Accepted XXX. Received YYY; in original form ZZZ}

\pubyear{2015}

\begin{document}
\label{firstpage}
\pagerange{\pageref{firstpage}--\pageref{lastpage}}
\maketitle

\begin{abstract}
Extreme emission line galaxies (EELGs) exhibit large equivalent widths (EW) in their rest-optical emission lines ([OIII]$\lambda5007$ or H$\alpha$ rest-frame EW$ > 750$\AA) which can be tied to a recent upturn in star formation rate (SFR), due to the sensitivity of the nebular line emission and the rest-optical continuum to young ($<10$Myr) and evolved stellar populations, respectively. 
By studying a sample of 85 star forming galaxies (SFGs), spanning the redshift and magnitude interval $3 <z<9.5$ and $-16>$ M$\rm_{UV}>-21$,  in the JWST Advanced Deep Extragalactic Survey (JADES) with NIRSpec/prism spectroscopy,
we determine that SFGs initiate an EELG phase when entering a significant burst of star formation, with the highest EWs observed in EELGs with the youngest luminosity-weighted ages ($<5$ Myr) and the highest burst intensity (those with the greatest excess between their current and long-term average SFR).
We spectroscopically confirm that a greater proportion of SFGs are in an EELG phase at high redshift in our UV-selected sample ($61\pm4\%$ in our $z>5.7$ high-redshift bin, compared to $23^{+4}_{-1}\%$ in our lowest-redshift bin $3<z<4.1$) due to the combined evolution of metallicity, ionisation parameter and star formation histories with redshift.
We report that the EELGs within our sample exhibit a higher average ionisation efficiency ($\log_{10}(\xi_\mathrm{ion}^\mathrm{HII}/{\rm erg^{-1}Hz}) =25.5\pm0.2$) than the non-EELGs. 
High-redshift EELGs therefore comprise a population of efficient ionising photon producers.
Additionally, we report that $53\%$ (9/17) of EELGs at $z>5.7$ have observed Lyman-$\alpha$ emission, potentially lying within large ionised regions. The high detection rate of Lyman-$\alpha$ emitters in our EELG selection suggests that the physical conditions associated with entering an EELG phase also promote the escape of Lyman-$\alpha$ photons.
\end{abstract}



%


\section{Introduction}
Extreme emission line galaxies (EELGs) exhibit significant line emission relative to their stellar continuum which manifest as large emission line equivalent widths (EWs). 
These large emission line equivalent widths are most commonly seen in the H$\alpha$ and [OIII]$\lambda5007$ lines in the rest-frame optical.
The nebular emission lines are often driven by the ionising photons produced in massive and short-lived O and B stars (or active galactic nuclei), whereas the surrounding rest-optical continuum includes the contribution from longer-lived and less massive stars (e.g., \citealt{Eldridge_Stanway22}). Hence the EW tells us about the ratio of the very young to the older stellar population, reflecting any change in star formation over time. 
Selecting EELGs within a galaxy population may identify galaxies going through an upturn or burst in star formation \citep[e.g.][]{Endsley23arXiv}, exhibiting large specific star formation rates (sSFR). 

Over the last decade EELGs have been observed locally (e.g., the "green pea" population, \citealt{Cardamone09, Amorin10, Izotov11, Brunker20, Kumari23}), at moderate redshifts ($z\sim2$, \citealt{van_der_Wel_2011, Atek_11, Amorin15, Maseda18, Mengtao19, Du20, Onodera20, Boyett22A}) and during the epoch of reionisation (EOR, \citealt{Smit_2015,deBarros_2019,Endsley20, Simmonds23}). 
These large EW systems are identified either directly from spectroscopy or from a photometric excess of flux between two adjacent imaging filters due to the contribution of line emission to one of them. 
We note that there is no set definition for what EW threshold identifies an EELG, with adopted thresholds varying between studies ([OIII]$\lambda5007$ EW thresholds ranging from $\sim100-1000$\,\AA\,  e.g., \citealt{Amorin15, van_der_Wel_2011, Mengtao19, Mengtao20, Du20}).
Studies over a wide range of redshifts have suggested that EELGs are more abundant at higher redshift \citep{Boyett22A}, perhaps due to changes in the characteristic star formation histories (SFHs) and other properties such as metallicity (e.g.,
\citealt{Matthee23}). However, prior to the James Webb Space Telescope (JWST), studies of EELGs at high redshifts ($z>4$) had been limited to photometric excess measurements, largely from Spitzer/IRAC, as the rest-optical H$\alpha$ and [OIII]$\lambda5007$ lines moved out of the wavelength coverage of the Hubble Space Telescope (HST) and ground-based spectroscopy (due to atmospheric opacity), and hence these measurements are highly dependent on assumptions of the underlying continuum shape and line ratios. 

The recent launch of JWST enables sensitive near-infrared spectroscopy out to 5.2$\mu$m with NIRSpec \citep{jakobsen22} permitting the direct measurement of the EWs of rest-optical emission lines out to high redshift (H$\alpha$ $z<7$, [OIII]$\lambda5007$ $z<9.5$). This allows the unambiguous identification of EELGs and characterisation of their physical properties. 
As spectroscopic samples grow, we will be able to constrain the fraction of the galaxy population in an EELG phase, and any evolution with redshift. 

In the first cycle of JWST observations, studies with a variety of instrument modes have been used to identify EELGs at high redshifts. Spectroscopic studies in slitless mode using NIRISS \citep{Willott22} and NIRCam \citep{Rieke23} have identified EELGs \citep{Boyett22B, Kashino23, Sun23a}. The use of NIRCam medium-band imaging to measure the flux excess between adjacent filters has also enabled EELG identification \citep{Withers23, Endsley23arXiv, Tacchella23}. These programmes are beginning to confirm that the galaxy properties of individual high redshift EELGs match what had been observed at cosmic noon (z$\sim2$) where studies have shown that galaxies with large EWs (rest-frame [OIII]$\lambda5007$ EW > 750\AA, for instance) are observed to exhibit lower masses, more compact morphologies, younger stellar populations, higher ionising photon production efficiencies, higher ionisation parameters and lower gas-phase oxygen abundances (metallicity) than typical star-forming galaxies (SFGs) found at the same epoch \cite[e.g.,][]{Mengtao19}. 

These properties make SFGs in an EELG phase effective ionising photon producers. 
However, at moderate redshifts, their abundance has been measured to be too low to dominate the ionising output of the star-forming galaxy population \citep{Boyett22A}, with the majority of the ionising output coming from SFGs in more typical modes of star formation. 
The bursty SFHs expected at higher redshifts and lower masses \citep{Ceverino18, Ma18, Faucher-Giguere18, Tacchella20}, which may make EELGs more common than at lower redshifts, 
suggests the potential for high escape fractions of ionising photons\footnote{Although we note that galaxies with near 100$\%$ $f_\mathrm{esc}$ would show very low EWs, since nebular emission lines are powered by the UV-ionising photons which do not escape into the intergalactic medium (IGM) but are reprocessed locally.}  \citep[$f_\mathrm{esc}$;][]{Katz23B}. To understand the nature and significance of EELGs at high redshift, it is critical to study a large spectroscopic sample.

Our sample originates from the JWST Advanced Deep Extragalactic Survey (JADES, \citealt{Bunker20, Rieke20, Eisenstein23}) which is obtaining spectra of thousands of galaxies between cosmic noon to within the EoR. In this paper, we use the JADES first data release \citep{Bunker23b, Rieke23b} which targets $\sim250$ galaxies in the Hubble Ultra Deep Field region (HUDF, \citealt{Beckwith06}) and the surrounding GOODS-South field (\citealt{Giavalisco04}).
We use the JADES NIRSpec and NIRCam data to look at SFGs over a redshift range $3<z<9.5$ to investigate the evolution of the abundance of these EELGs with redshift and study their physical properties in detail to determine what physical conditions initiate an EELG phase. We focus on the [OIII]$\lambda5007$ line as this typically has the highest EW in our spectroscopy, but we also consider cases where H$\alpha$ (accessible at $z\lesssim7$) may also have extreme EWs.

This paper is laid out as follows. We discuss the JADES observations and the selection of the parent spectroscopic sample in Section \ref{sec:obs}. We measure the EWs of emissions lines and identify a sub-sample of galaxies in an EELG phase in Section \ref{sec:method}. In Sections \ref{sec:line_trends} we examine any trends between the EWs
of [OIII]$\lambda5007$ and galaxy properties. Finally, in Section \ref{sec:discussion} we discuss what initiates an EELG phase and how the fraction of SFGs in an EELG phase evolves as a function of redshift. Where applicable, we use a standard $\Lambda$CDM cosmology with parameters \textit{H}$_{0}=70$ km/s/Mpc, $\Omega_{m}=$0.3, and $\Omega_{\wedge}=$0.7. All magnitudes are in the AB system \citep{Oke83}. When quoted, all EWs are in the rest-frame.

\section{Observations}\label{sec:obs}
\subsection{Spectroscopic and Imaging data}

The spectroscopic and photometric imaging data used in this study come from the DEEP tier of the JWST NIRSpec and NIRCam observations obtained as part of the JADES survey (programme ID 1210, 1180 P.I.: D. Eisenstein). The spectroscopic data used here have already been presented in \citet{Bunker23b} and we refer the reader to that paper for a detailed description of the target selection, observation strategy and data reduction (further reduction details will be provided in Carniani, in prep.). The NIRCam imaging used in this paper is described in \citet{Rieke23b}. We also refer the reader to \citet{Eisenstein23} for an overview of the JADES programme.  

Briefly, the spectroscopic observations were obtained using NIRSpec in the Multi-object spectroscopy (MOS) mode \citep{ferruit22}, with a three shutters nodding sequence. 
Three Micro-Shutter Assembly (MSA) configurations were constructed, each with three nod positions. 
Targets were assigned to the configurations following the prioritisation discussed in \citet{Bunker23b}. The highest priority targets were included in all three MSA configurations and received the maximum exposure time, while the remaining targets assigned to only one or two configurations received only 1/3 or 2/3 of the total observing time. This approach maximises the total number of galaxies observed whilst retaining the maximum exposure time for the highest priority sources. The prioritisation of targets that were selected and allocated may introduce a selection bias to our sample, which we discuss further in Section \ref{sec:selection}.

The configurations were observed with both the low-dispersion prism\footnote{The prism has an average resolving power $\frac{\Delta\lambda}{\lambda} = R\sim100$, but this varies with wavelength over the range $R=30-300$.} (total integrated exposure time of 100 ksec) and the $R\sim1000$ medium resolution\footnote{The set of three medium resolution gratings have an average resolving power $\frac{\Delta\lambda}{\lambda} = R\sim1000$, but this varies with wavelength over the range $R=300-1200$.} G140M/F070LP, G235M/F170LP and G395M/F290LP gratings (each with a total integrated exposure time of 25 ksec). The MSA was configured to avoid overlap in the prism spectra. To avoid spectra overlapping in the medium-resolution grating (where the dispersed light covers a greater pixel range on the detector) for the highest priority targets, a number of lower priority targets were removed from the configuration. The spectroscopic data were reduced using the pipeline developed by the ESA NIRSpec Science Operations Team and the NIRSpec GTO Team (Carniani, in prep), see \citet{Bunker23b} for specific details. 

For the imaging in this study, we utilise the JADES (\citealt{Rieke23b}) F090W, F115W, F150W, F200W, F277W, F335M, F356W, F410M and F444W NIRCam observations. Each of these have integrated exposure times between 24.7 and 60.5\,ksec (see \citealt{Rieke23b,Eisenstein23}).
The JADES F444W NIRCam imaging is supplemented with extended F444W imaging from the FRESCO survey \citep{Oesch23arXiv} and covers a larger footprint than the other filters. 
The JADES imaging over the HUDF is complimented by the JWST Extragalactic Medium-band Survey (JEMS, \citealt{Williams23}) F182M, F210M, F430M, F460M, and F480M observations. These have integrated exposure times between 13.9\,ksec and 27.8\,ksec. 
We note that the footprint in the different filters varies, and a sub-set of the galaxies targeted with NIRSpec have coverage only in F444W, whilst the majority are covered by all imaging filters. 

\subsection{Sample selection and selection bias}
\label{sec:selection}

\begin{figure*}
    \centering
    \includegraphics[width=\textwidth]{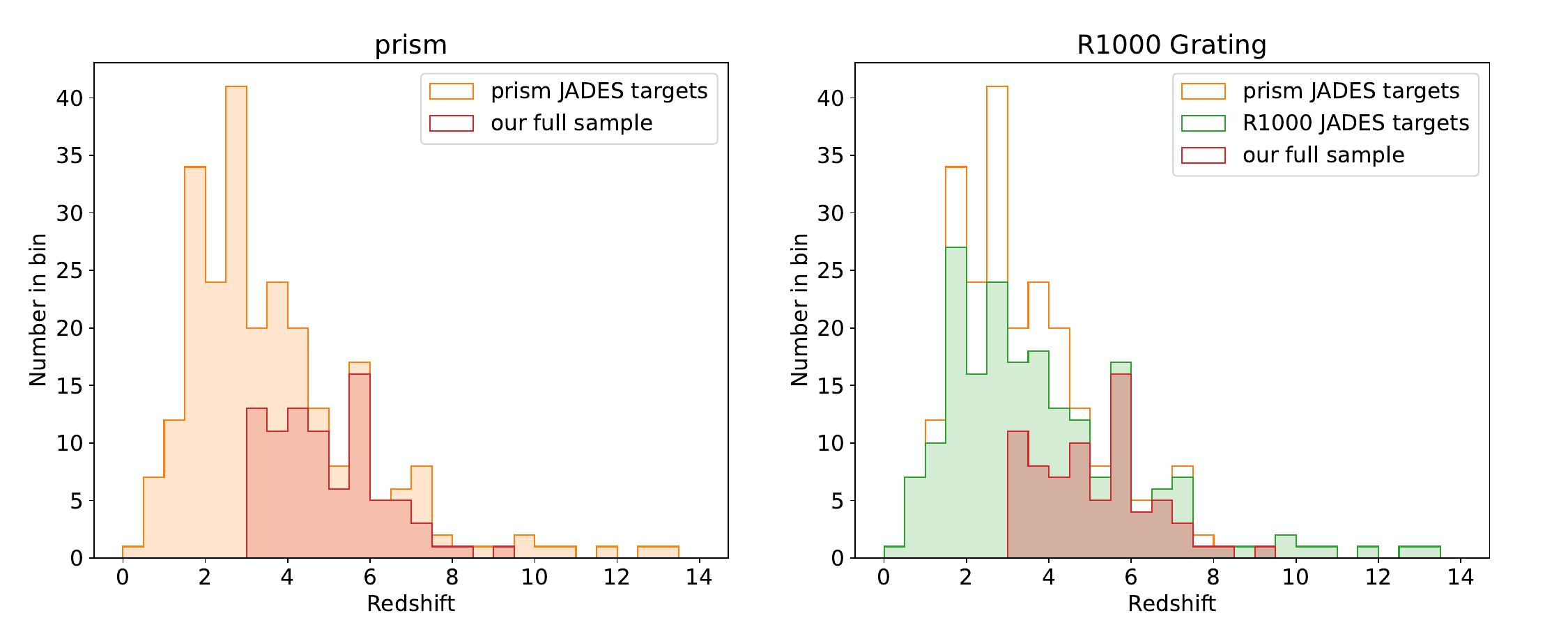}
    \caption{Redshift distribution of the JADES targeted galaxies (prism in orange, R1000 grating in green) and our defined full sample (red, $z>3$ galaxies with multiple emission lines, including [OIII]$\lambda5007$, detected at $>5\sigma$). Galaxies in our full sample with coverage in the prism and R1000 gratings are shown in the left and right panels, respectively. We have an upper redshift bound to our full sample where [OIII]$\lambda5007$ moves out of the wavelength coverage at $z\sim9.5$. When spectroscopic redshifts could not be determined using NIRSpec for the JADES targeted galaxies, we rely on the photometric redshifts used to selected these candidates (see \citealt{Bunker23b}).}
    \label{fig:redshift_dist}
\end{figure*}

Across the three MSA configurations, a total of 252 unique galaxies were observed in the prism, with a subset of 197 also observed in medium-resolution gratings (we refer the reader to \citealt{Bunker23b} for a detailed break-down of the spectroscopic survey). 

Of these, a total of 152 galaxies have identifiable redshifts based on at least one emission line detected at $>5\sigma$ in the prism spectra or have an identifiable Lyman spectral break. For the grating, 103 out of 197 galaxies observed had an identifiable redshift.

In this study, we will focus on the emission line sample at $z>3$, where the H$\beta$ and [OIII]$\lambda4959,5007$ emission lines are deblended at the prism resolution. We define a sample of 86 galaxies which are determined to have multiple detected lines at $>5\sigma$, including [OIII], in the prism spectra. These galaxies will become our sample for this study (hereafter "full sample").  

In Figure \ref{fig:redshift_dist}, we present the redshift distribution for the JADES targeted galaxies (presenting NIRSpec spectroscopic redshifts when determined otherwise photometric redshifts, see \citealt{Bunker23b, Eisenstein23})
and overlay the sub-set of galaxies which have multiple $>5\sigma$ emission line detections at $z>3$, which form the full sample in this study.  We note that [OIII]$\lambda5007$ passes out of the NIRSpec wavelength coverage at $z>9.5$, providing an upper redshift limit on our study.

This sample was pre-selected for observation through a target prioritisation strategy (as detailed in \citealt{Bunker23b}).
The selection of the highest redshift candidates, which were initially identified as dropout candidates from HST imaging at $z>5.7$ (these include I$_{775}$, Y$_{105}$ and J$_{125}$-band dropouts), used a rest-frame UV magnitude threshold longward of the Lyman-$\alpha$ break (rest-frame UV detection band AB magnitude $\lesssim29$) with prioritisation based on their photometric redshift. 
Over a sub-set of the area covered by the MSA, JWST NIRCam imaging was also available prior to the MSA design and we used this to refine the selection with better S/N at longer wavelengths\footnote{Those galaxies in \citet{Bunker23b} with an 8-digit ID starting with \lq100\rq\ were selected purely from HST imaging, while those with shorter IDs were selected from HST and JWST/NIRCam images.}.
We note that, as expected, all the high-redshift ($z>5.7$) EELGs we identify come from targets selected to be in the photometric redshift range $5.7<z<8.5$ (priority classes 4, 6.1 and 6.2 in \citealt{Bunker23b} depending on their HST or JWST/NIRCam magnitude) or $z>8.5$ (priority class 1 in \citealt{Bunker23b}), and of the galaxies targeted in each of these classes roughly half proved to be an EELG (see Section \ref{sec:indentify}).
As we are essentially selecting objects on the rest-UV continuum, in a spectral region without strong line emission (filters used for the rest-frame UV typically do not encompass CIII]$\lambda1909$), we are not biasing our magnitude limited sample towards strong line emitters, although we consider the selection effects due to SFH in Section \ref{sec:beta}. 

At lower redshifts ($z<5.7$), priority was given to obtaining a stellar mass-limited sample, by selecting on the longest wavelength broadband filter with good imaging data. In the case of HST selection this was the F160W HST/WFC3 (AB $<29$ mag), but for a subset of the NIRSpec footprint NIRCam imaging was available prior to designing the MSA configuration, enabling selection on the rest-frame optical F444W (AB $<27.5$ mag). The majority of our identified low-redshift EELG sample comes from targets selected to be in classes 7.5, 7.6 and 7.7 in \citet{Bunker23b} (corresponding to redshift slices $4.5<z<5.7$, $3.5<z<4.5$ and $2.5<z<3.5$ respectively), except for one object from class 6.2 (ID:8113 at $z=4.90$) where the spectroscopic redshift was slightly below the priority class expectation, one object in class 7.2 (ID:4270, $z=4.02$, up-weighted for scientific interest) which was selected as a potential quiescent galaxy from the photometry SED but we now can see was driven by line contamination rather than a strong Balmer/4000\AA\ break, and one object below the stated magnitude selection which was included as a filler (class 9, ID:10001916 at $z=4.28$). From the class 7.5, 7.6 and 7.7 targeted galaxies the fraction of EELGs we detect are 8/23, 6/31, 3/45, see Section \ref{sec:indentify}.

We note that the selection on the rest-frame optical means that over specific redshift ranges, the broadband flux may be boosted by the contribution from strong line emission (H$\alpha$, [OIII]+H$\beta$) which could bias the lower-redshift sample towards EELGs at faint magnitudes. We will discuss in Section \ref{sec:beta} how we can apply an apparent rest-UV magnitude cut to this sub-sample to address this selection effect. 

We also note that the rest-optical selection (at $z<5.7$) will include galaxies with less active star formation, which may account for the lower fraction of galaxies with detected emission lines at $z<6$, seen in Figure \ref{fig:redshift_dist}.

\section{Measurement of strong line equivalent widths}\label{sec:method}

To determine the EW for emission lines detected in the prism and R1000 grating spectroscopy, we require measurements of the line flux and the continuum flux density at the wavelength of the line. 

We use the line fluxes reported in \citet{Bunker23b} for the JADES DEEP spectroscopy, which relied on Gaussian modelling of the emission lines (with the width tied for blended lines) after continuum subtraction. All lines detected at a $>5\sigma$ significance are reported in the publicly available catalogue\footnote{\url{https://archive.stsci.edu/hlsp/jades}}, and where necessary (e.g., the line diagnostic diagrams in Section \ref{sec:line_trends}), we additionally make use of line fluxes below this significance threshold, measured the same way.

The continuum flux density at the wavelength of the lines can be determined either directly from the spectroscopy or from aperture photometry of the NIRCam imaging. We choose to measure the continuum flux density from NIRCam photometry as this provides a better constraint on the continuum for faint sources, where the continuum is often below the sensitivity of the spectroscopy. To determine the continuum flux density from the NIRCam imaging we first wish to avoid requiring a path loss correction (to account for slit losses associated with the NIRSpec MSA) and hence the continuum flux densities are derived from apodised photometry, with the aperture set to match the placement of the MSA shutter over the target.  A point-spread function (PSF) correction is then applied to account for the difference in PSF between the NIRCam and NIRSpec observations. \citet{Bunker23} show the consistency between the spectral flux through the microshutter and the apodised multiple filter NIRCam photometry over the same wavelength ranges.  The continuum flux density around the wavelength of the emission line is derived assuming a flat in $f_\nu$ continuum (matching the typical observed $\beta$ slopes in our full sample, 
see Section \ref{sec:beta}) after removing any contribution from spectroscopically detected line emission to the broadband filter, accounting for the wavelength-dependent filter transmission. The observed-frame EW for each line is then taken as the ratio of the individual line flux to the continuum flux density ($f_\lambda$) at the location of the line. This is then corrected to the rest-frame EW value by dividing by $(1+z)$.

The use of broadband photometry to determine the continuum flux density allows EWs to be determined for faint galaxies where constraints cannot be placed directly from spectroscopy. For a sub-set of continuum bright objects, where the continuum is well detected in the spectra, we check whether both methods recover consistent EW values, and present this work in Appendix \ref{sec:App:ew}. From this, we determine consistent EWs measurements for galaxies with a signal to noise per pixel in continuum greater than 3. It is clear that the uncertainty in the spectrum-derived EW increases as the continuum becomes fainter, and this supports our choice to use the broadband photometry to provide tighter constraints on the EWs. 

When computing the continuum flux density from the broadband photometry, we use the filter that contains the selected emission line. However, in a small number of cases, we do not have filter coverage at the wavelength of investigation. 
This accounts for 10/86 of our [OIII]$\lambda5007$ EW measurements. In these cases, we determine an estimate of the continuum flux density at the wavelength of the line using the nearest available filter. We find that using an adjacent filter provides a consistent EW estimate, albeit with a larger scatter driven by the variation of UV spectral slopes ($\beta$) in our sample. We discuss this further in Appendix \ref{sec:App:ew}.

Finally, we compare the EWs derived using line fluxes determined from the two spectral resolution modes (prism and R1000 grating). We determine good agreement for our EWs for lines detected in both the prism and R1000 spectra, with the R1000 providing higher resolving power for blended emission lines (e.g., the [OIII]$\lambda4959, 5007$ doublet becomes blended below $z\lesssim5.2$ in the prism).
However, we note a small systematic offset between the two modes with the R1000 fluxes roughly $8\%$ larger than the prism fluxes. This offset, due to the flux calibration of the spectroscopy, has been observed previously by \citet{Bunker23} who find that the prism flux calibration better replicates the broadband flux densities. Therefore in this study, we will adopt the prism EWs as our primary method. 
We note that because of this choice, we limit our sample to $z>3$ where the H$\beta$ and [OIII] lines are unblended in the prism. 

For one target (ID:7624), two galaxies fell within the NIRSpec MSA shutter (at $z=2.7\ \&\ z=4.8$). Despite observing strong emission lines at the expected wavelengths, the significant contribution of continuum flux from both systems means the emission line EW can not be determined from either the spectrum or the photometry. We therefore remove this source from our full sample (leaving 85 galaxies) and our stacking analysis.

\subsection{Identification of extreme emission line galaxies}
\label{sec:indentify}
There is no set EW threshold (or even specific emission line) for defining a star-forming galaxy as being in an EELG phase, with adopted thresholds varying from study to study. Commonly taken thresholds can be based on the EW of individual lines (e.g., [OIII]$\lambda5007$ or H$\alpha$) or complexes (e.g., [OIII]+H$\beta$) depending on the spectroscopic resolution of observations or if relying on broadband flux-excess. Threshold values for [OIII]$\lambda5007$ EW in literature have ranged from $\sim100-1000$\,\AA\,  \citep[e.g.,][]{Amorin15, van_der_Wel_2011, Mengtao19, Mengtao20, Du20}. 

In this paper, we will adopt [OIII]$\lambda5007$ EW $>750$\AA\ as our threshold, based on work by \citet{Boyett22A} and \citet{Mengtao19}. This choice is motivated by $z\sim2$ studies which show significantly different interstellar medium (ISM) characteristics (high ionisation parameter and ionisation efficiency as well as potentially high escape fractions of ionising radiation) above this threshold compared to the general SFG population at this epoch \citep{Mengtao19}. 

We identify 35 galaxies that meet this threshold for [OIII]$\lambda5007$ in our sample, with one additional galaxy consistent within $1\sigma$ of the [OIII]$\lambda5007$ threshold (ID 10001892) which also has a H$\alpha$ EW above $750$\AA\, and hence is an EELG. This makes up 42$\%$ (36/85) of our sample of emission line galaxies at $z>3$. This sub-sample of 36 galaxies will hereafter be referred to as the EELG sample. All other galaxies in our full sample below this threshold will therefore be referred to as the non-EELG sample (49 galaxies).
The emission line EWs for these EELGs are presented in Table \ref{tab:EELG}. We note that one EELG, ID:8083 has an underlying broad component of H$\alpha$ \citep{Maiolino23arXiv} and likely hosts AGN. One non-EELG in our full sample, ID:10013704 similarly shows broad H$\alpha$ \citep{Maiolino23arXiv}. In Table~\ref{tab:EELG} we flag these, and in Figures \ref{fig:EW_beagle} and \ref{fig:burst_intensity_alternative} we plot these two with different symbols. However, we retain these in our sample, since the broad-line contribution to the line fluxes is small, and the narrow lines and galaxy continuum may well be dominated by star formation rather than an AGN \citep{Maiolino23arXiv}.

\begin{landscape}
\begin{table}
\centering
\begin{tabular}{ccccccccccccc}
ID & z$_{\rm{spec}}$ & [OIII]$\lambda5007$ EW & H$\alpha$ EW & Ly$\alpha$ EW & M$_{UV}$ & $A_{1600}$ & $\beta_{\rm{obs}}$ &  O32 & R23 & $\log_{10}(\xi_\mathrm{ion}^\mathrm{HII}$/ & $\log_{10}\mathrm{(O/H)}$\\ 
& & \AA & \AA & \AA & & & & & & $\rm{erg^{-1}Hz})$ \\
\hline
10058975  & 9.44  &  $856\pm75$  &  --  &  <6 (P)  &  $-20.4\pm0.1$ &  $0.0\pm0.1$  &  $-2.59\pm0.01$  &  $41.3\pm8.0$  &  $6.0\pm0.2$  &  -- &  $7.4\pm0.1$  \\
\textbf{21842}  & 7.98  &  $1246\pm37$  &  --  &  $24\pm7$ (M)  &  $-18.7\pm0.1$ &  --  &  $-2.23\pm0.01$  &  --  &  --  &  -- &  $7.6\pm0.1$  \\ 
\textbf{10013682$^\ast$}  & 7.28  &  $2767\pm763$  &  --  &  $148\pm20$ (M)  &  $-15.7\pm1.4$ &  --  &  $-1.85\pm0.04$  &  --  &  --  &  -- &  $7.7\pm0.1$  \\ 
10013905  & 7.21  &  $844\pm77$  &  --  &  <32 (P)  &  $-18.8\pm0.1$ &  $0.0\pm0.8$  &  $-2.46\pm0.01$  &  $20.0\pm3.3$  &  $6.8\pm0.4$  &  -- &  --  \\ 
4297  & 6.72  &  $1971\pm39$  &  $1721\pm88$  &  <63 (P)  &  $-18.5\pm0.1$ &  $0.0\pm0.3$  &  $-2.29\pm0.02$  &  $12.1\pm2.0$  &  $8.9\pm0.6$  &  $25.6\pm0.1$ &  $8.0\pm0.2$  \\ 
3334  & 6.71  &  $1620\pm72$  &  $1558\pm307$  &  <31 (P)  &  $-18.1\pm0.1$ &  $1.5\pm0.6$  &  $-2.08\pm0.02$  &  $9.2\pm4.4$  &  $14.0\pm5.4$  &  $25.2^{+0.1}_{-0.2 }$ &  $7.8\pm0.1$  \\ 
\textbf{16625}  & 6.63  &  $1444\pm29$  &  $1978\pm97$  &  $39\pm17$ (M)  &  $-18.8\pm0.1$ &  $0.4\pm0.2$  &  $-2.29\pm0.01$  &  $>25.9$  &  $<5.1$  &  $25.6\pm0.1$ &  $7.3\pm0.1$  \\ 
\textbf{18846}  & 6.34  &  $1068\pm21$  &  $1185\pm24$  &  $39\pm4$ (M)  &  $-20.2\pm0.1$ &  $0.0\pm0.1$  &  $-2.53\pm0.01$  &  $38.8\pm7.3$  &  $6.4\pm0.2$  &  $25.3\pm0.1$ &  $7.5\pm0.1$  \\ 
18976  & 6.33  &  $839\pm30$  &  $620\pm43$  &  <20 (P)  &  $-18.7\pm0.1$ &  $0.0\pm0.3$  &  $-2.30\pm0.01$  &  $>16.2$  &  $<6.1$  &  $25.4\pm0.1$ &  $7.5\pm0.1$  \\ 
\textbf{19342}  & 5.98  &  $2090\pm72$  &  $1072\pm76$  &  $51\pm22$ (M)  &  $-18.7\pm0.1$ &  $0.0\pm0.2$  &  $-2.51\pm0.01$  &  $>22.1$  &  $<6.0$  &  $25.4\pm0.1$ &  $7.5^{+0.2}_{-0.1 }$  \\ 
\textbf{9422$^\ddagger$}  & 5.94  &  $3253\pm65$  &  $2214\pm44$  &  $130\pm23$ (M)  &  $-19.7\pm0.1$ &  $0.0\pm0.1$  &  $-2.31\pm0.01$  &  $38.7\pm4.3$  &  $7.7\pm0.2$  &  $25.6\pm0.1$ &  --  \\ 
\textbf{6002}  & 5.94  &  $870\pm20$  &  $619\pm26$  &  $44\pm20$ (M)  &  $-18.8\pm0.1$ &  $0.3\pm0.2$  &  $-2.52\pm0.01$  &  $8.9\pm1.7$  &  $8.6\pm1.1$  &  $25.2\pm0.1$ &  $7.7\pm0.1$  \\ 
\textbf{19606}  & 5.89  &  $962\pm21$  &  $941\pm36$  &  $84\pm10$ (M)  &  $-18.6^{+0.1}_{-0.2 }$ &  $0.3\pm0.3$  &  $-2.60\pm0.02$  &  $15.9\pm6.0$  &  $8.1\pm1.7$  &  $25.5\pm0.1$ &  $7.7\pm0.1$  \\ 
\textbf{10056849}  & 5.82  &  $1022\pm209$  &  $1249\pm256$  &  $77\pm20$ (M)  &  $-18.0\pm0.1$ &  $0.0\pm0.1$  &  $-2.52\pm0.01$  &  $>20.5$  &  $<4.4$  &  $25.6\pm0.1$ &  $7.4\pm0.1$  \\ 
10005113  & 5.82  &  $2107\pm103$  &  $1565\pm109$  &  <59 (P)  &  $-18.0\pm0.1$ &  $0.0\pm0.2$  &  $-2.23\pm0.02$  &  $>12.8$  &  $<4.9$  &  $25.6\pm0.1$ &  $7.8^{+0.3}_{-0.2 }$  \\ 
22251  & 5.80  &  $1318\pm26$  &  $857\pm18$  &  <16 (P)  &  $-19.0\pm0.1$ &  $0.7\pm0.2$  &  $-2.02\pm0.01$  &  $12.3\pm1.6$  &  $9.8\pm1.1$  &  $25.6\pm0.1$ &  $7.9\pm0.1$  \\ 
4404  & 5.78  &  $1035\pm21$  &  $793\pm26$  &  <16 (P)  &  $-19.3\pm0.1$ &  $0.0\pm0.1$  &  $-2.31\pm0.01$  &  $11.7\pm1.5$  &  $6.4\pm0.3$  &  $25.4\pm0.1$ &  --  \\ 
10016374  & 5.51  &  $1758\pm186$  &  $1337\pm143$  &  <19 (P)  &  $-18.7\pm0.1$ &  $0.4\pm0.2$  &  $-2.21\pm0.01$  &  $10.1\pm1.6$  &  $10.1\pm1.3$  &  $25.5\pm0.1$ &  $7.8\pm0.1$  \\
9743  & 5.45  &  $897\pm35$  &  $996\pm40$  &  <89(P)  &  $-17.4^{+0.7}_{-4.1 }$ &  $0.7\pm0.5$  &  $-1.30\pm0.05$  &  $>6.7$  &  $<7.3$  &  $26.0\pm0.5$ &  --  \\ 
10015338  & 5.07  &  $1639\pm187$  &  $1712\pm202$  &  <18.8 (P)  &  $-19.5\pm0.1$ &  $0.7\pm0.4$  &  $-2.42\pm0.01$  &  $>14.2$  &  $<8.1$  &  $25.3\pm0.1$ &  $7.6\pm0.1$  \\ 
8113  & 4.90  &  $943\pm19$  &  $739\pm17$  &  <49 (P)  &  $-18.1\pm0.2$ &  $0.5\pm0.2$  &  $-1.69\pm0.01$  &  $6.8\pm1.0$  &  $10.2\pm1.4$  &  $25.6\pm0.1$ &  --  \\ 
10005217  & 4.89  &  $1086\pm24$  &  $1546\pm37$  &  <34 (P)  &  $-18.1\pm0.1$ &  $0.0\pm0.1$  &  $-2.32\pm0.01$  &  $17.8\pm2.0$  &  $4.6\pm0.3$  &  $25.9\pm0.1$ &  --  \\ 
\textbf{7938}$^\star$  & 4.82  &  $891\pm18$  &  $620\pm12$  &  <8 (P$^\star$)  &  $-19.4\pm0.1$ &  $0.5\pm0.2$  &  $-2.40\pm0.01$  &  $11.6\pm1.9$  &  $11.2\pm1.3$  &  $25.2\pm0.1$ &  $7.8\pm0.1$  \\ 
18090  & 4.79  &  $976\pm20$  &  $768\pm15$  &  <22 (P)  &  $-18.9\pm0.1$ &  $0.3\pm0.2$  &  $-1.92\pm0.01$  &  $9.2\pm1.4$  &  $11.6\pm1.5$  &  $25.6\pm0.1$ &  $7.8\pm0.1$  \\ 
10001892  & 4.77  &  $663\pm176$  &  $1113\pm308$  &  <68 (P)  &  $-16.0^{+0.5}_{-0.8 }$ &  --  &  $-1.69\pm0.04$  &  --  &  --  &  -- &  --  \\
\textbf{8083}$^\dag$  & 4.67  &  $1808\pm36$  &  $1362\pm27$  &  $20\pm4$ (P)  &  $-18.7\pm0.1$ &  $1.2\pm0.1$  &  $-1.57\pm0.01$  &  $18.5\pm2.7$  &  $10.2\pm1.4$  &  $25.9\pm0.1$ &  --  \\ 
\textbf{10000626}  & 4.47  &  $763\pm42$  &  $923\pm45$  &  $185\pm71$ (P)  &  $-17.0^{+0.4}_{-0.6 }$ &  $0.0\pm0.6$  &  $-2.13\pm0.04$  &  $>5.3$  &  $<7.2$  &  $25.5^{+0.2}_{-0.3 }$ &  $7.3\pm0.1$  \\ 
10001916  & 4.28  &  $823\pm87$  &  $1322\pm202$  &  <373 (P)  &  $-16.7^{+0.5}_{-0.8 }$ &  --  &  $-1.80\pm0.08$  &  --  &  --  &  -- &  $7.2\pm0.2$  \\ 
7892  & 4.25  &  $820\pm16$  &  $677\pm14$  &  <14 (P)  &  $-18.5\pm0.1$ &  $0.0\pm0.2$  &  $-2.15\pm0.01$  &  $12.4\pm0.7$  &  $7.5\pm0.3$  &  $25.5\pm0.1$ &  $7.8\pm0.1$  \\ 
\textbf{6519}  & 4.24  &  $1461\pm31$  &  $849\pm18$  &  $46\pm7$ (P)  &  $-19.2\pm0.1$ &  $0.5\pm0.4$  &  $-2.25\pm0.01$  &  $20.0\pm5.8$  &  $12.0\pm2.8$  &  $25.3\pm0.1$ &  $7.7\pm0.1$  \\ 
\textbf{7507}  & 4.04  &  $765\pm20$  &  $738\pm15$  &  $44\pm8$ (P)  &  $-18.8\pm0.1$ &  $0.0\pm0.2$  &  $-2.44\pm0.01$  &  $>11.1$  &  $<6.0$  &  $25.3\pm0.1$ &  $7.5\pm0.1$  \\ 
4270  & 4.02  &  $1030\pm21$  &  $814\pm16$  &  <18 (P)  &  $-18.9\pm0.1$ &  $1.7\pm0.2$  &  $-1.68\pm0.01$  &  $6.1\pm1.2$  &  $10.3\pm2.0$  &  $25.8\pm0.1$ &  $8.0\pm0.1$  \\ 
10035295  & 3.58  &  $1255\pm43$  &  $870\pm25$  &  --  &  $-18.1\pm0.1$ &  $0.7\pm0.4$  &  $-1.92\pm0.01$  &  $32.7\pm10.1$  &  $13.4\pm3.4$  &  $25.7\pm0.1$ &  $7.6^{+0.3}_{-0.1 }$  \\ 
2651  & 3.58  &  $840\pm113$  &  $529\pm36$  &  --  &  $-16.3\pm1.0$ &  --  &  $-1.42\pm0.05$  &  --  &  --  &  -- &  --  \\ 
18322  & 3.16  &  $1363\pm41$  &  $1432\pm29$  &  --  &  $-16.9^{+0.4}_{-0.6 }$ &  $0.6\pm0.2$  &  $-2.19\pm0.05$  &  $9.6\pm1.9$  &  $9.4\pm1.4$  &  $25.9^{+0.2}_{-0.3 }$ &  --  \\ 
21150  & 3.09  &  $816\pm16$  &  $511\pm11$  &  --  &  $-19.2^{+0.5}_{-1.0 }$ &  $0.5\pm0.3$  &  $-2.36\pm0.01$  &  $5.4\pm1.0$  &  $10.3\pm1.7$  &  $25.3^{+0.2}_{-0.5 }$ &  $8.1\pm0.1$  \\
\hline 
\end{tabular}
\caption{ Extreme emission line galaxies identified within our sample, in descending redshift order, selected on a rest-frame [OIII]$\lambda5007$ EW $>750$\AA. We additionally include one galaxy (ID:10001892) consistent with this threshold within the measured uncertainties. Lyman-$\alpha$ EWs for our high redshift sample ($z>5.7$) are taken from \citet{Jones23arXiv} and are marked (P, M) depending if the line flux was determined from the prism or R1000 medium resolution grating. Lyman-$\alpha$ EWs for galaxies at lower redshifts are determined for this work using the same method as \citet{Jones23arXiv}. $\ast$ Galaxy ID: 10013682 is an extremely high EW Ly$\alpha$ emitter which was first discussed in \citet{Saxena23}. $\ddagger$ ID 9422 may be a nebular dominated galaxy (see \citealt{Cameron23arXiv}). $\bigstar$ We note that while we can only place an upper limit on the Lyman-$\alpha$ EW of galaxy ID: 7938 from NIRSpec spectroscopy, we confirm Lyman-$\alpha$ detection through legacy MUSE observations. $\dag$ flagged as a potential AGN \citep{Maiolino23arXiv}. The galaxy coordinates, photometry and emission line measurements for these galaxies are publicly available and have been presented in \citet{Bunker23} and \citet{Rieke23b}.}
\label{tab:EELG}
\end{table}
\end{landscape}

\subsection{Distribution of equivalent widths}
\begin{figure*}
    \centering
    \includegraphics[width=\textwidth]{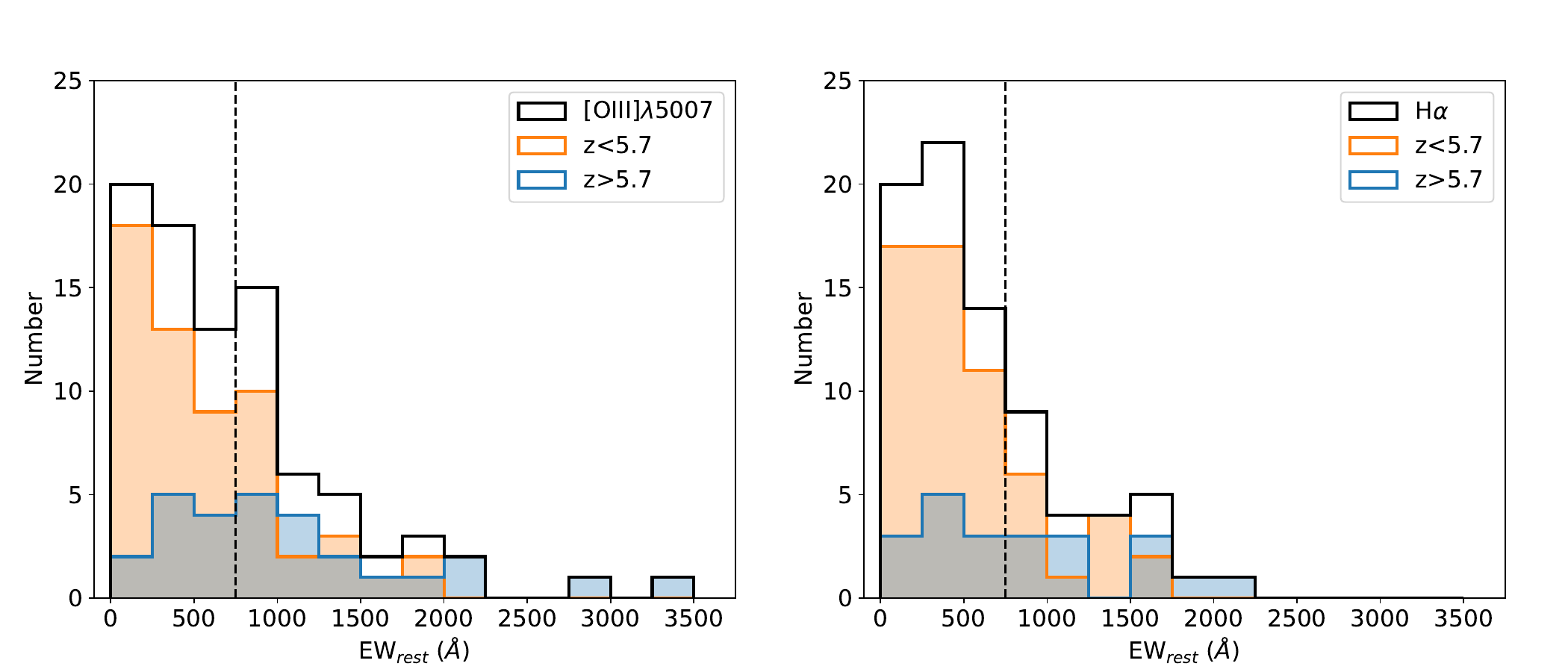}
    \caption{Distribution of equivalents widths. Left panel presents the [OIII]$\lambda5007$ EW distribution for the whole sample (black) and split into a high redshift ($z>5.7$, blue) and low redshift ($z<5.7$, orange) sub-samples. Right panel shows the same distributions for the H$\alpha$ EW. In each panel the vertical dashed line represents the EELG threshold of an EW $>750$\AA.}
    \label{fig:EW_dist}
\end{figure*}

In Figure \ref{fig:EW_dist}, we present the EW distribution of the [OIII]$\lambda5007$ and H$\alpha$ EWs.
We split our full sample into two redshift bins, those above and below $z=5.7$, dividing the EELG sample into two at the end of the EoR.
We expect that the galaxies within our sample at high redshift ($z>5.7$), which were selected based on the rest-UV, should avoid any selection bias towards high EW galaxies, whereas lower redshift galaxies within our sample are selected on the rest-optical and hence may preferentially include galaxies with large EWs in their rest-optical emission lines (see Section \ref{sec:selection}).
However, although there are even numbers of EELGs from the high and lower redshift samples, the lower-redshift sample contains a larger total number of galaxies (i.e. the EELG fraction is lower than in the higher redshift sample), suggesting significant evolution in the EELG fraction with redshift (discussed further in Section \ref{sec:disc:evolution}).

The EW distribution of H$\alpha$ and [OIII]$\lambda5007$ show similar profiles (see Figure \ref{fig:EW_dist}).
The high-redshift sample exhibits a flatter EW distribution which extends to high EW (EW$>2000$\AA), whereas at low-redshift the distribution peaks around low EW and while there is a high-EW tail this represents only a small fraction of the total population.

\section{Galaxy properties of EELGs}\label{sec:line_trends}

In this Section, we are now interested in characterising the physical properties of our EELG sample, both individually and averaged over two redshift bins split at $z=5.7$. We will compare these characteristics against our non-EELG sample to examine what attributes may be responsible for a galaxy entering an EELG phase. We will first discuss the combination of several spectra (stacking) in sub-section \ref{sec:stack}, then examine the galaxy properties in sub-sections \ref{sec:ISM_properties}-\ref{sec:beagle}. 

\subsection{Stacked spectra of EELGs}
\label{sec:stack}
\begin{figure*}
    \centering
    \includegraphics[width=\textwidth]{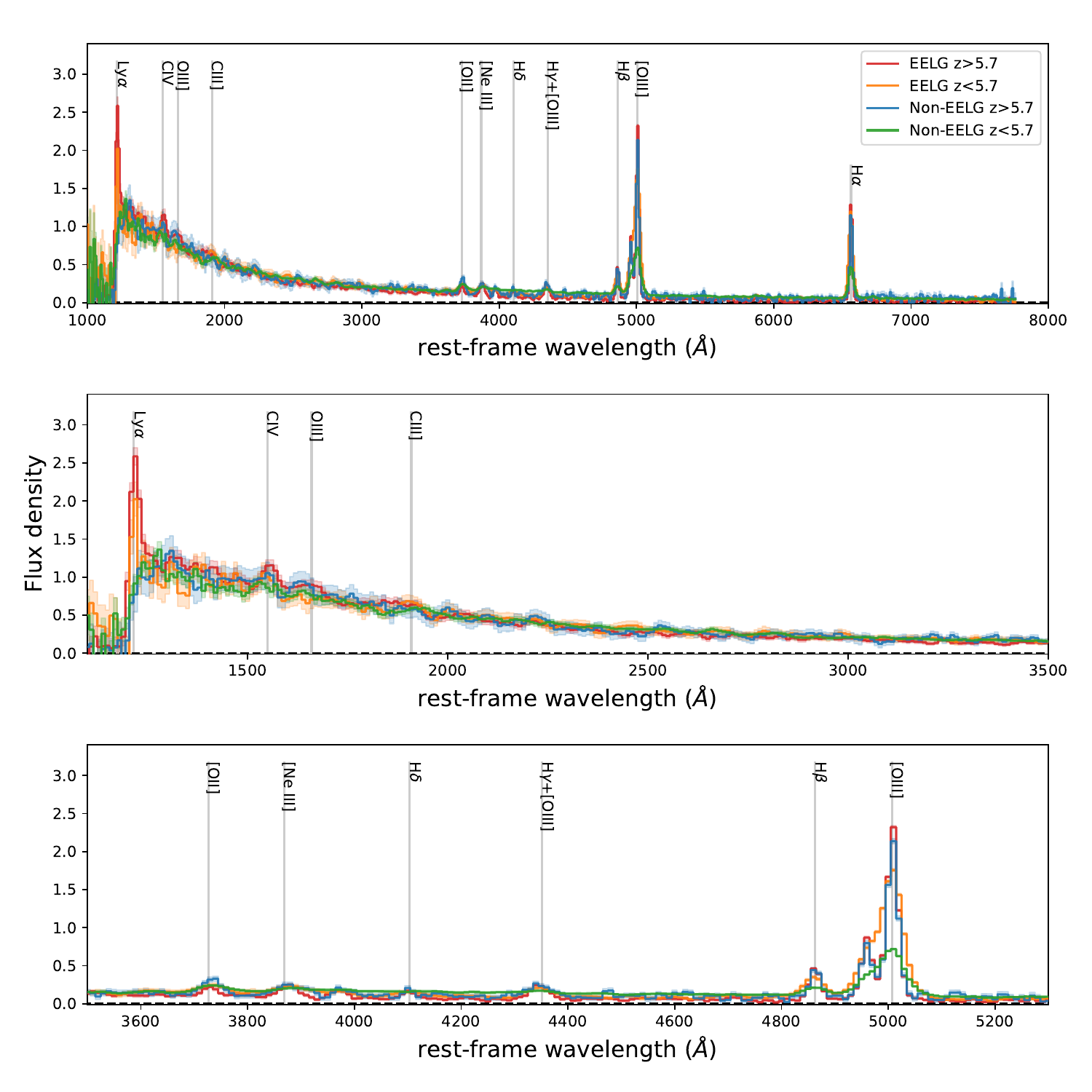}
    \caption{Stacks of the galaxy sub-samples split by redshift (at $z=5.7$) and equivalent width (at [OIII]$\lambda5007$ EW $=750$\AA). The $z>5.7$ EELG sample is shown in red and the $z<5.7$ EELG sample in orange. This redshift split naturally separates spectra where the [OIII] doublet is resolved or not. For the lower redshift sample, the spectral resolving power of the prism around [OIII] is lower. In blue and green we show the respective non-EELG samples at high and low redshifts. Top panel: The full spectral range. Middle panel: The rest-UV. Bottom panel: The rest-optical. }
    \label{fig:stacks}
\end{figure*}

\begin{table*}
\centering
\resizebox{.75\textwidth}{!}{%
\begin{tabular}{ccccc}
Stack             & high-z EELG & low-z EELG & high-z non-EELG & low-z non-EELG \\ \hline
N galaxies     &  17  &  19  &  11  &  37  \\ 
$[$OIII$]\lambda5007$ EW  &  1524$\pm$201  &  1482$\pm$123  &  572$\pm$59  &  221$\pm$7  \\ 
H$\alpha$ EW   &  1129$\pm$158  &  956$\pm$85  &  400$\pm$45  &  187$\pm$3  \\ 
Ly$\alpha$ EW   &  39$\pm$3  &  21$\pm$5  &  $<17$  &  $<29$ \\ 
H$\alpha$/H$\beta$  &  2.98$\pm$0.11  &  3.20$\pm$0.13  &  3.00$\pm$0.32  &  3.40$\pm$0.22  \\ 
A1600  &  0.15$\pm$0.14  &  0.42$\pm$0.15  &  0.18$\pm$0.41  &  0.65$\pm$0.24  \\ 
O32$^*$   &  25.8$\pm$4.7  &  19.3$\pm$4.1  &  6.1$\pm$1.6  &  6.01$\pm$1.11  \\ 
R23$^*$  &  7.6$\pm$0.7  &  8.3$\pm$0.9  &  9.8$\pm$2.4  &  9.47$\pm$1.51  \\ 
Ne3O2$^*$   &  2.9$\pm$0.6  &  1.0$\pm$0.3  &  $<1.4$  &  $<0.4$  \\ 
12+$\log_{10}\mathrm{(O/H)}$ & $7.53^{+0.07}_{-0.10}$ & $7.66\pm0.14$ & $7.95^{+0.32}_{-0.17}$ & $7.96^{+0.25}_{-0.19}$  \\
$\log_{10}(\xi_\mathrm{ion}^\mathrm{HII})$  &  25.49$\pm0.03$  &  25.66$\pm0.04$  &  25.50$^{+0.08}_{-0.10}$  &  25.37$^{+0.05}_{-0.06}$  \\ 
$\beta_{obs}$  &  -2.36$\pm$0.03  &  -1.97$\pm$0.04  &  -2.18$\pm$0.05  &  -1.87$\pm$0.02  \\ 
$\beta_{int}^*$  &  -2.45$\pm$0.11  &  -2.20$\pm$0.12  &  -2.28$\pm$0.28  &  -2.22$\pm$0.15  \\ 
\hline
\end{tabular}
}
\caption{Galaxy properties derived for our stacked spectra. Properties marked with ($^*$) have been corrected for dust. Upper limits are given at $3\sigma$. }
\label{tab:stack}
\end{table*}

In this Section, we stack the prism spectra for four different sub-samples of our dataset, split by redshift ($z=5.7$) and whether a galaxy meets the threshold to be considered an EELG ([OIII]$\lambda5007$ or H$\alpha$ EW $>750$\AA). There are 17 EELGs in our high-redshift sample and 19 in our low-redshift sample. We additionally have 11 high-redshift non-EELGs and 38 low-redshift non-EELGs (excluding ID:7624 as described in Section \ref{sec:method}) that we can use as a control sample, although we note that these still require at least two emission lines detected at $>5\,\sigma$.

We note that in our low-redshift non-EELG sample there is one galaxy (ID:4493, $z=3.59$) with an extremely red UV spectral slope of $\beta=+0.65\pm0.03$, the only galaxy with a positive $\beta$, which is an outlier at $\gg3\sigma$ from the full sample. We exclude this highly unusual galaxy from our stack, leaving 37 galaxies in our low-redshift non-EELG stack.

\begin{figure}
    \centering
    \includegraphics[width=\columnwidth]{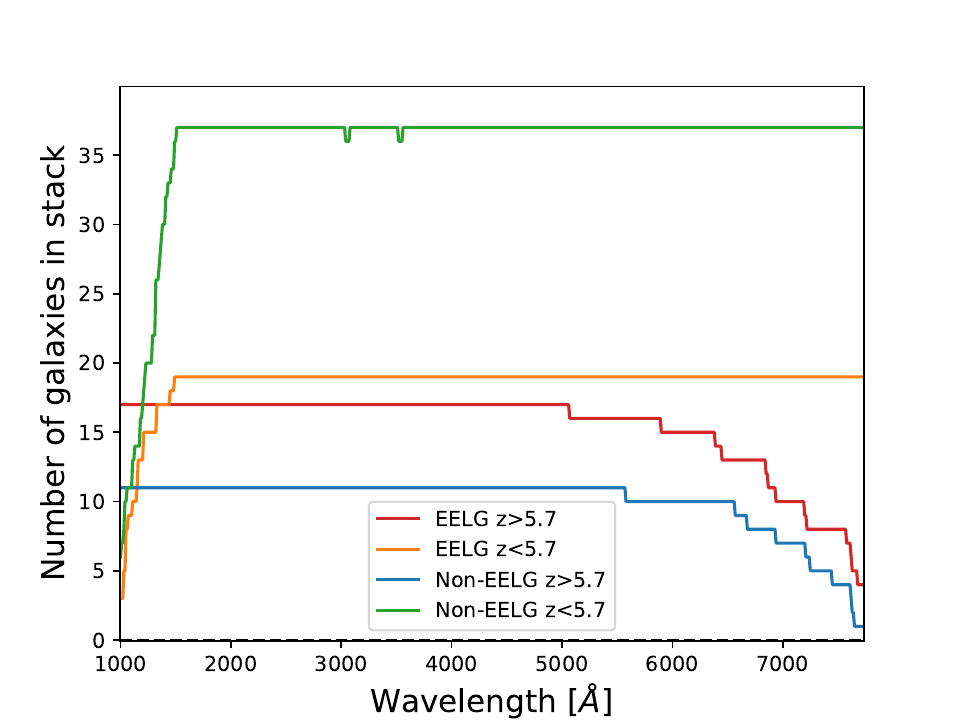}
    \caption{The number of galaxies contributing to each of our four stacks as a function of wavelength. The colours match those in Figure \ref{fig:stacks}. Each individual galaxy has a spectrum that covers a unique observed wavelength range, due to their redshift, the position of the dispersed light on the detector and any bad pixels/contamination. Therefore the number of galaxies contributing to each rest-frame wavelength in the stack varies. }
    \label{fig:stack_number}
\end{figure}

We de-redshift each individual prism spectrum to the rest-frame (with the redshift determined from the detected [OIII]$\lambda5007$ emission line), interpolating onto a common wavelength grid of 10\,\AA\ pixels, and normalise to the integrated 1400-1600\AA\, flux ($f_\lambda$), where possible determined from the NIRSpec prism spectrum (if the S/N is sufficient).
For redshifts below $z=4$ the NIRSpec spectrum was not sensitive to this portion of the rest-frame UV so we instead used the rest-UV flux inferred from the broadband photometry. At higher redshifts ($z>4$), if the spectroscopic continuum is only weakly detected in the rest-UV (integrated S/N$<3$ in the range 1400-1600\AA) we again adopt the M$\rm_{UV}$ inferred from the broadband imaging, as described in the Appendix \ref{sec:App:muv}, this only applied to 2 of our galaxies (ID 10009693 and 10009320).
We then average these UV normalised spectra; we note that because of the different redshifts of each galaxy, we do not have uniform coverage of all wavelengths (as shown in Figure~\ref{fig:stack_number}) and this is taken into account in the averaging. 
The top panel of Figure \ref{fig:stacks} presents the four sub-sets over the full wavelength range, and we show individual zooms around the Lyman-$\alpha$ and [OIII]+H$\beta$ wavelengths in the lower panels. In Appendix \ref{sec:App:weighted_stack} we consider a different stacking where we weight by the rest-frame UV luminosity and find consistent results. 

To measure the emission line fluxes from the stacks, we note that the combination of galaxies at different redshifts, and hence at different resolving power, means a Gaussian profile no longer provides an ideal fit. We instead measure the integrated line flux rather than attempting to fit a Gaussian profile. We report the galaxy properties of the stacks in Table \ref{tab:stack} and we will discuss the properties in the following sub-sections, along with a discussion of the distribution of individual galaxy measurements. 

\subsection{ISM properties from emission line diagnostics}\label{sec:ISM_properties}
In this section, we look at the physical properties of the individual galaxies (both for EELGs and non-EELGs) in our full sample and our stacked sub-samples. In sub-section \ref{sec:balmerdec} we examine the detected hydrogen emission lines and we consider how the dust properties,  [OIII]$\lambda5007$/H$\alpha$ flux ratio, H$\alpha$ and H$\beta$ EWs, and the detection of Lyman-$\alpha$ emission vary across our sample. In sub-section \ref{sec:ionisation} we examine how the ionisation parameter, metallicity and ionising photon production efficiency change with [OIII]$\lambda5007$ EW.  

\subsubsection{Hydrogen lines and the Balmer decrement}
\label{sec:balmerdec}
In many cases in our full sample (61/85), we detect both H$\alpha$ and H$\beta$ at $S/N>5$, and we will first use these lines to estimate the reddening of the nebular emission using the Balmer decrement, assuming an intrinsic H$\alpha$/H$\beta =2.86$ (\citealt{Osterbrock_06} with electron temperature and density, $T_e=10^4K$ and $n_e=10^2cm^{-3}$, respectively). At the highest redshifts ($z>7.1$) we lose H$\alpha$ from our wavelength coverage and instead we consider the H$\beta/$H$\delta$ decrement\footnote{The H$\gamma$ Balmer line is intrinsically stronger than H$\delta$ and hence could provide a more sensitive measurement of the dust reddening. However, we do not consider the H$\gamma$ emission line because it is blended in our spectroscopy with the [OIII]$\lambda4363$ emission line.} (with an intrinsic value of 3.88, \citealt{Osterbrock_06}). 
This applies to 6 galaxies in our full sample of which 4 are EELGs.
For 2 of these galaxies (both EELGs ID: 10013905 \& 10058975) we determine a H$\beta/$H$\delta$ decrement consistent with no dust, but for the other 4 no other Balmer lines are detected leaving the dust attenuation in these galaxies unconstrained (hence we will exclude these 4 targets from any dust corrected analysis). 

We use the Balmer decrement to infer the attenuation at other wavelengths, and we adopt the \citet{Calzetti00} attenuation law (with $R_v=4.05$) assuming that the reddening of the stellar continuum is related to that of the nebular emission lines by $E(B-V)_{\mathrm {stellar}}=0.44\,E(B-V)_{\mathrm{nebular}}$.
For the galaxies within our full sample where we can constrain the dust reddening, we measure the mean dust attenuation at 1600\AA\, to be A$_{1600}= 0.48\pm0.07$. The EELG sub-sample exhibits a marginally lower level of attenuation of A$_{1600}=0.38\pm0.08$.
In Section \ref{sec:beta} we compare our measured Balmer decrements with the rest-frame UV spectral slopes, another potential indicator of dust attenuation, and we discuss the dust properties of our sample in that section.
 
We note that if the reddening of the nebular lines and stellar continuum were identical at the same wavelength, the EW measurements would not be affected by the dust reddening, unlike the observed line ratio diagnostics. However, in a scenario in which very young stars (preferentially powering nebular line emission) are still enshrouded in their birth clouds (e.g., \citealt{Charlot00}), the EW would be dependent on the differential reddening where nebular lines are attenuated to a greater extent than the surrounding stellar continuum (e.g., in the \citealt{Calzetti00} law).  
In our sample we do not correct the EW for this differential extinction, but we consider the potential effect of this in Section \ref{sec:beta}. 

\begin{figure}
    \centering
    \includegraphics[width=\columnwidth]{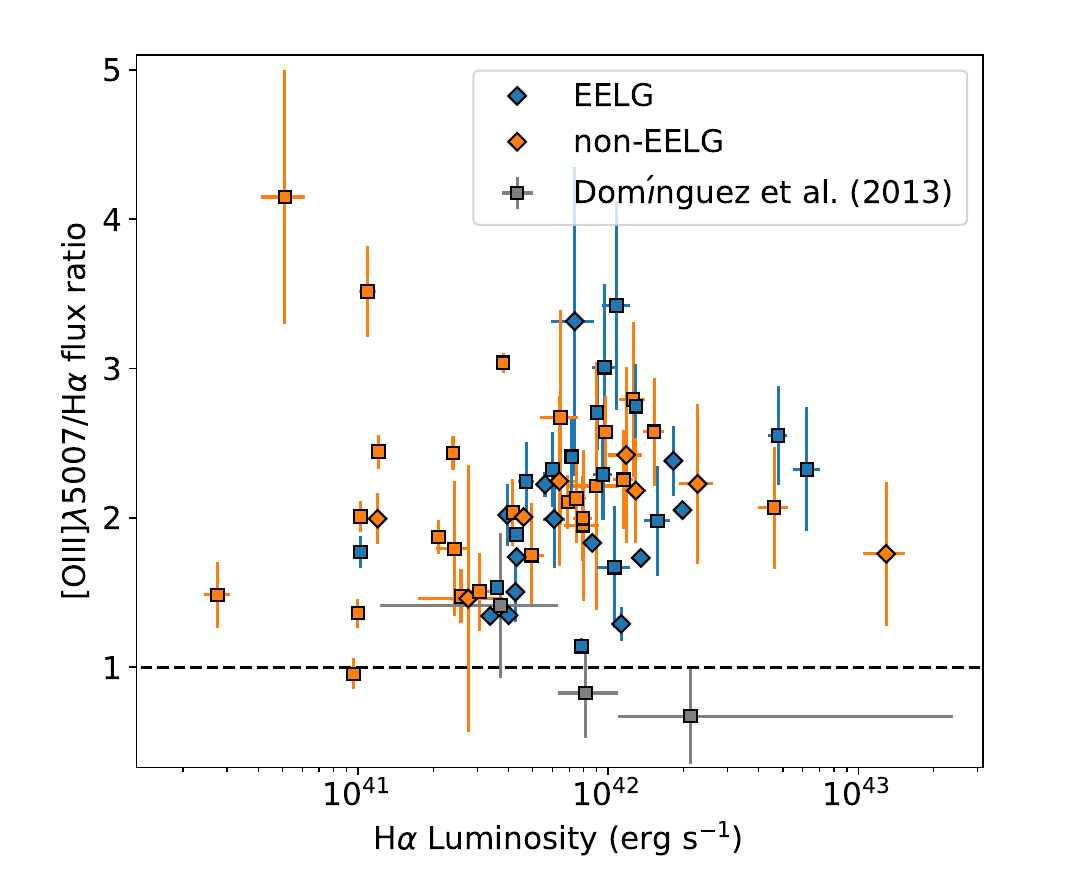}
    \caption{Dust corrected flux ratio of the [OIII]$\lambda5007$/H$\alpha$ line fluxes against the H$\alpha$ luminosity. 
    We overlay the lower redshift ($0.75<z<1.5$) \citet{Dom_nguez_2013} stacks in grey.
    The EELG sub-sample  ([OIII]$\lambda5007$ EW $>750$\AA) is coloured blue, with non-EELGs in orange. We distinguish our high- and low-redshift galaxies (split at $z=5.7$) as diamonds and squares, respectively. }
    \label{fig:OIII/Ha_ratio}
\end{figure}

We now consider the relative strength of rest-optical emission line fluxes within our sample. 
In our sample of EELGs, it is predominantly the [OIII]$\lambda5007$ line which has the highest equivalent width in these extreme systems. H$\alpha$ is usually the second strongest line in our NIRSpec spectroscopy. 
For the sub-set of 61 galaxies where both H$\alpha$ and H$\beta$ are detected above $5\,\sigma$ and a de-reddening correction can be applied, we plot the ratio of [OIII]$\lambda5007$/H$\alpha$ line fluxes in Figure \ref{fig:OIII/Ha_ratio}. As can be seen, in our sample [OIII]$\lambda5007$ is nearly always stronger than H$\alpha$, with an average flux ratio for the full sample of $f_{\rm{[OIII]}\lambda5007}/f_{H\alpha}=2.01\pm0.07$ after correction for dust reddening. Splitting our full sample into EELGs (coloured blue in Figure \ref{fig:OIII/Ha_ratio}) and non-EELGs (coloured orange), we see only a marginally higher ratio in the EELG sub-sample with a $f_{\rm{[OIII]}\lambda5007}/f_{H\alpha}$ flux ratio of $2.18\pm0.15$, compared to 
$1.93\pm0.08$ for our non-EELG sample.

The high [OIII]/H$\alpha$ flux ratio we observe at $z>3$ is in contrast to work at lower redshifts, where H$\alpha$ is often more luminous than [OIII]$\lambda5007$. For example, \citet{Dom_nguez_2013} at $z=0.75-1.5$ find a [OIII]/H$\alpha$ flux ratios spanning $1.3-0.5$, with the ratio dropping at higher H$\alpha$ luminosities (grey data points in Figure \ref{fig:OIII/Ha_ratio}). The significant increase we observe in the [OIII]$\lambda5007$/H$\alpha$ flux ratio at higher redshifts is likely attributable to decreasing metallicity and a higher ionisation parameter (e.g., \citealt{Kewley02}). Indeed, our observed flux ratio is consistent with the rising trend with redshift of [OIII]/H$\beta$ (which we correct to H$\alpha$ using the Case-B H$\alpha$/H$\beta$=2.86) at $z>3$ predicted in \citet{Cullen16}, who suggest that increasing ISM pressure and ionisation parameter is driving this.

\begin{figure*}
    \centering 
    \includegraphics[width=\textwidth]{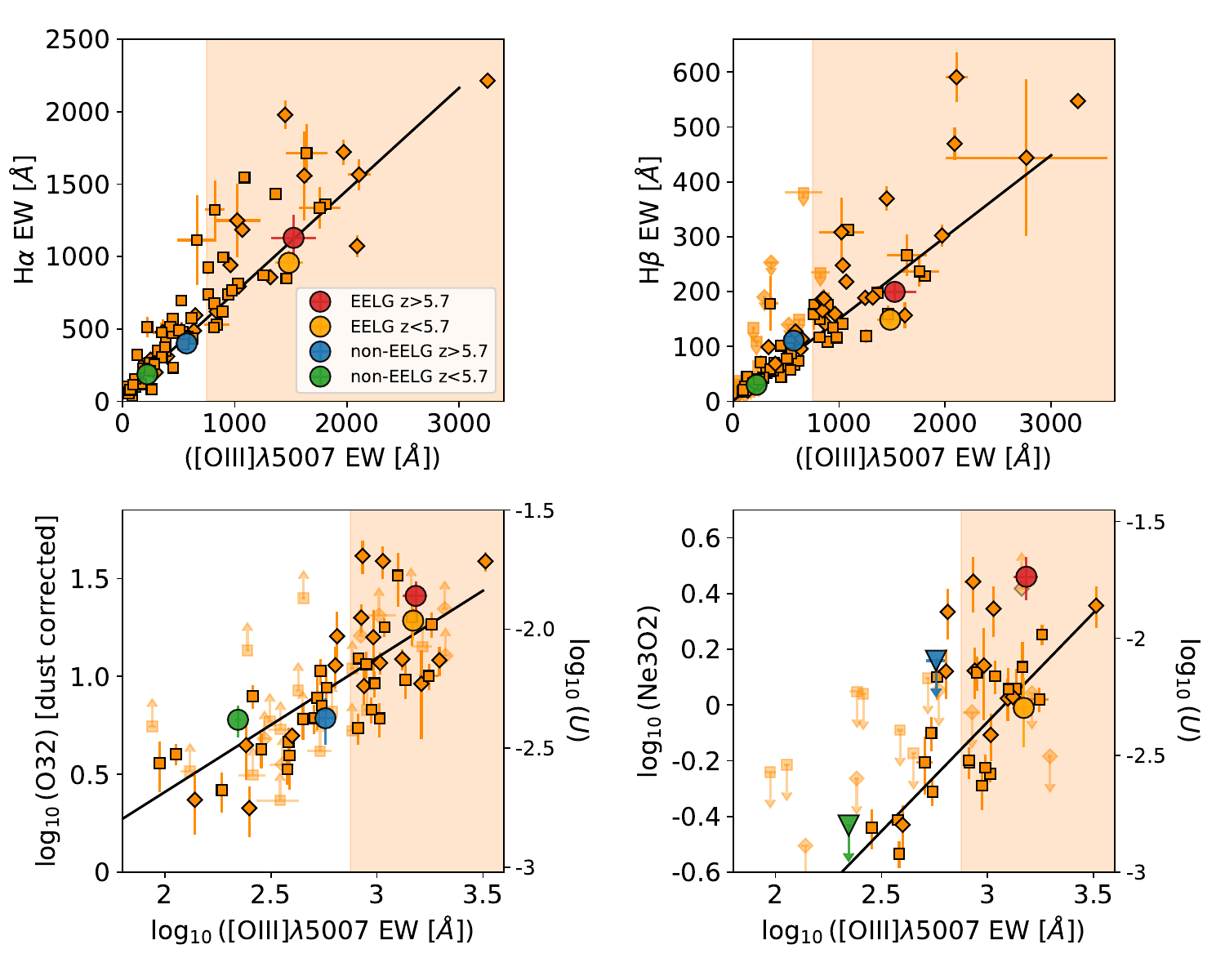}
    \caption{Top panels: Comparison of the measured [OIII]$\lambda5007$ EW with the H$\alpha$ EW and H$\beta$ EW.  Bottom panels: Comparison of the measured $\log_{10}$[OIII]$\lambda5007$ EW with the $\log_{10}$ O32 (Those with a black boarder are corrected for dust) and $\log_{10}$Ne3O2 diagnostics when all lines were detected at a $>3\sigma$ significance. Both O32 and Ne3O2 are related to the ionisation parameter $\log_{10}(U)$ and we provide an additional y-axis in each lower panel to present this, using the \citet{Witstok21} relations. In each panel, lines undetected at $3\sigma$ are replaced with a $3\sigma$ limit. We include shaded regions to mark the EELG EW threshold of [OIII]$\lambda5007 > 750$\AA, and the measured quantities for our four stacks are shown in large circles as in the legend in the top-left panel (coloured to match Figure \ref{fig:stacks}). In the bottom-right panel we change the stack symbol to triangles when Ne3O2 is given as a 3$\sigma$ upper limit. 
    We distinguish our high- and low-redshift galaxies (split at $z=5.7$) as diamonds and squares, respectively.
    }
    \label{fig:O3_ha_hb}
\end{figure*}

We can additionally consider how the EW of these strong Balmer rest-optical emission lines varies with [OIII]$\lambda5007$ EW. The hydrogen and [OIII] nebular emission lines are powered by the same sources, typically the hot ionising UV emission from hot massive stars. However, the EW of these lines will also be moderated by the stellar continuum at their respective rest-frame wavelengths.
In the top two panels of Figure \ref{fig:O3_ha_hb} we present the H$\alpha$ and H$\beta$ EWs against the [OIII]$\lambda5007$ EWs. 
The EW of both Balmer lines show a clear positive trend with the [OIII]$\lambda5007$ EW, with best fit relations
\begin{equation}
    \begin{split}
        \rm{H}\alpha\, \text{EW/\AA} = (0.71\pm0.01) \rm{[OIII]}\lambda5007 \text{EW/\AA} + (42\pm1)
    \end{split}
\end{equation}
and 
\begin{equation}
    \begin{split}
        \rm{H}\beta\, \text{EW/\AA} = (0.15\pm0.01) \rm{[OIII]}\lambda5007 \text{EW/\AA} + (1.1\pm0.9)
    \end{split}
\end{equation}
We additionally find that the EELG stacks (defined as having [OIII]$\lambda5007$EW$>750$\AA) at high and low redshift exhibit consistent H$\alpha$ EWs, which are both considerably larger than those exhibited by the non-EELG stacks (see Figure \ref{fig:O3_ha_hb}) .
Here we see that the high sSFRs that drive extreme EWs in the hydrogen nebular emission lines (e.g., \citealt{Marmol16}) also contribute to the high [OIII]$\lambda5007$ EWs in our sample. The high H$\alpha$ and H$\beta$ EWs exhibited in our EELG sample likewise reflect that the stellar populations in these galaxies will likely be young (the EW of these hydrogen lines starts decreasing within 10Myrs for an instantaneous burst, e.g., \citealt{Leitherer99}). 

Finally, we turn to the Lyman-$\alpha$ rest-UV emission line, which our NIRSpec prism spectroscopy has sensitive coverage of down to $z>4$.
Although Lyman-$\alpha$ is known to be resonantly scattered and highly suppressed in most galaxies, it has been suggested that extreme galaxies such as EELGs (with potentially high ionisation of the ISM, and starburst-driven outflows) might be able to clear escape channels for Lyman-$\alpha$ photons \citep{Ostlin05, Herenz17}. We now look for any evidence of Lyman-$\alpha$ emitters (LAEs) in our stacks and in individual EELGs, and we report strong evidence that Lyman-$\alpha$ does indeed escape from EELGs at high redshift.

In the middle panel of Figure \ref{fig:stacks} we zoom in on the rest-frame UV in the stacked data to study the emergence or otherwise of Lyman-$\alpha$. 
We first consider the two EELG stacks.
The high-redshift EELG stack has the strongest Lyman-$\alpha$ emission (Ly$\alpha$ EW = $39\pm3$\AA), 
with the low-redshift EELG stack also showing Lyman-$\alpha$ in emission, but at a slightly weaker level (Ly$\alpha$ EW = $21\pm5$\AA). 
However, we note that the lower sensitivity of NIRSpec at shorter wavelengths ($<1\mu m$) means that in our lower redshift stack we are less sensitive to weak  Lyman-$\alpha$ emission close to the spectral break \citep[see][]{Jones23arXiv}.

Turning to the non-EELG stacks, there is no Lyman-$\alpha$ emission and both redshift bins have the appearance of a damping wing \citep{Miralda-Escude98}. It is interesting that we observe a similar softening of the Lyman-break in both non-EELG redshift stacks.  In our high-redshift stack, where we sample galaxies during the EoR, we may anticipate Lyman-damping due to a high hydrogen neutral fraction in the IGM, however, in the low-redshift stack we would not expect damping from the IGM because of the near-zero neutral fraction during this epoch ($z<5.7$). Softening of the spectral break may instead be due to either damped Lyman-$\alpha$ from HI in the surrounding environment (ISM and CGM) of the galaxies \citep[e.g.,][]{Heintz23arXiv} or be an artefact, resulting from the softening of the Lyman-break due to convolution with the Line Spread Function (LSF) of NIRSpec (see \citealt{Jones23arXiv}).

Individually, we find that out of 17 EELGs at $z>5.7$, $53\%$ (9/17) have detected Lyman-$\alpha$ emission, as determined by \citet{Jones23arXiv, Saxena23arXiv, Witstok23arXiv}. This would suggest that selections based on high [OIII]$\lambda5007$ EW galaxies are also preferentially selecting LAEs. 
This LAE fraction is higher than among the high-redshift non-EELGs, where only $18\%$ (2/11) have detected Lyman-$\alpha$ emission\footnote{Galaxy IDs 3968 and 18179. These galaxies have measured [OIII]$\lambda5007$ EWs of $295\pm19$\AA\, and $241\pm6$\AA, respectively, and Ly$\alpha$ EWs of $57\pm17$\AA\, and $41\pm9$\AA, respectively.}. These two non-EELG Lyman-$\alpha$ emitters show moderate [OIII]$\lambda5007$ EWs ($\sim300$\AA), in both cases the relative strength of the nebular emission lines (being driven by recent $<10$Myr star formation) has potentially decreased due to strong rest-optical continuum emission from an older stellar population. 

Below $z\sim4$, where we have 4 EELGs in our sample, Lyman-$\alpha$ does not lie within the wavelength window where we have sensitive spectral coverage with NIRSpec.
Between $4.0<z<5.7$ we have a sample of 15 EELGS with 4 showing Lyman-$\alpha$ in the NIRSpec prism spectroscopy. One of these is the potential AGN identified by \citet{Maiolino23arXiv} to have a broad component to the H$\alpha$ emission line (Galaxy ID: 8083).
We additionally look at the MUSE Data release II IFU spectroscopy of the HUDF region \citep{Bacon23} and find one additional EELG galaxy (Galaxy ID 7938, $z=4.815$) where MUSE detects Lyman-$\alpha$ emission which is not visible in our low dispersion prism spectrum. 
This leaves us with an LAE fraction of 5/15 ($33\%$) of the EELG galaxies in the redshift range $4.0<z<5.7$, nominally lower than the fraction at higher redshifts ($z>5.7$) of $53\%$ (9/17), although we note that the combination of sensitivity and spectral resolution of the prism means that NIRSpec's abilty to detect Lyman-$\alpha$ decreases towards the short wavelength end (affecting the lower redshifts).
Hence, the LAE fraction among our low-redshift EELG sample may be a lower limit and the difference in the fractions of LAEs in the low- and high-redshift EELG samples may not reflect genuine evolution. 
However, what we can say is that at high redshift, roughly half of EELGs are Lyman-$\alpha$ emitters, whereas for the non-EELGs at these high redshifts only $18\%$ of our sample have Lyman-$\alpha$ in the NIRSpec spectroscopy. This is a remarkable result which we will discuss further in Section \ref{sec:disc:LAE}.

Finally, we note 2/19 ($11\%$) non-EELGs at $4<z<5.7$ show Lyman-$\alpha$ emission, the lowest LAE fraction in any of our sub-samples. Like the high-z non-EELG LAEs these two\footnote{Galaxy IDs 5759 and 7304. These galaxies have measured [OIII]$\lambda5007$ EWs of $523\pm13$\AA\, and $233\pm21$\AA, respectively, and Ly$\alpha$ EWs of $63\pm10$\AA\, and $34\pm9$\AA, respectively.} also show rest-optical continuum emission from an evolved stellar population. 

\subsubsection{Ionisation parameter, Excitation state and Metallicity}
\label{sec:ionisation}
To investigate how the ionisation state of the ISM changes with [OIII]$\lambda5007$ EW within our full sample we employ the O32\footnote{line flux ratio between [OIII]$\lambda4959,5007$ and [OII]$\lambda3727,3729$} and Ne3O2\footnote{line flux ratio between [NeIII]$\lambda3869$ and [OII]$\lambda3727,3729$} diagnostics (e.g., see \citealt{Mengtao19}),
which are sensitive to the ionisation state of the ISM which is shaped by the ionising flux and hardness of the ionising spectrum.
O32 provides a flux ratio between two different ionisation states of the same elements, while Ne3O2 provides a flux ratio between two alpha-elements in different ionisation states. 
The Ne3O2 diagnostic utilises two lines close in wavelength and is therefore insensitive to the dust attenuation and does not require any correction, whereas the large baseline of O32 (3727\AA-5007\AA) means we must take into account reddening in our measurements (as described in Section \ref{sec:balmerdec}).  
Although the Ne3O2 ratio is sensitive to the Ne/O abundance ratio, these are
both alpha elements and this abundance ratio is expected to vary only minimally \citep[e.g.][]{Kobayashi20}.
Together they provide a complimentary estimate of the ionisation parameter. 

In the bottom-left panel of Figure \ref{fig:O3_ha_hb} we plot the O32 diagnostic against the [OIII]$\lambda5007$ EW, with galaxies corrected for dust reddening.
We note that when H$\alpha$ and H$\beta$ are detected and we can measure the Balmer decrement, the average reddening suggests minimal dust attenuation between 3727\AA\, and 5007\AA.
We additionally plot the Ne3O2 ratio against the measured [OIII]$\lambda5007$ EW of the full sample in the lower-right hand panel.
Both O32 and Ne3O2 are related to the ionisation parameter, and in both these Figure panels we additionally provide the ionisation parameter derived following the \citet{Witstok21} relations.

We find a clear positive correlation in both O32 and Ne3O2 with [OIII]$\lambda5007$ EW, obtaining best fit relations: 
\begin{equation}
    \begin{split}
        \log_{10}(\mathrm{O32}) = &(0.69\pm0.03)\times \log_{10}([OIII]\lambda5007\, \text{EW/\AA}) \\
        &-(0.96\pm0.10)
    \end{split}
\end{equation}
and 
\begin{equation}
    \begin{split}
            \log_{10}(\mathrm{Ne3O2}) = &(0.78\pm0.12)\times \log_{10}([OIII]\lambda5007\, \text{EW/\AA}) \\ 
            &-(2.4\pm0.4)
    \end{split}
\end{equation}

This correlation suggests that galaxies with greater [OIII]$\lambda5007$ EW have more highly ionised ISMs. This result is not surprising considering young stellar populations are responsible for the nebular emission. The youngest stellar populations retain the most massive stars with the hardest ionising spectra, 
and have higher ionising photon fluxes,
making the ISM highly ionised. EELGs show the greatest nebular emission relative to the continuum and therefore are likely to be the youngest systems (see Section \ref{sec:beagle})

When we turn to our stacked sub-samples in Figure \ref{fig:O3_ha_hb}, the high redshift EELG stack has the highest O32 and Ne3O2 ratios, with the low-redshift EELG stack showing marginally lower values. Both EELG stacks show significantly higher values for all quantities in Figure \ref{fig:O3_ha_hb} compared to the non-EELG stacks. 

\begin{figure}
    \centering
    \includegraphics[width=\columnwidth]{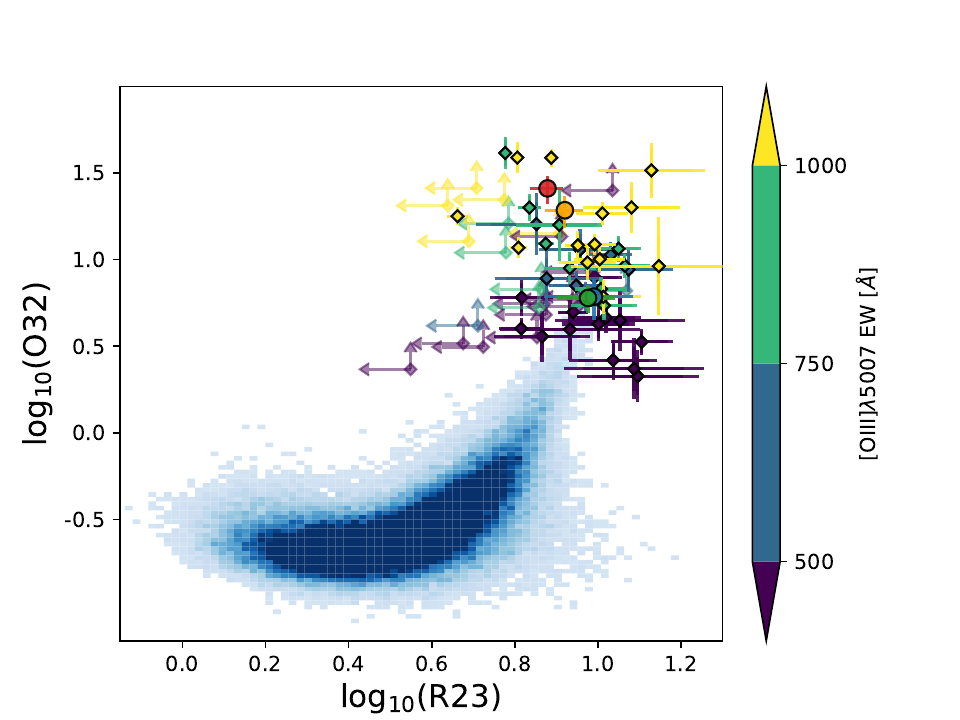}
    \includegraphics[width=\columnwidth]{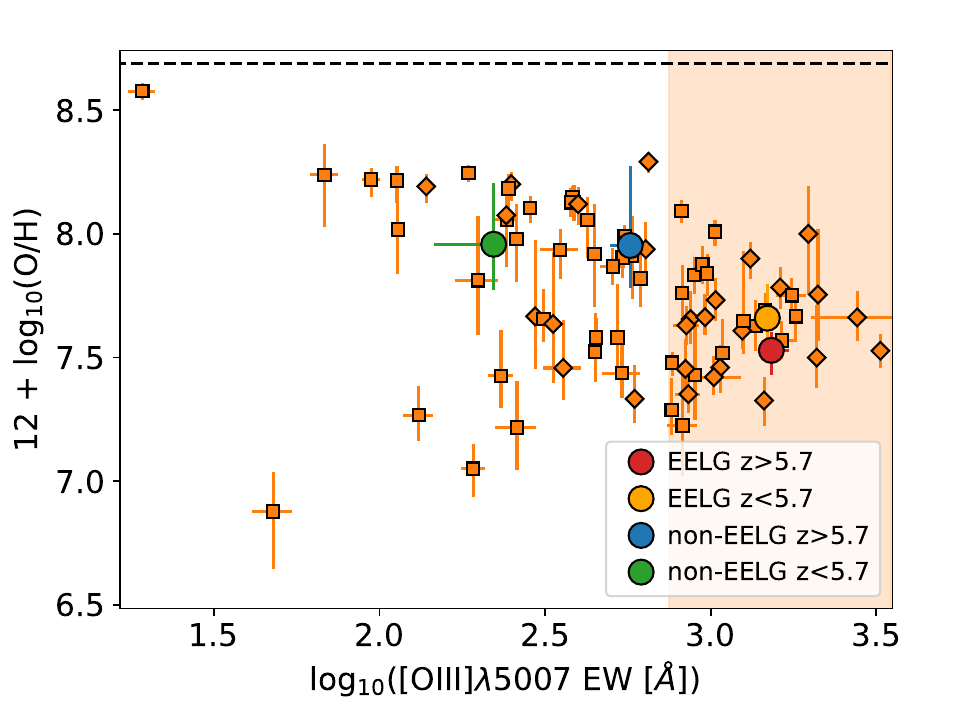}
    \caption{Top panel: The ionisation and excitation conditions of the full sample as inferred from the dust corrected O32 and R23 strong line diagnostics (when either [OII] or H$\beta$ is non-detected we provide a $2\sigma$ limit on the diagnostics). Each data point is coloured by the [OIII]$\lambda5007$ EW. The local SDSS galaxy population is presented in the blue shaded region.
    Bottom panel: The Oxygen abundance relation with $\log_{10}$([OIII]$\lambda5007$ EW), with metallicities taken from \citet{Curti23barXiv}. We mark solar metallicity (8.69 in these units \citealt{Asplund09}) as a horizontal dashed line. We exclude 10 galaxies where only [OIII] and H$\alpha$ were detected as metallicity constraints could not be placed using only these lines. 
    We include a shaded region to mark the EELG EW threshold of [OIII]$\lambda5007 > 750$\AA, and in both panels the measured quantities for our four stacks are shown in large circles (coloured to match Figure \ref{fig:stacks}). We distinguish our high- and low-redshift galaxies (split at $z=5.7$) as diamonds and squares. }
    \label{fig:r23_o32}
\end{figure}

To further examine the state of the ISM we employ the dust-corrected O32 and R23\footnote{Line flux ratio between [OIII]$\lambda4959,5007$ + [OII]$\lambda\lambda3727,3729$ and H$\beta$,  a proxy for the Oxygen abundance (metallicity).}  diagnostic diagram (Figure \ref{fig:r23_o32}), where galaxies of different ionisation and excitation states occupy distinct regions .  
For our full sample, we determine the O32 and R23 emission line diagnostics for galaxies, which we correct for reddening (see Section \ref{sec:method}). Where individual lines are undetected at  $3\sigma$, they are represented as $3\sigma$ limits. These are plotted in Figure \ref{fig:r23_o32}. We find that our full sample ($3<z<9.5$) lies above the local ($z<0.1$) SDSS galaxy locus\footnote{We use archival data from the Sloan Digital Sky Survey (SDSS; \citealt{York00}) Data-Release 7 \citep{Abazajian09}. We set a $\geq5\sigma$ detection requirement on each emission line for inclusion in Figure \ref{fig:r23_o32} and we reject AGN following the method set out in \citet{Runco21}}, to higher O32 and R23 values. This is the region numerous studies have found local ``high redshift analogue" galaxies to occupy \citep[e.g., Green peas and Blueberries,][]{Yang17a, Yang17, Cameron23}.
We find a trend that galaxies with larger [OIII]$\lambda5007$  EWs tend to sit further from the local SDSS galaxy distribution, and have much higher O32.

We observe no clear turn-over in the R23 diagnostic for the highest EWs, suggesting that while these galaxies are all metal-poor (sub-solar), they have not entered the extremely metal-poor branch of the R23 diagnostic \citep{Curti20} where the oxygen lines lose strength relative to the hydrogen recombination lines due to the extremely low range of oxygen abundances. It is therefore unsurprising that the R23 diagnostic does not reveal a clear trend with EW as we are probing the turn-over region of the R23 diagnostic and our resulting measurements of metallicity will be poorly constrained by R23.

More robust oxygen abundance measurements for our full sample have been presented in \citet{Curti23barXiv}. These measurements rely on multiple strong-line calibrations utilising the rest-optical emission lines, and metallicity estimates from spectral model fitting. For 10 objects in our full sample (including 1 EELG, ID 2651) metallicity constraints could not be provided due to too few key lines being detected (typically only [OIII] and H$\alpha$ were detected in these cases).
The \citet{Curti23barXiv} metallicities for our EELGs lie in the range $7.3 < 12 + \log_{10}(O/H) < 8$ (where solar metallicity is $8.69$ in these units, \citealt{Asplund09}), suggesting that while these systems are all sub-solar (all $<0.2Z_\odot$) none show extremely low metallicity (below $<0.04Z_\odot$). The metallicities for our sample are shown in the lower panel of Figure \ref{fig:r23_o32} against the [OIII]$\lambda5007$ EW.

Within our sample we find that galaxies with higher [OIII]$\lambda5007$ EWs have typically lower metallicities, albeit with some degree of scatter.
This trend is additionally seen in the stacked sub-samples in the lower panel of Figure \ref{fig:r23_o32}, with the metallicity measured to be lower in the stacks with the highest [OIII]$\lambda5007$ EW. 

This trend is expected since [OIII]$\lambda5007$ is a collisionally excited line and the luminosity increases with the electron temperature, related to the metallicity (due to the limitation of the metal line cooling mechanism), allowing the EW to increase at sub-solar oxygen abundance (up to a point when there are too few oxygen atoms - i.e., we expect monotonic behaviour at very low metallicity where the EW will decrease).
We note that the highest EW EELGs in our sample have metallicities of $\sim0.1Z_\odot$, with maximum EW found in galaxies with measured metallicities above and below this value never reaching as high. This maximum-EW turn-over with metallicity is consistent with photoionisation models, such as in \citet{Tang21}, that predict only marginal increase in the [OIII]+H$\beta$ EW between 0.3-0.1$Z_\odot$, with a turn-over around $0.1Z_\odot$, after which the EW begins to decrease with further decrease in metallicity. This turn-over may explain the few extremely metal-poor galaxies with low EWs in Figure \ref{fig:r23_o32}. 

\subsection{UV properties from spectral modelling}\label{sec:spectral_modelling}

Going beyond the rest-optical emission lines, the rest-frame UV properties of a galaxy are sensitive to recent star formation and the presence of hot O/B type stars whose emission peaks at shorter wavelengths.
The UV continuum is sensitive to star formation on 100Myr timescales, whereas nebular lines are sensitive to shorter timescales of 10Myr (see Figure \ref{fig:app:EW_history}).
To examine the non-ionising UV continuum we model the rest-frame UV spectrum for each target in our full sample, using our NIRSpec prism spectra (which have been corrected for slit losses). 

\subsubsection{$\beta$ slopes and M$\rm_{UV}$}
\label{sec:beta}

\begin{figure*}
    \centering
    \includegraphics[width=\textwidth]{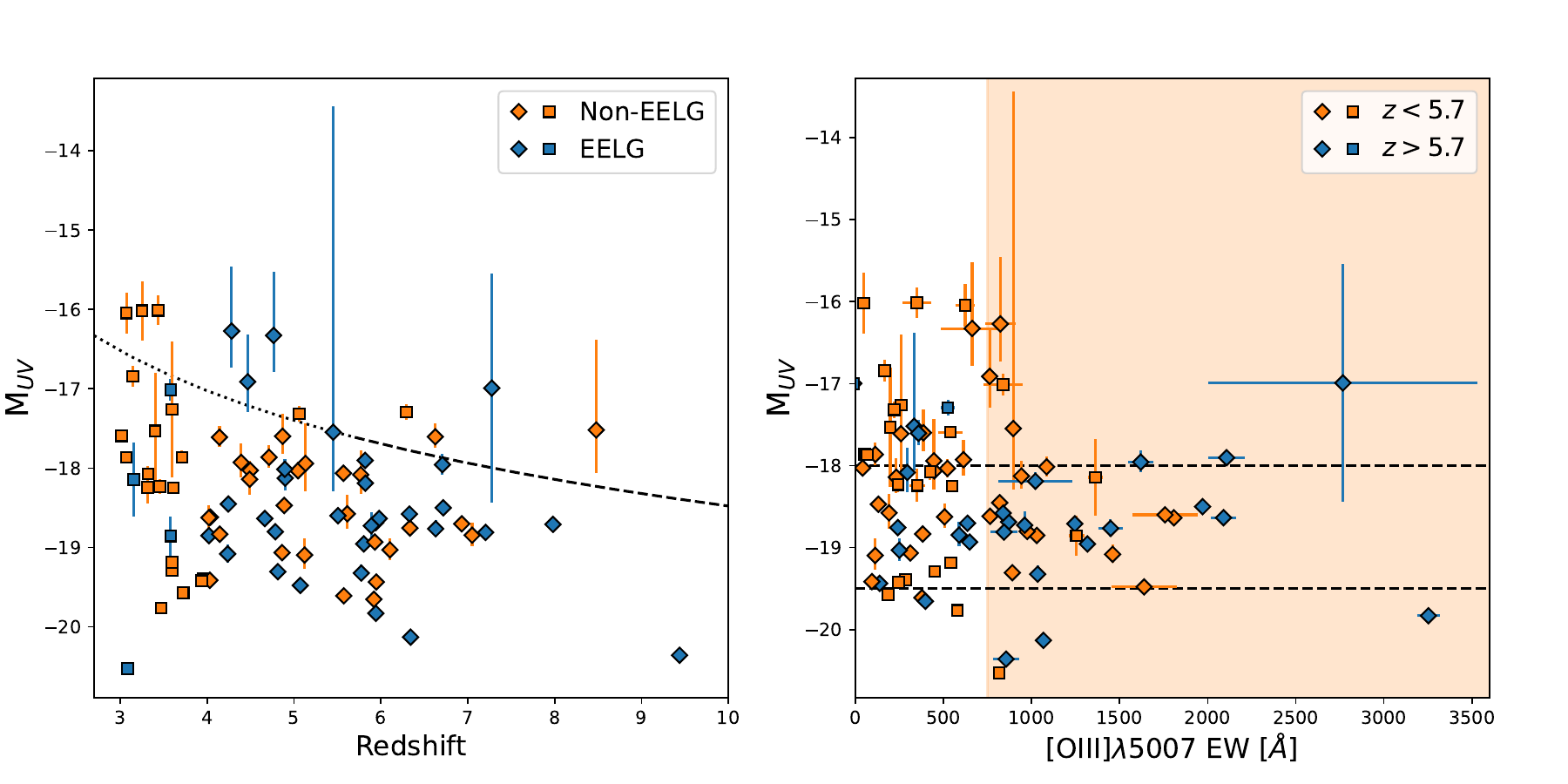}
    \caption{UV absolute magnitude determined for our full sample. Left panel: M$\rm_{UV}$ plotted against spectroscopic redshift. We overlay as a black line the apparent m$\rm_{UV}=29$ magnitude used in the selection of the high-redshift targets, plotted as dashed ($z>5.7$) and dotted ($z<5.7$) lines to indicate the UV selection within our full sample. We use galaxies below this line in our redshift comparison to obtain a consistent selection across our sample. The EELG sub-sample are shown in blue and the non-EELGs in orange.
    Right panel: M$\rm_{UV}$ plotted against the [OIII]$\lambda5007$ EW. We additionally overlay dashed lines at M$\rm_{UV}$ = -18.0 and -19.5, which correspond to the  UV absolute magnitude thresholds used in \citet{Endsley23arXiv}. Galaxies in our high-redshift sub-sample ($z>5.7$) are shown in blue, whilst those at lower redshifts are orange. 
    In both panels, galaxies are presented as diamonds when M$\rm_{UV}$ was measured directly from the spectroscopy and as squares when M$\rm_{UV}$ was measured from photometry.}
    \label{fig:muv_z}
\end{figure*}

\begin{figure*}
    \centering
    \includegraphics[width=\textwidth]{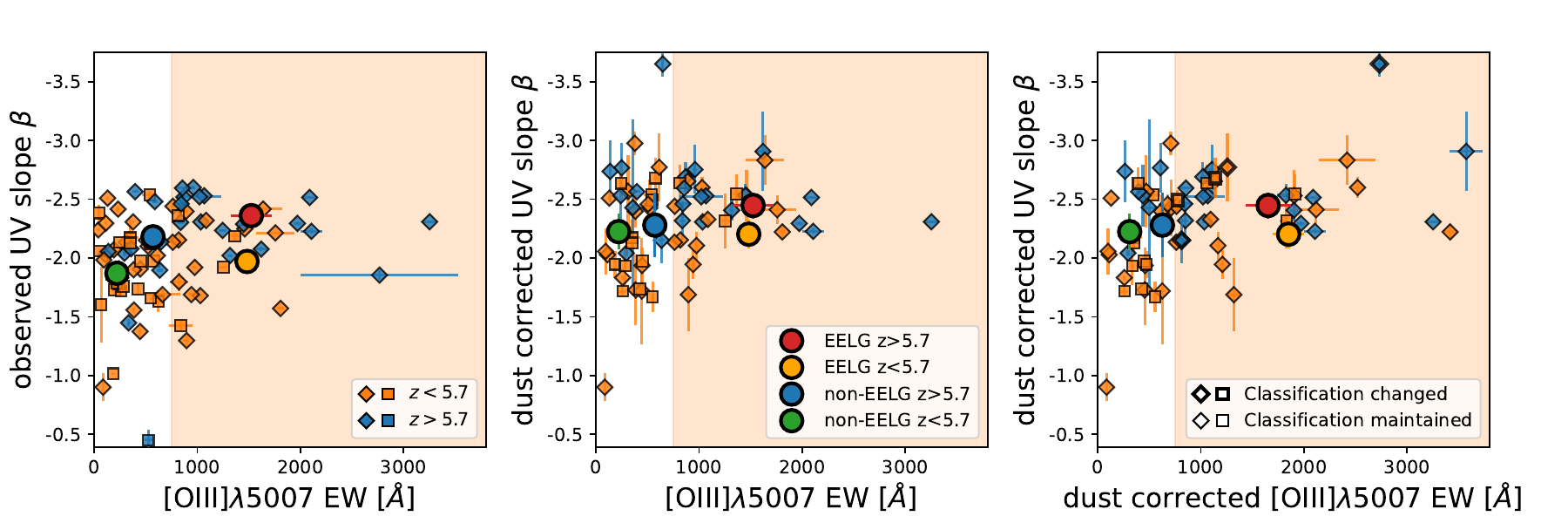}
    \caption{The UV spectral slope ($\beta$, where $f_\lambda \propto \lambda^{\beta}$) plotted against the [OIII]$\lambda5007$ EW. Left panel: The observed $\beta$ slope. Centre panel: $\beta$ slope corrected for dust attenuation. Right panel: $\beta$ slope plotted against [OIII]$\lambda5007$ EW, both corrected for dust attenuation (taking into account the differential extinction of the nebular line relative to the continuum as in \citealt{Calzetti00}). 
    In each panel we mark galaxies as diamonds where the $\beta$ slope was measured directly from the prism spectroscopy with diamonds, and plot as squares those where we rely on the spectrum at a slight longer rest-frame wavelength and photometry for our $\beta$ measurement (typically for those only at $z<4$). High and low redshift sub-samples are identified as blue and orange datapoints, respectively. 
    In all panels, we include a shaded region to mark the EELG EW threshold of [OIII]$\lambda5007 > 750$\AA, and the measured quantities for our four stacks are shown in large circles (coloured to match Figure \ref{fig:stacks}). In the right panel we show in bold the individual galaxies where their classification changes from non-EELG to EELG due to the dust correction.
    }
    \label{fig:beta}
\end{figure*}

We determine the absolute magnitude in the rest-frame UV (M$\rm_{UV}$ at 1500\AA) for our sample, which we measure over a rest-frame 50\AA -wide boxcar filter centred on 1500\AA\ in our prism spectra. In Appendix \ref{sec:App:ew} we also consider the photometry from the NIRCam filter best approximating the rest-UV (uncontaminated by the Lyman-break) using the Kron magnitude to get an indication of the total flux, and we find in most cases good agreement in M$\rm_{UV}$. 
We also determine the UV spectral slope ($\beta$, where $f_{\lambda}\propto\lambda^{\beta}$), from our prism spectra over a rest-frame range 1340\AA\ to 2400\AA\, (we mask the regions contaminated by the CIV$\lambda1550$, HeII$\lambda1640$, [OIII]$\lambda1660$ and CIII]$\lambda1990$ emission lines). 
We note that, as described in \citet{Bunker23b}, we have employed wavelength dependent slit loss corrections, accounting for the differing PSF and placement of the galaxy in the MSA shutter, ensuring we are not biasing UV slope measurements to bluer values with preferentially more flux lost at longer wavelengths.
The uncertainties on the UV slope are measured via a Monte Carlo based approach, whereby the power law fit is performed 500 times with the uncertainties on each pixel randomly sampled from a Gaussian distribution. The resulting standard deviation of the UV slope measurement from 500 iterations is used as the uncertainty, and we note that this error reflects the statistical uncertainty but does not include systematic uncertainty.
We note that for 24 galaxies in our full sample at $z<4$, the rest-frame 1500\AA\, moves to the extreme blue end of our spectral coverage where the sensitivity reduces. In these cases, we move our $\beta$ fitting spectral range to slightly longer rest-frame wavelengths [2000-3500\AA] and use this fit to estimate M$\rm_{UV}$ at 1500\AA. For these objects, we additionally measure M$\rm_{UV}$ from the JADES aperture photometry\footnote{We adopt the `KRON' Diameter \citet{Rieke23b} circular apertures} of the legacy HST/ACS F775W imaging (see \citealt{Rieke23b}), which covers the rest-frame 1500\AA\, at $z\sim3.5-4$. There were 2 galaxies at $z>4$ (ID: 10009693 and 10009320) with a S/N$<3$ in the integrated spectrum over 1400-1600\AA\,(rest-frame), and as in Section \ref{sec:stack} we use instead the NIRCam photometry in the filter best approximating 1500\AA\, (rest) to determine M$\rm_{UV}$. 

The M$\rm_{UV}$ and $\beta$ slope measurements for our full sample are shown against the [OIII]$\lambda5007$ EW in Figures \ref{fig:muv_z} and \ref{fig:beta}.
Our full sample covers a broad M$\rm_{UV}$ range, between $-16 > $M$\rm_{UV} > -20$. As noted in Section \ref{sec:selection}, at $z>5.7$ the HST-Deep spectroscopy is essentially a rest-UV-based selection (a proxy for star formation rate, hereafter SFR) with a magnitude cut of $AB\sim29$\,mag in the filter covering wavelengths just redward the Lyman-$\alpha$ break, but uncontaminated by the break. We plot in Figure~\ref{fig:muv_z} the corresponding M$\rm_{UV}$ selection at $z>5.7$ as a dashed line, and as can be seen our spectroscopic measurements of the UV luminosity broadly agree with the range expected from initial pre-selection of spectroscopic targets. Our selection is sensitive to all galaxies brighter than M$\rm_{UV}=-18$ over our entire redshift range for the EELG selection (out to $z=9.5$), and we are sensitive to galaxies with fainter UV luminosities at lower redshifts (as we impose a cut in apparent magnitude).

At lower redshifts, the spectroscopic target selection was based on redder rest-frame wavelengths (a better proxy for stellar mass rather than SFR). This means that the full sample does not have an M$\rm_{UV}$ cut, although in Figure~\ref{fig:muv_z} we show where such a cut would have fallen adopting the same rest-UV apparent magnitude limit as in the $z>5.7$ selection (dotted line). About 12\% of the $z<5.7$ sample (7/57) are UV-faint ($m_{\rm{UV}}>29$), and can be excluded to make a fairer comparison with the higher redshifts. We note that applying an apparent magnitude UV cut will not recover galaxies with extremely blue UV spectral slopes $\beta \ll -2$, which would have been selected in a UV selection but may have been lost in our longer-wavelength selection.

We now consider the beta slopes and present these in Figure \ref{fig:beta}. The full sample covers a broad range of $\beta$ slopes ($-0.5 > \beta > -3$).
The average observed $\beta$ slope of the 36 EELGs is $\beta=-2.2$ with a standard deviation of $\sigma=0.3$, so the average is $\beta_{\rm EELG}=-2.2\pm0.1$ (where we quote the standard error on the mean). This UV spectral slope is steeper than that for the 49 non-EELGs, which have $\beta_{\rm nonEELG}=-1.9\pm0.1$ (again, quoting the standard error on the mean from the standard deviation of $\sigma=0.5$). Hence the EELGs have marginally more blue UV spectral slopes at the $\approx 3\sigma$ level. 

The UV spectral slope can depend on the level of dust attenuation and the SFH. Nebular continuum emission (e.g. \citealt{Topping22, Cameron23arXiv}) may also affect the measured UV slope, particularly with the very high EW lines in EELGs, making UV slopes redder than expected from very young ionizing stars alone.
We correct our $\beta$ slope measurements for dust reddening using the measured Balmer decrement (see section \ref{sec:balmerdec}). In Appendix \ref{sec:App:dust_beta} we present a simple relation for the dust correction to the observed $\beta$ slope as a function of dust attenuation ($A_{1600}$). We apply this relation to our sample to correct the UV-slope $\beta$ for dust and we compare the trends with the [OIII]$\lambda5007$ EW in the central panel of Figure \ref{fig:beta}. As noted in Section \ref{sec:balmerdec}, we are unable to constrain the dust attenuation for 4 galaxies and these objects are excluded from the dust corrected panels in the Figure.

We find the EELG sample occupies a smaller and slightly bluer range of de-reddened $\beta$ slopes than the non-EELGs. One non-EELG (ID:10013704) with a reddening corrected $\beta<-3.5$, has been identified as a potential AGN \citep{Maiolino23arXiv} which may drive the observed blue slope $\beta=-2.14\pm0.01$ and large measured Balmer decrement.

Interestingly we find many objects with very blue slopes ($\beta<-2.5$) that are not classified as EELGs. These objects are also expected to host a young stellar population and therefore produce copious quantities of ionising photons, however they show low [OIII]$\lambda5007$ EWs ($<500$\AA) and also low EWs of the Balmer lines $H\alpha$ and H$\beta$. 
In the right hand panel of Figure \ref{fig:beta} we additionally correct our [OIII]$\lambda5007$ EWs for dust reddening to understand whether this population of blue $\beta$ slopes at low EW may be driven by differential dust attenuation of the emission line relative to the continuum (e.g., \citealt{Calzetti00}). We find that while more galaxies exhibit dust corrected [OIII]$\lambda5007$ EWs$>750$\AA, there still remains a sub-sample of galaxies with blue slopes and low EWs. 
Galaxies with young stellar populations but low EWs have been previously identified at high redshift ($z\sim5-7$, e.g., \citealt{Endsley23A}), where the lack of high EWs has been attributed to either extremely-low oxygen abundance or a high escape fraction of ionising radiation. Extremely low metallicities and high $f_\mathrm{esc}$ can both reduce the luminosity of the [OIII]$\lambda5007$ line. However, we do not see strong H$\alpha$ emission in these galaxies with low [OIII]$\lambda5007$ EWs and blue $\beta$ slopes, which dis-favours the extremely low metallicity scenario. Hence these particular galaxies may be ``Lyman Continuum leakers" with non-negligible escape fractions of ionising photons. 

\begin{figure}
    \centering
    \includegraphics[width=\columnwidth]{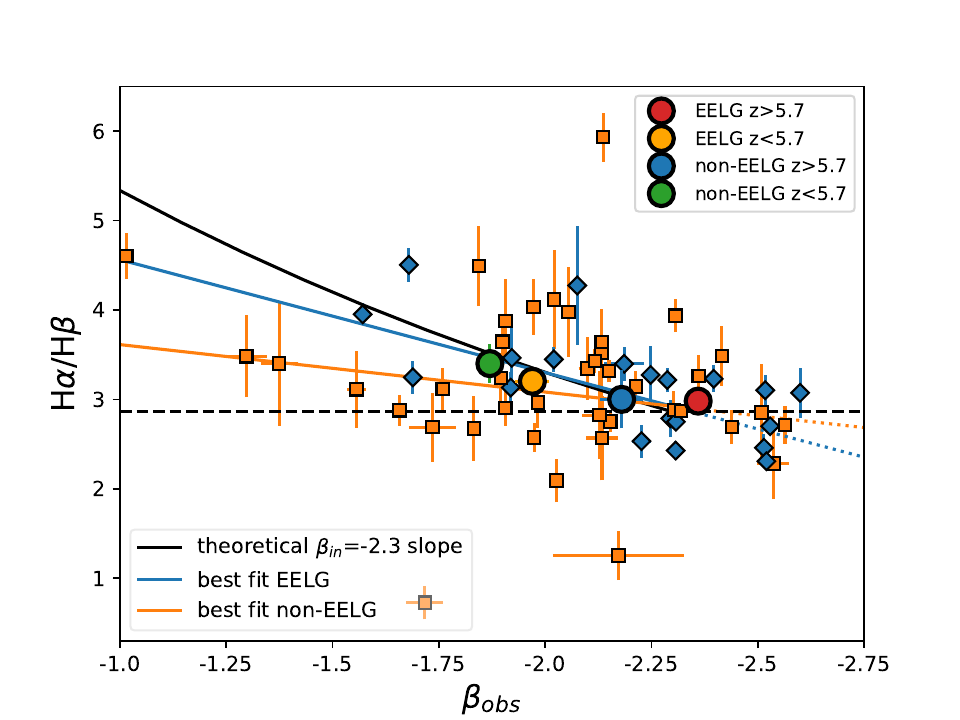}
    \caption{The observed Balmer decrement plotted against the observed UV-slope ($\beta$). Our full sample is split into EELGs (blue) and non-EELGs (orange), and are modelled by best fit relations shown in the matching colours.
    Both sub-samples favour an intrinsic UV-slope of $\beta\sim-2.3$. When the theoretical curve is plotted (black) we find good agreement with the EELG sub-sample and poorer agreement with the non-EELG sub-sample. The measured quantities for our four stacks are shown in large circles (coloured to match Figure \ref{fig:stacks})}
    \label{fig:balmer_dec}
\end{figure}

To further examine the effect of dust and SFH on the observed $\beta$ slope, we compare the $\beta$ slope with the Balmer decrement for both the EELG and non-EELG samples in Figure \ref{fig:balmer_dec}. We determine best fit relations
\begin{equation}
    H\alpha/H\beta = (-1.3\pm0.3)*\beta + (5.8\pm0.8)
\end{equation}
for the EELG sample and 
\begin{equation}
    H\alpha/H\beta = (-0.5\pm0.3)*\beta + (4.1\pm0.6)
\end{equation}
for the non-EELG sample.
For the EELG sample (shown in blue in Figure \ref{fig:balmer_dec}), we find an anti-correlation between the observed $\beta$ slope and the Balmer decrement, which closely tracks that expected for an intrinsic (i.e. de-reddened) UV spectral slope of $\beta=-2.3$ (\citealt{Wilkins11}, shown in black in Figure \ref{fig:balmer_dec}). 
However, the non-EELG population displays a higher scatter and typically redder observed $\beta$ slopes that do no exhibit such a tight anti-correlation with the Balmer decrement. 
We find this apparent in the measure Spearman rank coefficients for the two sub-samples, with the EELG sample showing a strong anti-correlation with a Spearman rank coefficient of -0.74 (p-value of 2e-4) while the non-EELG sample correlation is weaker with a Spearman rank coefficient of only -0.23 (p-value of 0.16).
This is consistent with the picture where galaxies with modest emission line EWs have a variety of SFHs leading to a range of intrinsic beta slopes, whereas to have extreme EWs a galaxy must be dominated by recent intense star formation leading to a blue intrinsic $\beta$ slope. 

This is seen clearly in the stacks in Figure \ref{fig:balmer_dec}, where the high-redshift EELG stack (shown in red) exhibits the steepest UV spectral slope of $\beta=-2.36\pm0.03$ while the shallowest slope is found in the low-redshift non-EELG stack (in green) with a $\beta=-1.87\pm0.02$.
This trend reflects the older stellar ages (and/or potentially more dust attenuation) in the lower redshift non-EELG sample. 

In the rest-optical, the non-EELG low-redshift stack is the only one to show evidence for a Balmer break (see Figure \ref{fig:stacks}), with an excess of continuum flux above 3645\AA, confirming the change in galaxy properties between the samples (see also Section~\ref{sec:beagle})
This is not unexpected as the low-redshift sample comes from a mass-limited survey which will include less actively star forming galaxies. For this low redshift sample, the Universe is older than the high redshift sample, so there has been more opportunity for older stellar populations to form and contribute flux at rest-optical wavelengths, reducing the EW of the emission lines.

\subsubsection{Ionisation efficiency}

\begin{figure}
    \centering
    \includegraphics[width=\columnwidth]{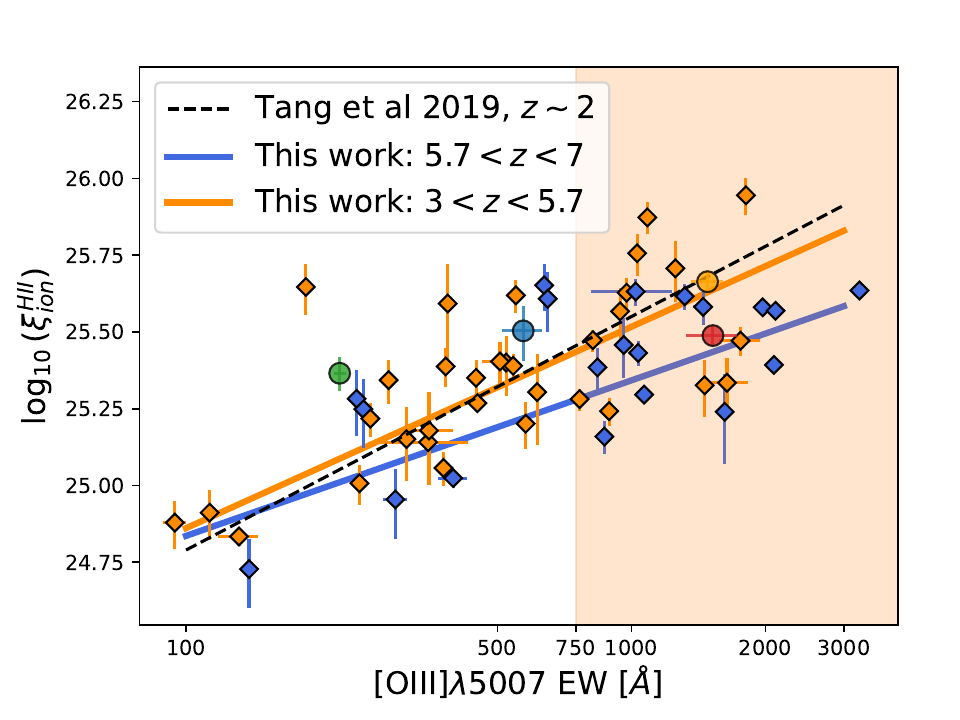}
    \caption{$\log_{10}(\xi_\mathrm{ion}^\mathrm{HII}/{\rm erg^{-1}Hz})$ ionisation efficiency (corrected for dust attenuation) plotted against the [OIII]$\lambda5007$ EW, split between the high-redshift (blue, $z>5.7$) and low-redshift (orange, $z<5.7$) samples. We overlaid \citet{Mengtao19} $z\sim2$ best fit relation. The shaded region marks the EELG EW threshold of [OIII]$\lambda5007 > 750$\AA\, and the measured quantities for our four stacks are shown in large circles (coloured to match Figure \ref{fig:stacks}).}
    \label{fig:xi_ion}
\end{figure}

We determine the ionisation efficiency ($\xi_\mathrm{ion}^\mathrm{HII}$, the hydrogen ionising photon production rate per unit UV luminosity at 1500\AA, corrected for dust attenuation) of our full sample using the dust corrected H$\alpha$ flux and M$\rm_{UV}$ for each galaxy.
We follow the method of \citet{Mengtao19} to determine $\xi_\mathrm{ion}^\mathrm{HII}$.
Here we define 
\begin{equation}
  \xi_\mathrm{ion}^\mathrm{HII}[{\rm erg^{-1}Hz}] = \frac{N(\mathrm{H}^0)[{\rm s^{-1}}]}{\mathrm{L}_{\mathrm{UV}}^{\mathrm{H\,II}}[{\rm erg\,s^{-1}Hz^{-1}}]}
\end{equation}

where we determine the UV luminosity from the measured M$\rm_{UV}$, and determine the hydrogen ionising photon production rate $N$(H$^0$)[s$^{-1}$] from the measured H$\alpha$ luminosity L(H$\alpha$)[erg s$^{-1}$] using the linear scaling factor
(1.36$\times10^{-12}$ erg) from \citet{Osterbrock_06}. 
We split our sample by redshift ($z=5.7$) and plot ionisation efficiency as a function of [OIII]$\lambda5007$ EW, along with the best fit relations, on Figure \ref{fig:xi_ion}. We over plot the $z\sim2$ best fit from \citet{Mengtao19} and we find a comparable relation with larger $\xi_\mathrm{ion}^\mathrm{HII}$ being found at higher [OIII]$\lambda5007$ EWs. 
This correlation should not be surprising as the calculation of $\xi_\mathrm{ion}^\mathrm{HII}$ and EW are closely related quantities: $\xi_\mathrm{ion}^\mathrm{HII}$ comes from the ratio of H$\alpha$ flux to the rest-UV continuum and EW is related to ratio of the [OIII]$\lambda5007$ flux to the rest-optical continuum. We note that the best fit relation for our lower redshift sample lies at higher $\xi_\mathrm{ion}^\mathrm{HII}$ than our high-redshift sample. 
If we consider $\xi_\mathrm{ion}^\mathrm{HII}$ simply as the total number of photons with energies higher than 13.6 eV, then at higher metallicities, these photons will be absorbed by many other metals such as C, N, Ne, Ar, etc, in addition to O, leading to a weaker increase in [OIII]$\lambda5007$ EW with increasing $\xi_\mathrm{ion}^\mathrm{HII}$. Likewise, \citet{Curti:2017} find that the [OIII]/H$\beta$ ratio is strongly dependent on metallicity. Therefore, evolution of the oxygen abundance with redshift would mean that the expected [OIII]$\lambda5007$ EW would vary for the same $\xi_\mathrm{ion}^\mathrm{HII}$ between our two sub-samples.
We determine a best fit relation
\begin{equation}
    \begin{split}
        \log_{10}(\xi_\mathrm{ion}^\mathrm{HII}/{\rm erg^{-1}Hz}) = & (0.59\pm0.03)\log_{10}(\rm{[OIII]}\lambda5007 \,\rm{EW}/{\text\AA}) \\
        & + (23.6\pm0.1)
    \end{split}
\end{equation}
for the high-redshift sample and 
\begin{equation}
    \begin{split}
        \log_{10}(\xi_\mathrm{ion}^\mathrm{HII}/{\rm erg^{-1}Hz}) = & (0.73\pm0.03)\log_{10}(\rm{[OIII]}\lambda5007 \,\rm{EW}/{\text\AA}) \\
        &+ (23.4\pm0.1)
    \end{split}
\end{equation}
for the low-redshift sample. 

We find that the average $\xi_\mathrm{ion}^\mathrm{HII}$ in the EELG sample (high and low redshift combined) is $\log_{10}(\xi_\mathrm{ion}^\mathrm{HII}/{\rm erg^{-1}Hz})$ of $25.5\pm0.2$, 
which is higher (although within $1\sigma$) than the average of $25.2\pm0.3$ in the non-EELG sample, consistent with the harder ionising spectra (bluer UV spectral slopes) of the EELGs (in line with results from \citealt{Simmonds24} using the JADES photometric dataset). We find this pattern is repeated in Figure \ref{fig:xi_ion} in the stacks of EELGs (both high and low redshift) which are either consistent or have higher ionisation efficiencies than the non-EELG stacks.

This result reaffirms that EELGs, including those at high redshift, are efficient ionising photon producers, an important characteristic as a potential source of the UV-ionising background during the EoR. 

\subsection{Inferred galaxy properties from spectral fitting}\label{sec:beagle}

In addition to studying the emission line diagnostics and rest-UV spectral properties, we can also model the full spectrum of each galaxy using spectral evolution synthesis codes to infer the stellar mass and the SFH. In this section we are interested in what characteristics of the modelled SFHs can be associated with a galaxy that has entered an EELG phase. 

\subsubsection{BEAGLE SED fitting}

We model the full prism spectrum ($0.6\lesssim\lambda_{obs}/\mu m\lesssim5.2$) for each galaxy using \texttt{BEAGLE} \citep[version 0.27.10,][]{Chevallard16} to constrain the galaxies' physical properties and SFHs. 
Briefly, for each galaxy the modelled redshift is set to the spectroscopic value, 
and the maximum formation redshift is set to $z=30$ (corresponding to an age of the Universe of 100 Myr for the adopted cosmology).
We utilise the photoionisation models from \citet{Gutkin16}, which are computed with Cloudy \citep{Ferland17} and are based on a more recent version of the \citet{Bruzual03} stellar population synthesis models (for more details see Chevallard in prep). We adopt the \citet{Charlot00}  two-component dust model, fixing to 0.4 the fraction of dust attenuation arising in stellar birth clouds. 
The specific allowed ranges for the variable \texttt{BEAGLE} parameters are laid out in Table \ref{tab:beagle}. 

We employ a two-component SFH model. This choice reflects that EELGs may be experiencing an upturn in SFR or a short-lived starburst. Our model SFH is comprised of a delayed exponential component\footnote{$SFR\propto t\times exp(-t/\tau)$, a function of time ($t$) and the e-folding time ($\tau$), where we fit for the $e$-folding time, $\tau$, the maximum stellar age, and the normalisation of the SFR} and a constant SFR burst of variable duration, where the two components are truncated, i.e they are not simultaneous.
The choice of a recent burst component is related to the nebular emission properties (continuum and line), which are powered by massive O-type stars which typically move off the main sequence before they reach an age of $\sim10$Myr, and  the disruption timescale of stellar birth clouds \citep{Murray10, Murray11}. For the single stellar models used here\footnote{Single-stellar denotes that these population models do not consider stars in binary pairs.}, more than 99\% of the ionising flux is emitted by stars younger than 10Myr.

The full \texttt{BEAGLE} data products used in this paper will be presented in Chevallard in prep. We note that when observing the rest-optical caution should be taken as bright emission from young stars may hide an older and fainter/redder stellar population. 

\begin{table*}
\centering
\resizebox{.75\textwidth}{!}{%
\begin{tabular}{ccl}
Parameter       & Range                              & Description     \\ 
\hline 
$\log_{10}(\tau/\rm{yr})$          & [6, 12]          & Star formation timescale of smooth component                      \\
$\log_{10}(Z_{\rm{ISM}}/Z_\odot)$               & [-2.2, 0.4] & Metallicity, $Z_\odot = 0.0152$                                                                                       \\
$\log_{10}(M_*/M_\odot)$               & [6, 12]     & Mass (integral of SFH) \\
$\log_{10}(t_{\rm{max}}/\rm{yr})$        & [6, 10.8]      & Maximum stellar age                                                                                                    \\
$\log_{10}(\rm{SFR/M_\odot \rm{yr}^{-1}})$        &   [-4, 4] & Star formation rate of burst \\
$\log_{10}(t_{\rm{burst}}/\rm{yr})$   & [6, 7.5] & Duration of burst component \\
z               &   [z$_{\rm{spec}},\sigma=0.01$]$^\star$                     & Redshift                                                                                                               \\
$\rm{z}_{form}$ & $\rm{z}_{form} = 30$  &  formation redshift \\
$\log_{10}(U)$               & [-4, -1]            &  Galaxy-wide ionization parameter                                                                                           \\
$\log_{10}(\xi_\mathrm{d})$           & [0.1, 0.5]        &  Galaxy-wide dust-to-metal mass ratio                                                                                          \\
$\tau_{\rm{v, eff}}$  & [0, 6]             & V-band attenuation optical depth                                                                                                                     \\
$\mu$           & $\mu = 0.4$                      & \begin{tabular}[l]{@{}l@{}}Relative V-band attenuation from \\ diffuse ISM to total\end{tabular}                                                               
\end{tabular}
}
\caption{\texttt{BEAGLE} free parameter variable ranges set for spectral model fitting. Uniform priors are set over each range, except for the redshift($^\star$) which follows a Gaussian prior with mean set to $z_{\rm{spec}}$ and $\sigma=0.01$.}
\label{tab:beagle}
\end{table*}

We determine galaxy properties for 84/85 galaxies in our full sample, with only one galaxy\footnote{NIRSpec ID: 3184} unable to be fit by \texttt{BEAGLE} due to low signal-to-noise in the continuum. We present the full range of SFH property constraints in Tables \ref{tab:beagle_results_update_1} and \ref{tab:beagle_results_update_2} in Appendix \ref{sec:App:BEAGLE}.
In Figure \ref{fig:EW_beagle} we present the \texttt{BEAGLE} best fit total stellar mass (including stars and stellar remnants), and the mass-weighted- and luminosity-weighted (v-band)-ages of the stellar population as a function of the [OIII]$\lambda5007$ EW.

\begin{figure*}
    \centering
    \includegraphics[width=\textwidth]{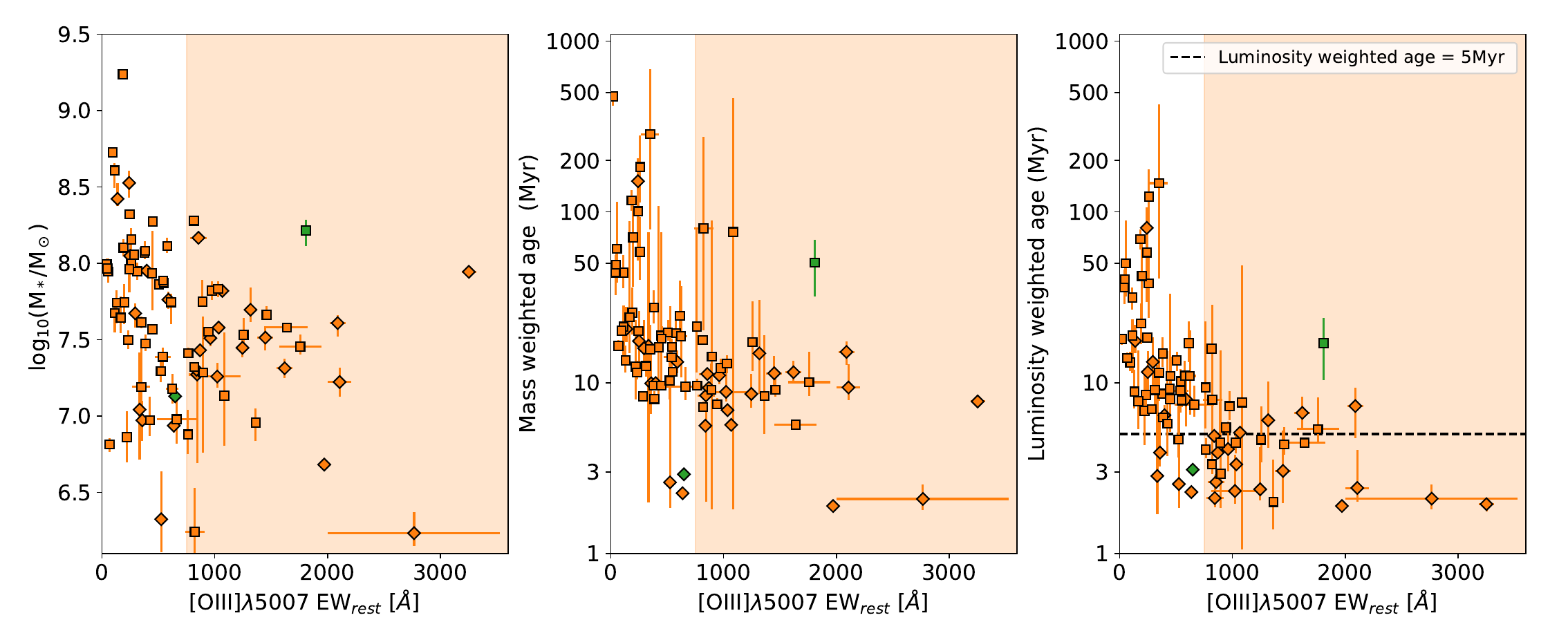}
    \caption{Total mass (Left panel), Mass-weighted-age (central panel) and luminosity-weighted-age (right panel) against the [OIII]$\lambda5007$ EW of our full sample. Two potential AGN are shown in green. The shaded region in each panel marks the EELG EW threshold of [OIII]$\lambda5007 > 750$\AA. Here, EELGs show typically lower masses, have lower mass-weighted-ages and typically luminosity-weighted-ages below $5$Myr (marked on the right panel). We distinguish our high- and low-redshift galaxies (split at $z=5.7$) as diamonds and squares, respectively.}
    \label{fig:EW_beagle}
\end{figure*}

We highlight the region identifying EELGs (shaded region in Figure \ref{fig:EW_beagle}) and find that the EELG sample on average exhibits lower masses (typically $\lesssim10^8M_\odot$) and lower ages ($\lesssim5-10$Myr, weighted either by mass or luminosity) than the non-EELG sample. 
While both EELG and non-EELG sub-samples cover a broad range of masses and luminosities, the dynamic range of ages in the EELG sub-sample is considerably smaller than in the non-EELG sample.

\begin{figure*}
    \centering
    \includegraphics[width=\textwidth]{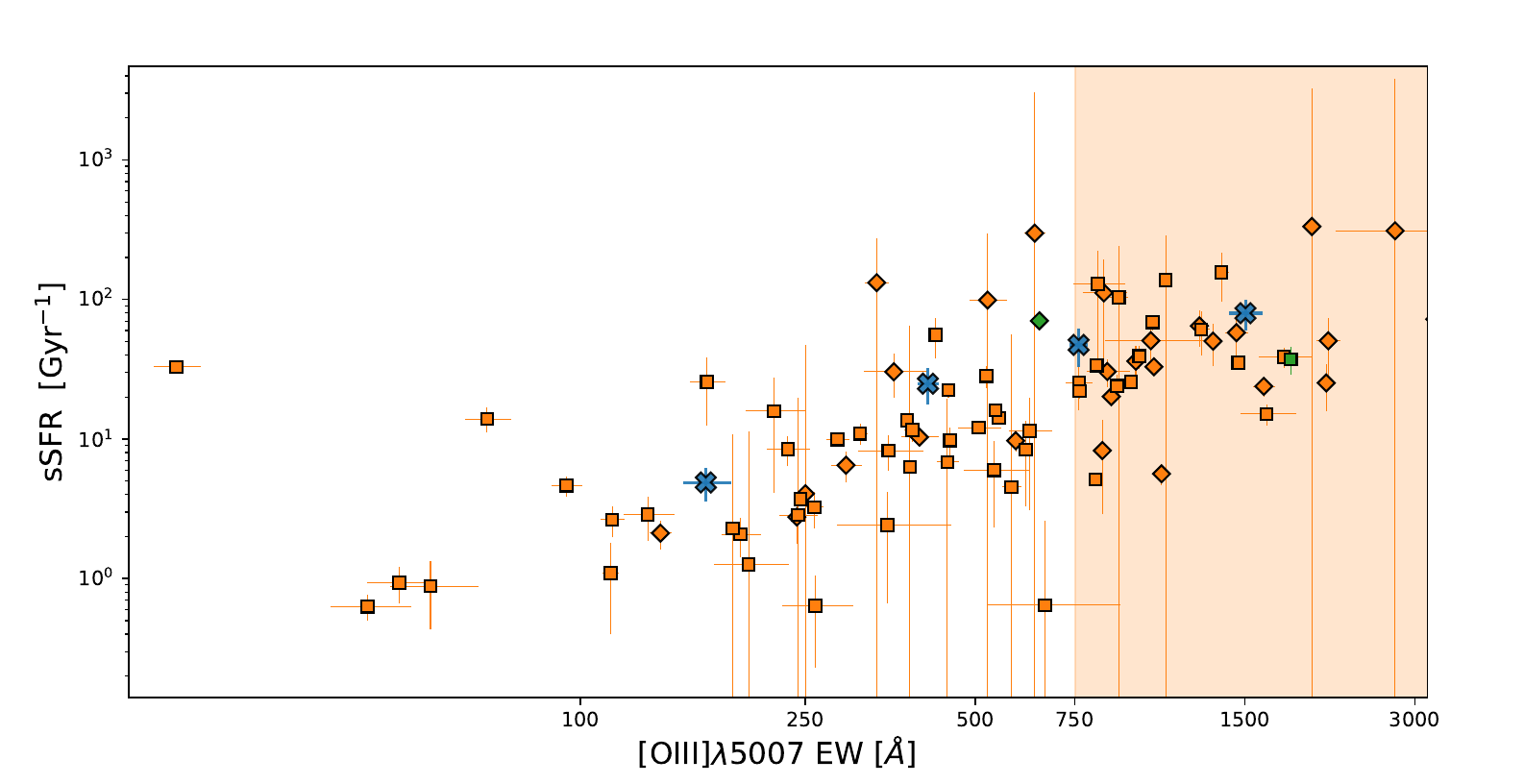}
    \caption{The sSFR dependence on the [OIII]$\lambda5007$ emission line EW (utilising the 3Myr short-term SFR to determine sSFR).  The shaded region marks the EELG EW threshold of [OIII]$\lambda5007 > 750$\AA. 
    Two potential AGN are shown in green and we distinguish our high- and low-redshift galaxies (split at $z=5.7$) as diamonds and squares, respectively.
    We additionally split our sample into 4 equal EW bins and plot the average sSFR for each bin as blue crosses. 
    }
    \label{fig:sSFR}
\end{figure*}

To consider this further we present in Figure \ref{fig:sSFR} the sSFR for our sample as a function of EW, utilising the SFR average over the most recent 3Myr. Galaxies with high EWs typically have high sSFR, with EELGs having sSFR typically above 10-20\,Gyr$^{-1}$ (meaning mass-doubling times of $\lesssim 100$\,Myr). 
This relation is fundamental as the high sSFR is needed in EELGs for the nebular lines to dominate over the rest-optical stellar continuum (to produce a large EW), which in turn drives the mass-weighted-age to low values as seen in Figure \ref{fig:EW_beagle}.

On the other hand, the larger dynamic range among non-EELGs reflects the greater variety of SFHs in this sub-sample. 
Likewise, we find that this pattern is repeated in the luminosity-weighted-ages, with EELGs showing ages below $\sim5$Myr\footnote{Only 2 non-EELGs have luminosity-weighted-ages below the 5Myr dividing line; one is the potential AGN ID:10013704 with an [OIII]$\lambda5007$ EW=$649\pm13$\AA\, and the other is ID:10013609 which has a an  EW=$637\pm28$.}. Here, the young, massive stars necessary to produce the ionising radiation that power the nebular emission lines outshine the light from older stellar populations in the rest-frame V-band, heavily skewing the luminosity-weighted-ages.
We find greater discrimination between EELGs and non-EELGs in the luminosity-weighted-ages than in the mass-weighted figure (due to the high  luminosity of the highest mass and shortest lived O- and B-type stars), and we note that all bar one non-EELG is consistent within $1\sigma$ with an age $>5$Myr and all EELGs are consistent within $1\sigma$ with ages younger than this.

\subsubsection{Star formation histories and SFR duty cycle}
\begin{figure*}
    \centering
    \includegraphics[width=\textwidth]{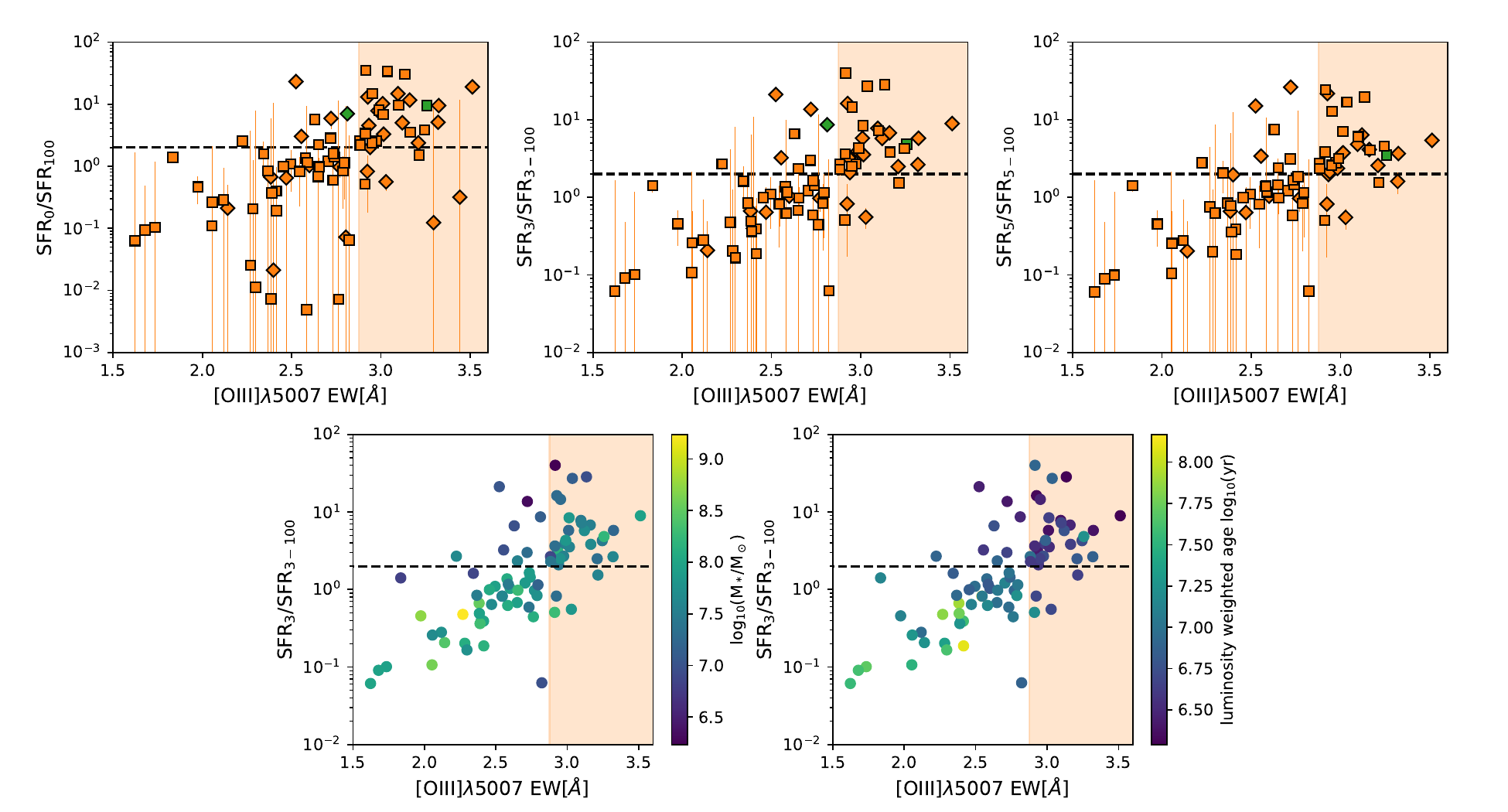}
    \caption{Burst fractions against $\log_{10}($[OIII]$\lambda5007 EW)$ for our full sample. 
    Top row, from left to right: the ratio of the average SFR over the most recent burst period (instantaneous, 3Myr, 5Myr) to the average SFR over the remaining period that makes up 100Myr (100Myr, 97Myr, 95Myr). Bottom row, left to right: the 3Myr burst plot now colour coded by total mass and luminosity weighted age. 
    Horizontal lines at an SFR ratio = 2 have been included to indicate a rough split between the EELG and non-EELG populations. The shaded region marks the EELG EW threshold of [OIII]$\lambda5007 > 750$\AA. We distinguish our high- and low-redshift galaxies (split at $z=5.7$) as diamonds and squares, respectively.
    }
    \label{fig:burst_intensity_alternative}
\end{figure*}

We can use the SFH of each galaxy, returned by \texttt{BEAGLE}, to assess their duty cycle of star formation (i.e., what fraction of time a galaxy spends in an active star forming phase), which is a measure of the `burstiness' of star formation. 
We are interested in whether EELGs show an upturn or burst in their SFR in the last few million years, the period corresponding to the short lifetime of the massive O-type stars which provide the source of the ionising radiation which power the nebular emission. Some of this information is encapsulated in the sSFR (Figure~\ref{fig:sSFR}), but we caution that the stellar masses returned in the fitting may miss very old stellar populations which do not significantly contribute to the observed light. Hence to analyse the recent variability (burstiness) of the star formation, that the observations are more sensitive to than the overall stellar mass,
in Figure \ref{fig:burst_intensity_alternative} we show the ratio of the intensity of the very recent SFR\footnote{For the ``short-term" SFR we consider the instantaneous SFR, as well as that averaged over the most recent 3\, Myr or 5\,Myr, and for the ``long-term" SFR we consider the remaining SFH extending back 100Myr.} to the average SFR over a longer timescale preceding this (extending to 100\,Myr\footnote{We note that many of our galaxies have luminosity-weighted ages younger than 100\,Myr (see Table~\ref{tab:beagle_results_update_1}), but for each galaxy we average the SFR over the entire preceding 100\,Myr for consistency, even if the galaxy formed more recently.}), and compare this with the equivalent width of [OIII]$\lambda5007$. 
We adopt a comparison period of 100\,Myr because this is the timescale over which the rest-frame UV continuum can be used as an SFR indicator (Section~\ref{sec:spectral_modelling}), and hence if star formation is intrinsically bursty on shorter timescales than this it will be revealed through a comparison of SFR$_{100}$ to the SFR inferred from the emission lines (which respond to star formation within $<10$\,Myr).

The majority of EELGs have recent SFRs which are more than twice the longer-term average (over the past $\sim$100Myr), with many showing a factor of 10 increase in SFR, implying that a recent sharp upturn in the SFH is required to produce an EELG. In these panels we draw on a horizontal line at the SFR$_{\rm short-term}/$SFR$_{\rm long-term} = 2$, this roughly splits the EELG and non-EELG samples albeit with some scatter. We remind the reader that the 750\AA\, [OIII]$\lambda5007$ EW is arbitrary and we do not necessarily expect two fully distinct samples to emerge (i.e., the population is continuous rather than bimodal).

As can be seen from Figure \ref{fig:burst_intensity_alternative}, galaxies with high short-term/long-term SFR ratios are typically EELGs, but we now briefly discuss those with high ratios and lower EWs.
In the lower panels of Figure \ref{fig:burst_intensity_alternative}, we repeat the 3Myr/3-100Myr-average SFR ratio and colour code by the total mass (left panel) and the luminosity-weighted-age (right panel) of the galaxies. Here we notice that the non-EELGs with SFR ratios above our horizontal line are low mass and have young luminosity-weighted-ages, yet exhibit low EWs. These highly-bursty but low-EW outliers could be due to unaccounted escape of ionizing photons (which is not included in the \texttt{BEAGLE} fits, where $f_{\rm esc}$ is set to 0), or alternatively we are catching them very shortly after an EELG phase ($\sim10$Myr) when the emission lines are dying away.

We note considerable scatter around these relations,
some of which results from the signal-to-noise of the observations but also potentially the parametric form of the SFH we impose for the \texttt{BEAGLE} fitting which may not be fully appropriate for a more complicated actual SFH. 
The scatter in Figure \ref{fig:burst_intensity_alternative} indicates that it may not be the SFH alone 
which governs the EW of the rest optical lines, but other factors such as dust attenuation and metallicity may play a role.

The high SFR$_{\rm{short-term}}$/SFR$_{\rm{long-term}}$ ratio we see in our EELG sample (with a median value of 4.2 compared to 0.8 for non-EELGs for the 3Myr short-term timescale) suggests that we are catching the EELG galaxies in an ``on phase" of active star formation which is short compared to the ``off phase" of less-active star formation (where the galaxy will not be an EELG in this part of the duty cycle). The high excess of SFR at recent times compared to the long-term average suggests we are witnessing a short duration burst in EELGs.
This is supported by the high sSFR we observe in our EELG sample in Figure \ref{fig:sSFR}.

\begin{figure}
    \centering
    \includegraphics[width=\columnwidth]{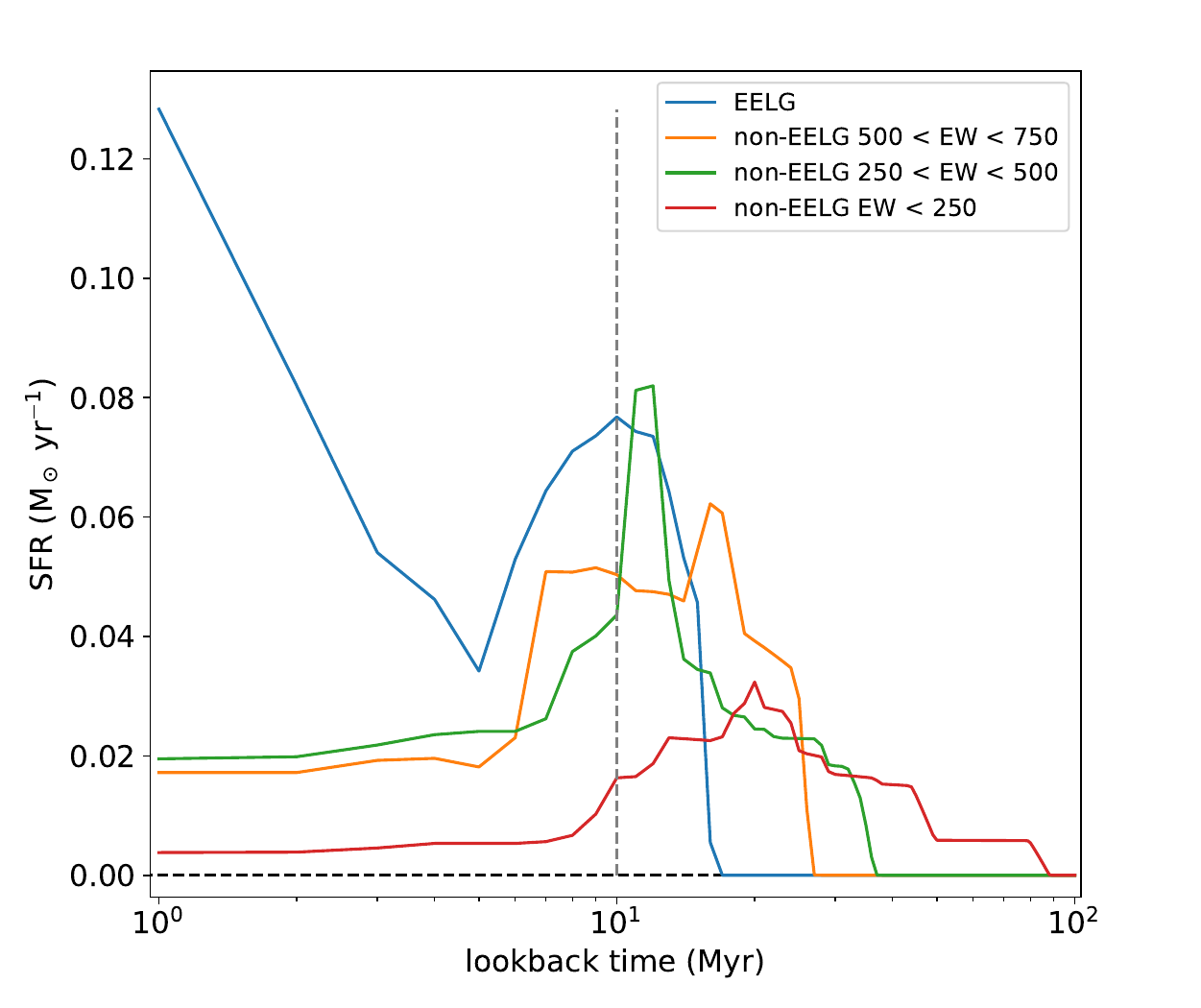}
    \caption{Median stacks of the \texttt{BEAGLE} SFHs for 4 [OIII]$\lambda5007$ EW bins ($EW<250$\AA, $250<EW<500$\AA, $500<EW<750$\AA, $EW>750$\AA), each stack normalised to have equal mass formed.}
    \label{fig:average_sfh}
\end{figure}

To examine the SFH as a function of EW in our sample, we stack the SFHs over the last 100Myr of each galaxy into four [OIII]$\lambda5007$ EW bins. To produce each stack we take the \texttt{BEAGLE} SFH with a 1Myr timestep and normalise each individual SFH so the mass formed in the last 100Myr is equal, then for each EW bin we stack the subsets performing a median average. We re-normalise the stacks to have equal mass and present these stacks in Figure \ref{fig:average_sfh}. We find that the stacks of the higher the EW bins have a greater proportion of star formation occurring in the most recent 10Myr, as expected for the time period over which the nebular emission occurs. 
For the bins of lower EW, the stars are on average older with star formation in the last 10Myr lower compared to activity at earlier times. 
Interestingly, we note that the moderate EW stacks ($250<EW<500$ and $500<EW<750$) exhibit a peak of star formation just before a  lookback time of 10Myr. Such a bump in a SFH at this lookback time would generate a blue UV spectral slope $\beta$ yet yield a low EW of nebular emission lines. This SFH may reflect a scenario where star formation has recently been quenched in these galaxies. 
Alternatively perhaps, these galaxies may be actively star forming at this current epoch but may have high escape of ionising photons and hence have lower EW. 

\begin{figure*}
    \centering
    \includegraphics[width=\textwidth]{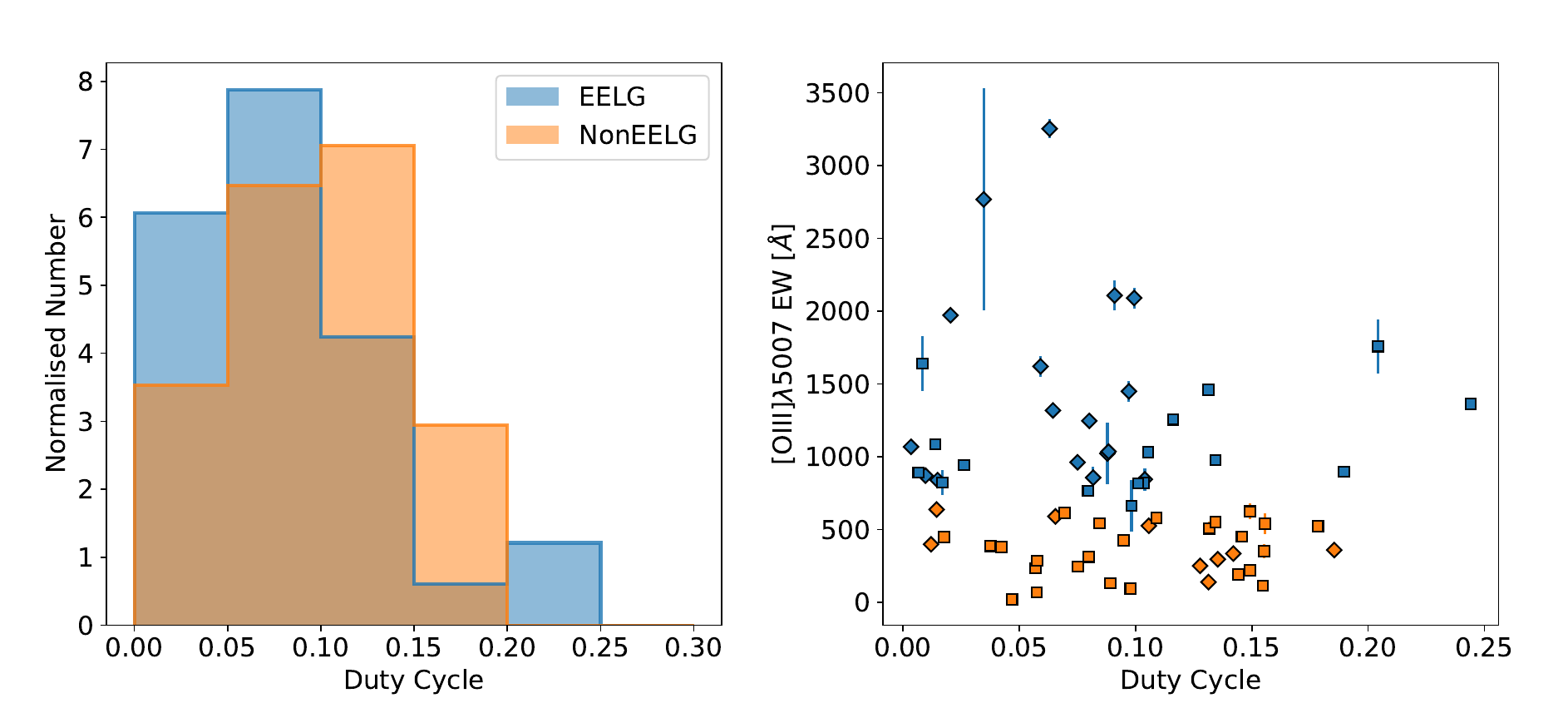}
    \caption{Duty cycle, defined as the fraction of time over the last 100Myr that a galaxy has spent in an \lq\lq active\rq\rq phase, when the SFR exceeds three times the 100Myr average. Left: Normalised histogram of the duty cycle for EELGs (blue) and non-EELGs (orange). Right: The distribution of [OIII]$\lambda5007$ EW as a function of duty cycles, colour coded as the left-hand panel. We distinguish our high- and low-redshift galaxies (split at $z=5.7$) as diamonds and squares, respectively.   
    We remove galaxies from these figure that never exceed the active phase SFR threshold.}
    \label{fig:Duty_cycle}
\end{figure*}

Finally, using our \texttt{BEAGLE} SFHs we can measure a duty cycle for each galaxy. 
Here, we define the duty cycle as the fraction of the time (over the last 100Myr) where the SFR of the galaxy exceeds our approximate threshold for an EELG (when the measured SFR exceeds the long-term average SFR by a factor of 3). The fraction of the time spent in this “active phase” is shown in Figure \ref{fig:Duty_cycle}.
What is apparent is that EELGs tend to have short duty cycles, spending an average of 5-10$\%$ of the last 100Myr in an active phase. The non-EELGs show a marginally more extended distribution of duty cycles, although we caution that the parametric form of the SFH we adopt (a recent burst, preceded by a delayed exponential) may be less appropriate for systems which have undergone a recent down-turn in star formation (due perhaps to quenching). Hence the duty cycle inferred for the non-EELGs is less reliable.  
It is likely that a sub-set of non-EELGs in our sample might actually be observed to be EELG if we were to catch them at the right time in their SFH, but the more extended distribution duty cycles might imply there may also be a population there that has a much smoother variation of SFR (although again we caution that the non-EELG duty cycle determination may be subject to uncertainty due to the parametric form of the SFH adopted). However, overall our results indicate that star formation at high redshift has a significant contribution from bursty systems.

\subsection{EELGs as bursty star formers: a repeating burst model}

The BEAGLE fits we performed involved an ongoing starburst of variable duration following a delayed exponential SFH. 
At high-redshift bursty star formation plays an increasing role \citep{Looser23, Dressler24, Sun23b}, to assess whether EELGs could be such bursty galaxies caught in an on-phase we introduce a toy model
for the SFH, whereby there are repeated identical ``top hat'' bursts of star formation which are regularly spaced in time. We vary the duration and frequency of these bursts, and consider the equivalent width of the [OIII]$\lambda5007$ emission as a function of time, assessing for each burst duration and frequency the fraction of the time that our EELG criterion of $EW>750$\,\AA\ would be met. 

\begin{figure*}
    \centering
    \includegraphics[width=\textwidth]{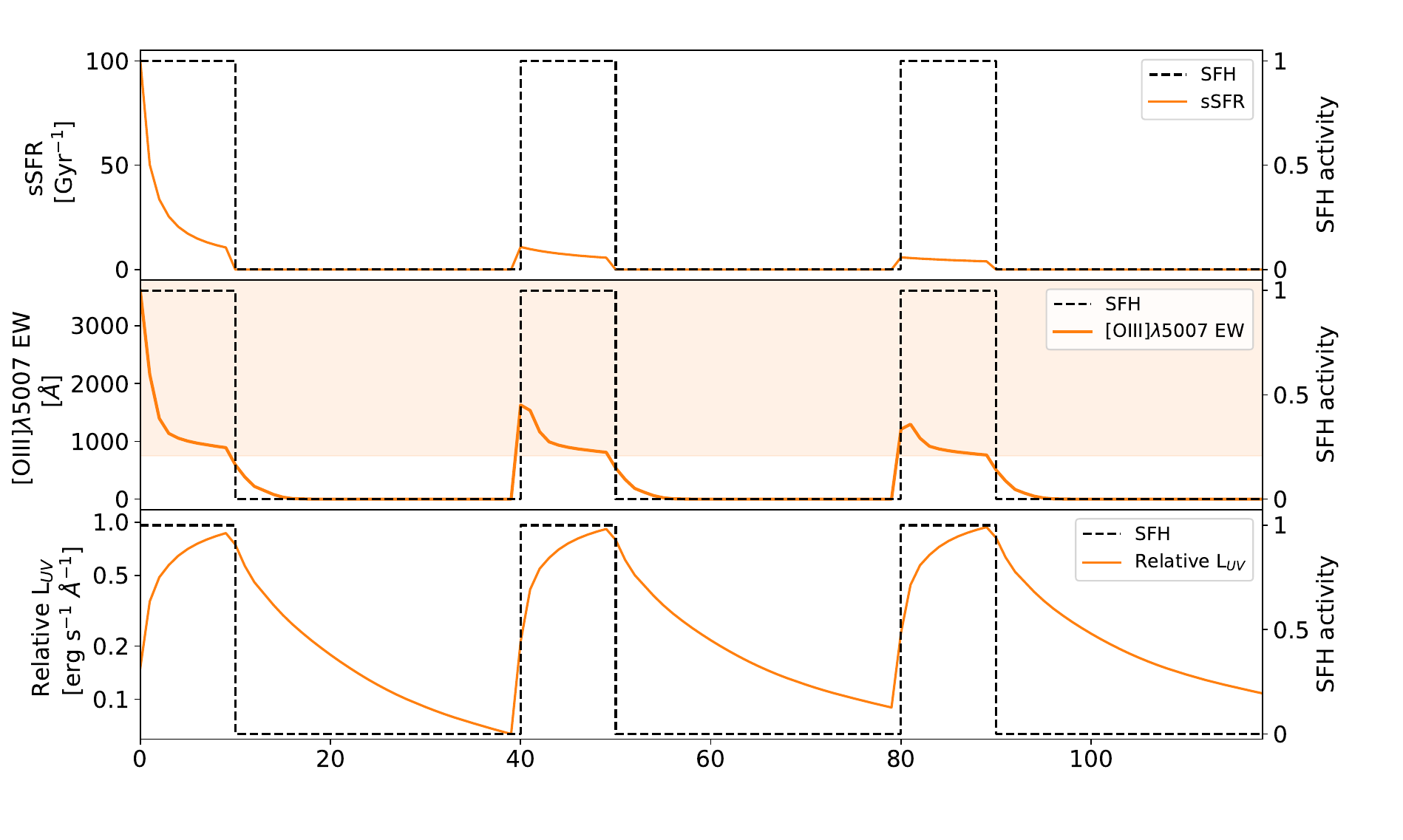}
    \caption{The evolution of the sSFR (Top panel), [OIII]$\lambda5007$ EW (Middle panel) and UV luminosity (Bottom panel) following a repeating ``top hat'' SFH of burst duration T$_{\rm burst}=10$Myr every T$_{\rm repeat} = 40$Myr (shown in black in both panels). Our fiducial model adopts log$_{10}$(U)=-2 and 10\% solar metallicity. 
    The right y-axis in each panel shows the normalised SFR (dashed black line) with the evolving parameter shown in orange following the left y-axis. In the central panel the shaded region marks the EELG EW threshold of [OIII]$\lambda5007 > 750$\AA. 
    }
    \label{fig:app:EW_history}
\end{figure*}

In Figure \ref{fig:app:EW_history} we show an example of the evolution of the EW([OIII]) for a burst of 10\,Myr duration which repeats every 40\,Myr (i.e. a star formation duty cycle of 25\%). We adopt a 10$\%$ solar metallicity and $\log_{10}(U)=-2$ fiducial model (aligning closely to the properties of our EELG sample, see section \ref{sec:ionisation}, the lower panel of Figure \ref{fig:r23_o32} and the lower-left hand panel of Figure \ref{fig:O3_ha_hb}).
We find that galaxies following this SFH continue to meet the EELG threshold even after several successive bursts. However, the maximum EW achieved in each successive burst decreases due to the older stellar population starting to build up (as reflected in the sSFR in Figure \ref{fig:app:EW_history}) and significantly contributing to the continuum around 5007\,\AA , diluting the EW of the [OIII].

\begin{figure}
    \centering
    \includegraphics[width=\columnwidth]{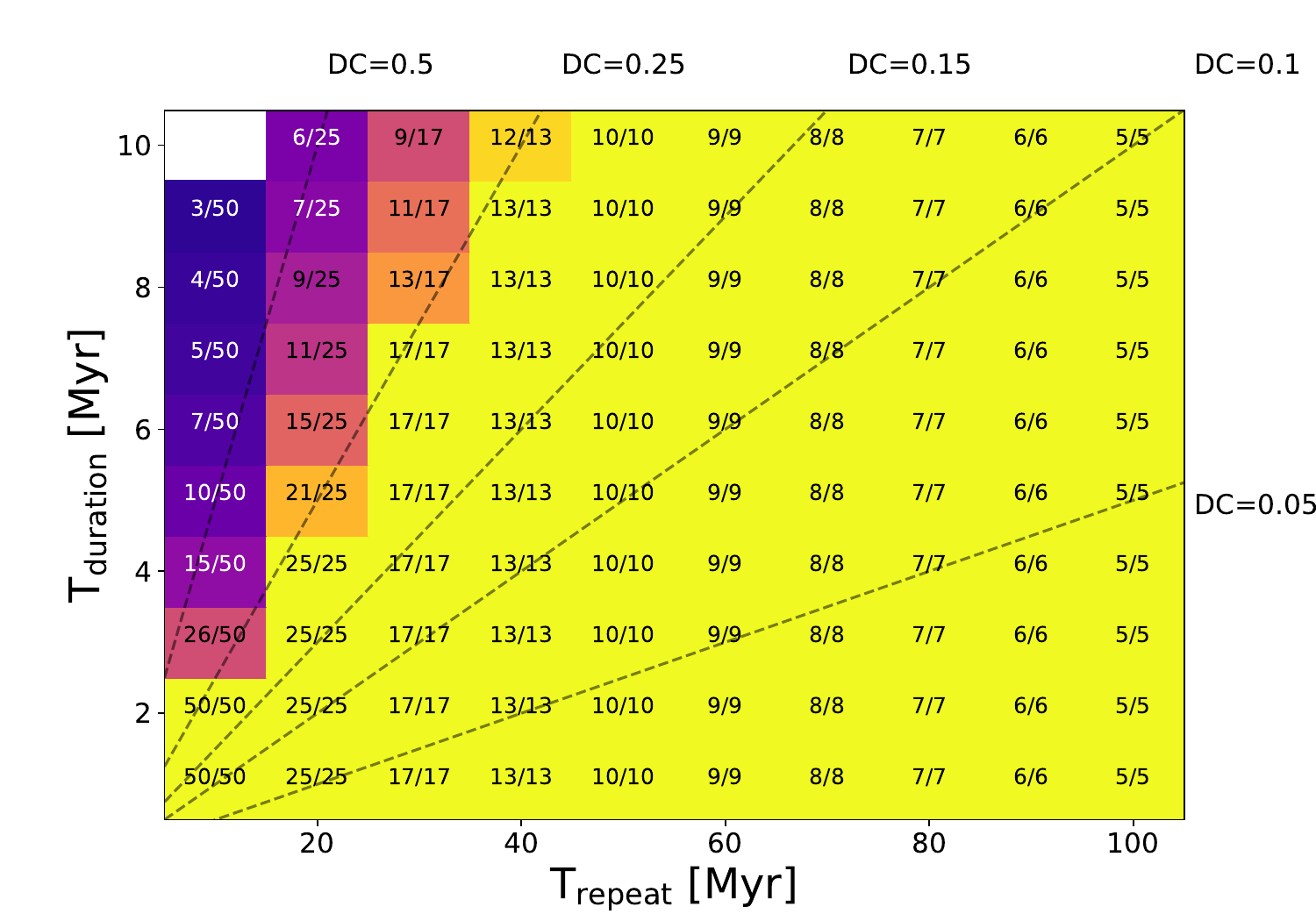}
    \includegraphics[width=\columnwidth]{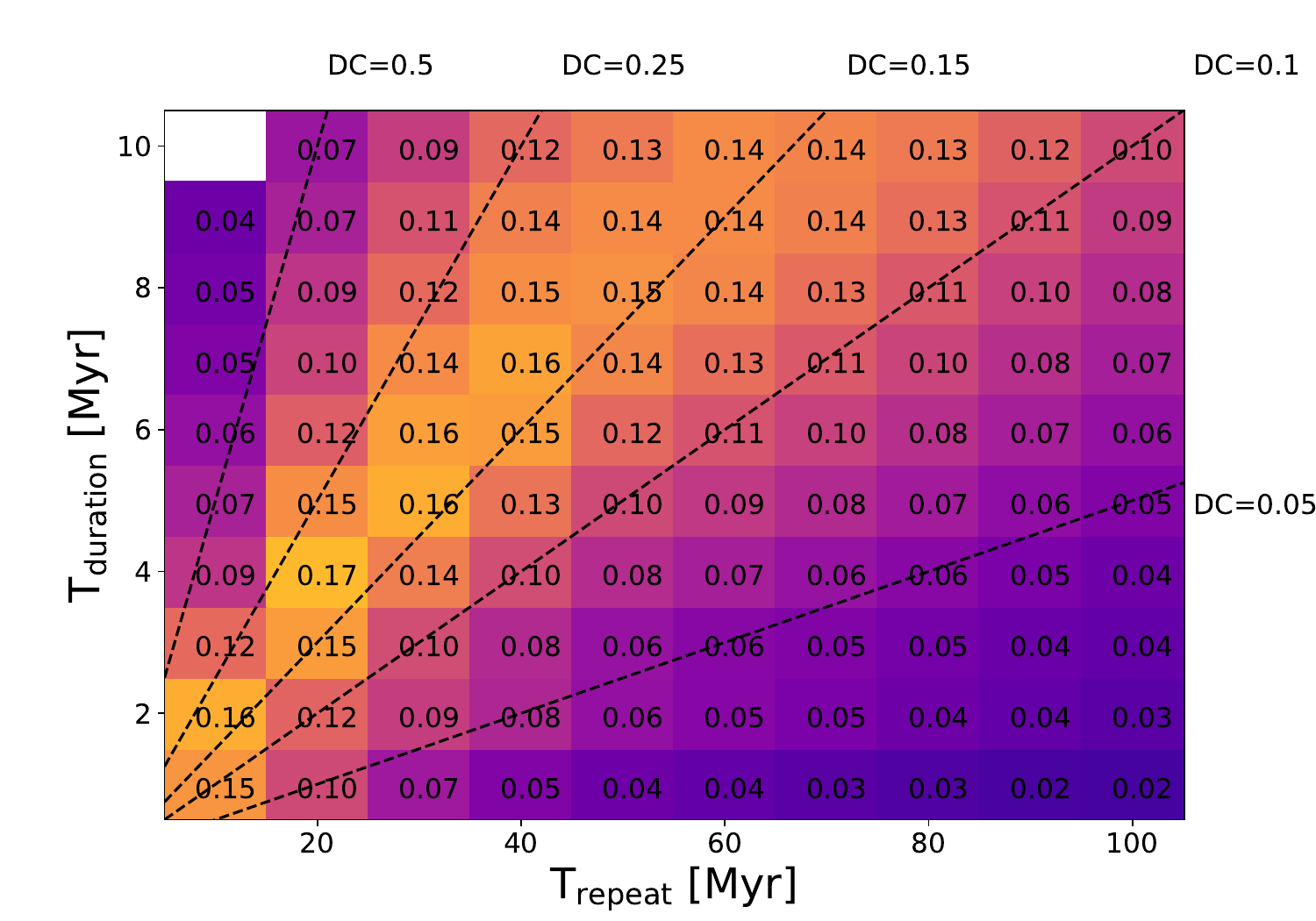}
    \caption{EELG on-phase as part of a repeated ``top hat'' burst toy model for a grid of burst duration (T$_{\rm duration}$) and frequency (how often bursts repeat, T$_{\rm repeat}$). Toy model run using our fiducial case (0.1$Z_\odot$, $\log_{10}$(U)=-2). Top: The fraction of repeated bursts over a 500Myr period in which a galaxy achieves an EELG-phase. Bottom: The fraction of time (over a 500Myr period) a galaxies spends in an EELG-phase. In both panels dashed diagonals present lines of constant star forming duty cycles (DC=T$_{\rm duration}$/T$_{\rm repeat}$). }
    \label{fig:app:time_spent}
\end{figure}

We consider different burst durations and frequencies up to an age of 500\,Myr (which is about the age of a galaxy at the median redshift of our EELG sample, $z\approx 5.7$, which formed at $z\approx 10$), and in the top panel of Figure \ref{fig:app:time_spent} we show the fraction of the repeating bursts up to 500\,Myr which reach an EELG phase. 
The bottom panel of Figure \ref{fig:app:time_spent} shows the total fraction of time over the 500\,Myr of repeating bursts that a galaxy would be observed as an EELG.

For a star forming duty cycle (DC = T$_{\rm duration}$ / T$_{\rm repeat}$) of less than 0.1, every burst of star formation results in a short lived EELG-phase. 
For higher DCs than this, the accumulation of low mass stars (that contribute to the stellar continuum at 5007\AA\, but not to the line emission) prohibits the galaxy from reaching high EW in subsequent bursts, meaning that the galaxy no longer can enter an EELG phase after several cycles of star formation. 

The region of parameter space in our toy model that delivers the greatest time spent as an EELG over 500Myr has a DC$\sim0.15$. Towards higher DC, the EELG-phase is prohibited in later bursts (as a significant older population of stars has built up, diluting the line emission), and at lower DC the galaxy spends such little time in a burst that the total accumulation of time as an EELG is short. Therefore, we are most likely to catch an EELG with a duty cycle of 10-20$\%$ (model dependent).

Although the toy-model of repeating bursts presented here is not exactly the same as the SFH adopted for the BEAGLE SED fitting, both approaches indicate that star formation duty cycles of $\approx 10-20$\% are likely to yield galaxies with an EELG phase.

Taken at face value, it seems curious that EELGs comprise 42\% of our spectroscopic sample (rising to 62\% for the $z>5.7$  sub-sample), while from the SFH we infer a short ``on phase'' where a young galaxy will spend only $\sim 10$\% of its time as an EELG. This can be reconciled when we consider that our high-redshift sample is selected in the rest-frame UV. In our toy model, we assume top hat bursts with zero star formation in the off-phase. As can be seen from Figure \ref{fig:app:EW_history}, after a burst of duration $\sim10$Myr the UV flux at 1500\AA\, our preferred model  will drop by a factor 5 over the first 10Myr and by a factor 10 after 20Myr. 
With such UV dimming it may not enter our UV-selected sample, as seen in Figure \ref{fig:muv_z}. This implies that galaxies in an EELG phase can be boosted into our sample during an on-duty phase, but would be missed during the off-duty phase. This also explains why the EELGs have lower stellar masses on average than the non-EELGs (which would be detectable in our spectroscopy even if they have much lower sSFRs than the EELGs). Lower mass galaxies are expected to be much more numerous (the stellar mass function may be Schechter in shape, e.g., \citealt{Weibel24}), but only those in a transitory phase of intense star formation will be bright enough to appear in our spectroscopic sample.
 
\section{Discussion}\label{sec:discussion}

Having identified a sample of EELGs and examined how their galaxy properties (line diagnostics, UV-properties, SFHs) compare to non-EELGs, we can now assess what conditions are required for a galaxy to enter an EELG phase. We will discuss what physical characteristics drive a galaxy to have extreme equivalent width line emission in Section \ref{sec:disc:mechanism}. We will then consider these conditions in the context of the high rate of Lyman-$\alpha$ emission line detections in our EELG sample in Section \ref{sec:disc:LAE}. Finally, we will discuss in Section \ref{sec:disc:evolution} how the fraction of SFGs in an EELG phase changes with redshift and what evolution in galaxy properties may drive this. 

\subsection{What causes SFGs to enter an EELG phase?}
\label{sec:disc:mechanism}
From our sample of EELGs, we have determined consistent characteristics that all EELGs exhibit, and what observational features are distinct from non-EELGs. We determine that galaxies can enter an EELG phase if they meet the following conditions:
\begin{enumerate}
    \item  \textbf{They contain a young stellar population,
    with the highest EWs coming from stellar populations caught in the first few Myr.} This ensures the presence of massive, hot O-type stars in the stellar population which drive the rest-UV nebular emission through their ionising UV emission. The blue rest-UV spectral slopes ($\beta$) typically seen in EELGs are consistent with this very young stellar population dominating the light, and the SED fitting also finds very low mass-weighted ages (compared with non-EELGs).
    
    \item \textbf{These young stellar populations must exhibit low levels of chemical enrichment and dust.}
    Many EELGs are consistent with no dust -- we note that differential dust reddening between the nebular emission lines (the star-forming regions) and the surrounding continuum from older stars will decrease the observed EW.
    In terms of metallicity, our EELG sample conforms to the trend that the highest EWs are being produced in galaxies with low oxygen abundance (sub-solar, although not extremely metal poor). 
    This reflects that the collisionally excited [OIII]$\lambda5007$ emission line luminosity increases with the electron temperature, which increases at lower oxygen abundances due to the restriction in the metal line cooling mechanisms (in extremely metal poor environments the lack of oxygen atoms will reduce the luminosity). The EELGs also exhibit high O32 ratios (and correspondingly high ionisation parameters), typical of low-metallicity galaxies at high redshift.
    
    \item \textbf{These young stellar populations must be part of a rising star formation history} - either a very recent burst of star formation activity, or a more extended SFH which has recently seen a rapid upturn in SFR. 
    Typically, EELGs in our full sample exhibit a greatly elevated SFR in the recent past compared with the long term average. With EELGs typically having a short-term/long-term SFR ratio $>3$.
    
    \item{\textbf{The mass created in the recent burst (in the last 10Myr and responsible for the nebular emission) must make up a significant fraction of the total mass}. This is seen in the young mass- and luminosity-weighted-ages among the EELG sub-sample which are below 10Myrs. These young ages reflect that there is no significant older stellar population that would produce continuum in the rest-optical strong enough to dilute the EW.
    }
 
\end{enumerate}

From our sample, these are all necessary conditions for an EELG. We note there are galaxies that meet these conditions and yet are not EELGs. Hence, these appear to be necessary but not sufficient conditions for galaxies to be in an EELGs phase. For example, we note that galaxies meeting all of these conditions but also exhibiting very high escape fractions of ionising photons will show low EWs, as the Lyman continuum photons leak out before generating recombination and nebular lines. 
Therefore it is plausible that there remains a population of ideal ionising photon producers that are also ultra-efficient emitters (specifically Lyman-leakers) that will not be captured by an EELG selection. 

We note that one EELG in our sample (3$\%$, 1/36, ID:8083) exhibits a broad H$\alpha$ emission line and a narrow [OIII]$\lambda5007$ emission line, suggesting that the emission we are observing is partially coming from a broad-line region around an AGN (see \citealt{Maiolino23arXiv}). None of the other EELGs in our sample exhibit a broad line component to any recombination line. Therefore in at least one system the ionising radiation field is not solely generated by star formation but may have a contribution from other hard ionising sources (e.g., an AGN).   

\subsection{Connection to LAEs} \label{sec:disc:LAE}

As we reported in Section \ref{sec:balmerdec}, the fraction of galaxies with detected Lyman-$\alpha$ emission is higher in our EELG samples than in our non-EELG samples, with $53\%$ (9/17) of high-redshift EELGs determined to also be Lyman-$\alpha$ emitters. 
This measurement is inline with \citet{Chen23arXiv} who determined that 9/24 (38$\%$) galaxies with [OIII]+H$\beta$ EW$> 1000$\AA\, (roughly [OIII]$\lambda5007$ EW $> 750$\AA\, following the \citealt{Mengtao19} relations) have Lyman-$\alpha$ detections, which they use to determine that $50\pm11\%$ of $z\sim6$ galaxies exhibit high Lyman-$\alpha$ escape fraction ($f_\mathrm{esc}^{Ly\alpha}>0.2$).
The production of Lyman-$\alpha$ photons may be expected to be higher in EELGs considering that within these active star-forming galaxies, the young stellar populations will produce ionising photons which should yield Lyman-$\alpha$ recombination line emission.
However, typically Lyman-$\alpha$ photons are strongly suppressed within the galaxies through resonant scattering in an optically thick environment. 
In the case of LAEs where Lyman-$\alpha$ emission does escape, our observations are aligned with sight-lines exhibiting a lower column density of neutral hydrogen that exists within these systems and hence allows the Lyman-$\alpha$ photons to escape in our direction. 
The high LAE fraction (53$\%$) in our high-redshift EELG sample, suggests that these EELGs have a low covering fraction of neutral gas, creating channels of lower column density, which may have been generated by a bursty SFH (which may ionise and/or expel gas, e.g., \citealt{Katz23B}),
although we note that there must be some remaining neutral gas in the ISM to produce the line emission that we observe in our EELGs.

Furthermore, for the Lyman-$\alpha$ photons that are produced in a galaxy and escape from the interstellar and circum-galactic medium (ISM and CGM), our observation of any Lyman-$\alpha$ emission additionally depends on the transmission of these photons through the local IGM surrounding the galaxy. 
At high-redshifts these LAEs likely lie within large ionised bubbles, either created by their own stellar populations or from close proximity neighbours. 
The large ionised bubbles allow the transmission of Lyman-$\alpha$ photons such that they can travel unimpeded until they are redshifted to longer wavelengths before they interact and are scattered by any neutral hydrogen in the IGM.
Although a visual inspection of the 9 high-redshift EELG-LAEs shows that all are compact isolated systems, the JADES survey and FRESCO data \citep{Oesch23arXiv} have revealed large-scale structures at various redshifts in GOODS-South, and several of our EELGs which exhibit Lyman-$\alpha$ emission lie within known overdense regions at $z=5.8-5.9$ \citep{Helton23arXiv, Stanway04, Stanway07} and $z=7.3$ \citep{Helton23arXiv, Endsley23arXiv}, as can been seen in Figure \ref{fig:redshift_dist}. As noted by \citet{Witstok23arXiv}, these overdensities may lead to a highly-ionised IGM and enhanced transmission of Lyman-$\alpha$.  
However, not all of our EELG-LAEs coincide with these known overdense regions and it is unlikely that these systems created their own ionised bubbles considering how young the stellar populations are. Instead, they may indicate the presence of previously undetected overdense regions populated by galaxies below our observational sensitivity. 

In our low-redshift EELG sample the fraction of EELGs with detected Lyman-$\alpha$ emission is lower at 33$\%$ than the high-redshift EELG sub-sample. 
This is consistent with the findings of \citep{Stark10, Stark11} who determine the fraction of LBGs that are Lyman-$\alpha$ emitters ($X_{Ly\alpha}$) increases with redshfit from $z=4$ to $z=6$ (the end of the EoR).
Since the IGM plays an increasing role as $z=6$ is approached (and into the EoR), the evolution of the true Lyman-$\alpha$ emission, after IGM absorption has been corrected for, may even be more dramatic.

\subsection{Evolution of the EELG fraction}
\label{sec:disc:evolution}

\begin{figure*}
    \centering
    \includegraphics[width=\textwidth]{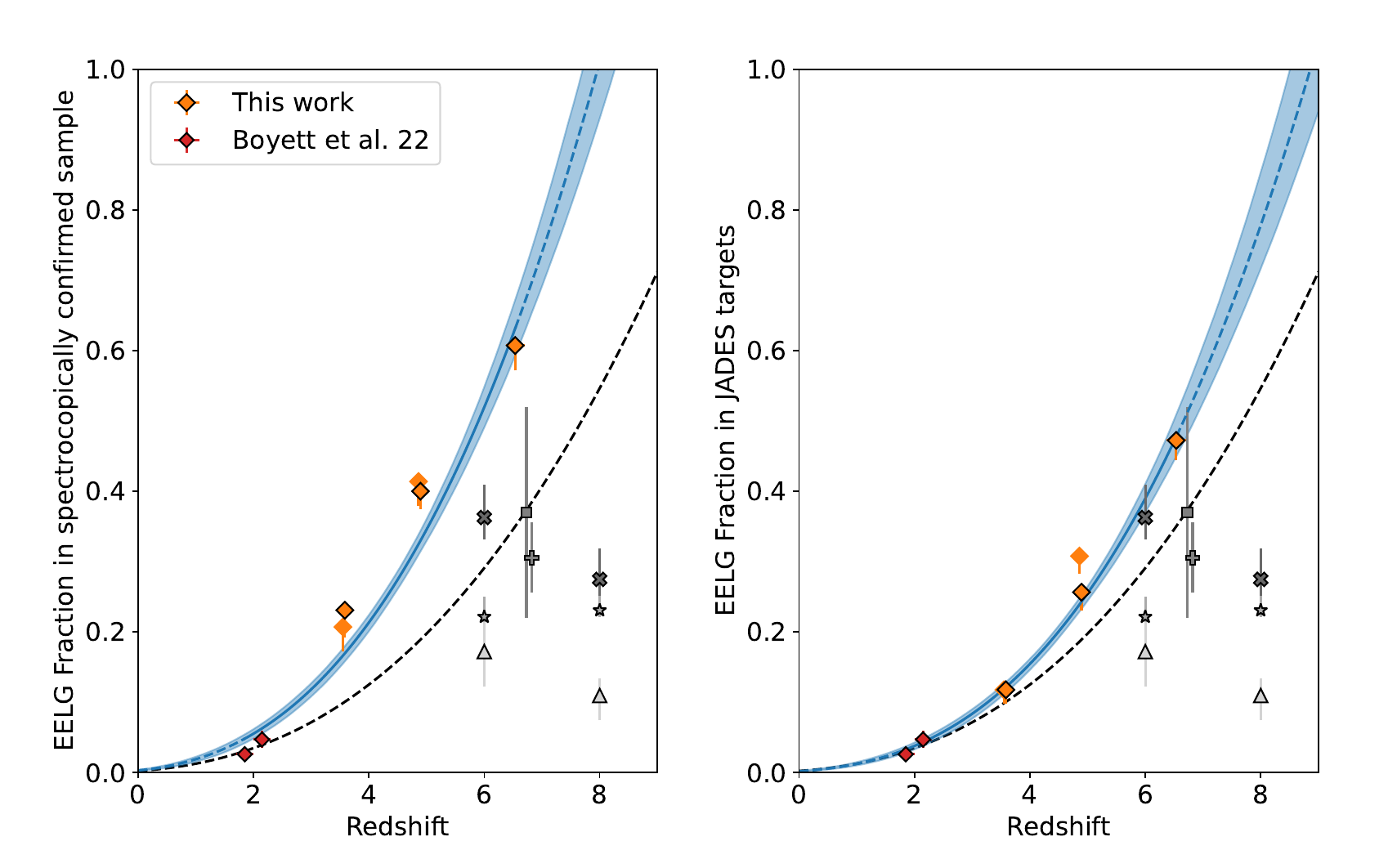}
    \caption{The observed fraction of galaxies in an EELG phase ([OIII]$\lambda5007$ EW$>750$\AA) as a function of redshift. In the left panel we plot the EELG fraction within our full sample (those with spectroscopically confirmed redshift), and in the right hand panel we plot the EELG fraction additionally considering JADES targets which did not have multiple emission line detections and were excluded from our full sample (under the assumption that the photometric redshift is correct and the lack of extreme emission lines means these are not EELGs).
    In both panels, our full sample is split into three redshift bins of roughly equal size (orange diamonds, $\blacklozenge$). The two lower redshift bins ($z<5.7$) are shown before and after an apparent UV magnitude cut at mag$=29$ (shown without and with a black border, respectively) is applied to match our high redshift bin. 
    We overlay the spectroscopically determined $z\sim2$ EELG fraction from \citet{Boyett22A} (red diamonds, $\blacklozenge$) and combine these with the data points in our work to determine the best fit power law relation and $1\sigma$ uncertainty (blue line and shaded region). 
    We additionally overlay in grey EELG fractions determined from photometric surveys. The \citet{Endsley20} $z\sim7$ IRAC/Spitzer EELG fraction is shown as a square ($\blacksquare$), with the \citet{Boyett22A} best fit relation using this point shown as the black dashed line. 
    Crosses, stars and triangles (\ding{54}$, \bigstar, \blacktriangle$) present the \citet{Endsley23arXiv} JADES/NIRCam EELG fractions for their bright, faint and very faint M$\rm_{UV}$ samples.
    The plus (\ding{58}) shows the \citet{Endsley23A} CEERS/NIRCam EELG fraction.
    }
    \label{fig:abundance}
\end{figure*}

As we have discussed, galaxies may spend periods of their SFH in an EELG phase when they meet a set of ideal conditions. We can determine the fraction of galaxies in our survey which are in an EELG phase (hereafter, the EELG fraction) and look at how this fraction may change with redshift. 

Our full JWST/NIRSpec sample allows us to measure the EELG fraction both in the EoR and at lower-redshifts ($3<z<6$).
In the left hand panel of Figure \ref{fig:abundance} we show the fraction of galaxies in our full sample that exhibit an [OIII]$\lambda5007$ EW $>750$\AA\, split into 3 roughly equal size redshift bins (our high-redshift sample at $z>5.7$ becomes one bin, and we divide our low-redshift sample equally into two bins at $3<z<4.1$ and $4.1<z<5.7$). We note that at $z>5.7$ we have a UV selected sample and we create a similar cut for the two lower redshift bins by removing galaxies fainter than an apparent UV AB magnitude $= 29$ (See sample selection in section \ref{sec:selection}  and the M$_{\rm{UV}}$ distribution of our sample in Figure \ref{fig:muv_z}).  
To compute the uncertainty on our EELG fractions we randomly vary the EW of each individual galaxy according to their measurement uncertainty and determine the fraction that have an EW above our EELG threshold, we repeat this $n=1000$ times and report the standard deviation on the resultant EELG fraction distribution as the fraction uncertainty.  

To make a fair comparison over a range of redshifts, we include studies based on a similar UV selection. Prior to JWST, measurements of the EELG fraction in rest-frame UV selected surveys had been made at $z\sim2$ using HST/grism spectroscopy (\citealt{Boyett22A}, for M$\rm_{UV}<-19$\,mag) and at $z\sim7$ using Spitzer/IRAC photometric flux excess \citep{Endsley20}. 
Comparing these two samples, \citet{Boyett22A} determined that the EELG fraction increased with redshift by a factor of 10, from $\sim4\%$ at $z\sim2$ to $\sim40\%$ at $z\sim7$. We overlay the \citet{Boyett22A} relation for the redshift evolution of the EELG fraction in Figure \ref{fig:abundance}. We additionally overlay two recent photometric [OIII]+H$\beta$ EW distribution studies by \citet{Endsley23A, Endsley23arXiv} at $6<z<9$, and here we determine the EELG fraction adopting the [OIII]-H$\beta$ EW relation from \citet{Mengtao19}. We note that the $z\sim6$ and $z\sim8$ samples from \citet{Endsley23arXiv} have been split into a bright (M$\rm_{UV}<-19.5$), faint ($-19.5 < $M$\rm_{UV} < -18$, ) and a very faint (M$\rm_{UV}>-18$) sub-sample.

We find that the EELG fraction in our spectroscopically confirmed sample increases with redshift, from $0.23^{+0.04}_{-0.01}$
in our lowest redshift bin ($3<z<4.1$) to $0.61\pm0.04$ 
in our highest bin ($z>5.7$), after we have applied the UV-selection (see left panel in Figure \ref{fig:abundance}). 
We measure consistent EELG fractions (within $1\sigma$) with and without applying our additional UV selection criteria, for both the lowest redshift and intermediate redshift bins. Comparing our results to literature values, the EELG fractions in our full sample exceed the measurements determined in photometric surveys at similar redshifts. We fit a power law $(1+z)^P$ relation to the spectroscopic EELG fractions from our full sample and those from \citet{Boyett22A}, and determine a best fit $P=2.6\pm0.1$ and the expected normalisation of the EELG fraction at $z=0$ to be $0.30\%\pm0.01\%$. This relation means that half of all galaxies in a UV-magnitude selection such as ours for which we have spectroscopic redshift are in an EELG phase at $z>6$. 

The larger EELG fraction in our full sample compared to high-redshift photometric surveys may be down to our initial sample selection requirement of multiple detected emission lines in our NIRSpec spectroscopy. This restriction will remove galaxies from our sample that have low EWs, where the corresponding line emission is below the sensitivity of our observations. To correct for this bias, in the right hand panel of Figure \ref{fig:abundance}, we plot the EELG fraction for the same bins where we now consider the complete list of JADES targets for which spectra were obtained. If we make the assumption that the 133 galaxies with expected redshifts $z>3$ but without multiple NIRSpec line detections (which were removed during our sample selection) have correct photometric redshifts, then the EELG fractions reduce. 
In the lowest redshift bin ($3<z<4.1$) the EELG fraction changes from $0.23^{+0.04}_{-0.01}$ using the sample with robust redshifts from multiple emission lines to $0.12\pm0.02$ using the full JADES target list. In the middle redshift bin ($4.1<z<5.7$) the change in the fraction is from $0.41\pm0.03$ to $0.31\pm0.03$. In the highest redshift bin ($5.7 < z <9.5$) the fraction changes from $0.61\pm0.04$ based on the robust redshift sample to $0.47\pm0.03$ using the entire JADES targets. 
While this assumption, that the photometric redshift is correct in the absence of a spectroscopic redshift, may not be robust, and many targeted galaxies may not have line detections for reasons beyond low EW (e.g., incorrect photometric redshifts, placement within the slit), it does demonstrate that by considering non-detections the EELG fraction is lower. In our lowest redshift bin there is better agreement with the \citet{Boyett22A} best fit relation. However, in our highest redshift bin we still have a larger EELG fraction than expected from the extrapolation of the \citet{Boyett22A} relation. We also find good agreement with the $z\sim6$ \citet{Endsley23arXiv} \lq bright\rq\, sample (M$\rm_{UV}<-19.5$). 
We again determine a best fit power law model, with a measured $P=2.8\pm0.1$ and the expected normalisation of the EELG fraction at $z=0$ to be $0.18\%\pm0.01\%$. This initial JADES spectroscopy from DR1 used for this paper is largely based on a selection from HST imaging, and was by design and necessity (given the short wavelengths) primarily a rest-frame UV-based selection. Hence we have compared the EELG fraction in our UV-based selection with comparable selections in the literature, accounting for the impact on completeness introduced by only a subset of our spectroscopically-targetted galaxies having multiple emission lines for a robust redshift. JADES spectra taken later in the survey benefit from a selection based on JADES JWST/NIRCam imaging in the GOODS fields (as well as HST photometry), enabling use to select a sample based on longer rest-frame wavelengths which are more sensitive to the underlying stellar mass rather than the recent SFH. In a future paper we will look at the evolution of the EELG fraction with redshift in a stellar-mass limited sample, as well as in a rest-frame UV sample significantly larger than that used in this paper (which will enable consistent UV absolute magnitude cuts to be used in many redshift bins while retaining large numbers of galaxies for robust statistics).

To consider why the observed EELG fraction within our UV-selected sample may change with redshift, we revisit the conditions that must be satisfied for a galaxy to enter an EELG phase. 
Many of the galaxy properties that we have based criteria on are known to evolve with redshift.
At fixed stellar mass, simple galaxy models predict that the sSFR is expected to increase with redshift \citep[e.g.,][]{Tacchella18}, driving higher EWs.  
The average metallicity and dust content is also lower in the galaxy population at higher redshift \citep[e.g.,][]{Langeroodi22arXiv, Curti23barXiv}, further boosting the [OIII] emission line luminosity. 
The evolution of these galaxy properties means the average observed [OIII] EW in the galaxy population will increase with redshfit and hence a greater proportion of the galaxy population at high redshift will meet the conditions needed to enter an EELG phase. 

Although harder to constrain through observations, the star formation duty cycle within the galaxy population is also expected to change with redshift. Evolution in the galaxy merger rate, galaxy dynamical timescale and rate of gas accretion, all affect the fraction of time a galaxy spends in an active phase of star formation \citep{Kimm15, Ceverino18, Faucher-Giguere18, Furlanetto22}. These changes may allow galaxies to enter an active star forming phase more frequently at high redshift, again increasing the fraction of objects observed in an on-phase as EELGs. 

Interestingly, \citet{Endsley23arXiv} measure a decrease in the EELG fraction for each of their M$\rm_{UV}$ bins between $z\sim6$ and $z\sim8$. This suggests that the evolution turns over at very high redshift, which is not captured by our best fit power law relation. Such a turn over may be due to a further decrease in the metallicity limiting the [OIII] emission line luminosity or other changes in the galaxy population properties (e.g., $f_\mathrm{esc}$). From this initial DR1 data release, we do not have a large enough sample at $z\sim8$ to place statistically significant constraints on the EELG fraction to investigate the presence of a turn-over. However, the full JADES spectroscopic survey should have sufficient size to test this, and this will be the subject of a future paper.

Finally, we note that because of the greater fraction of galaxies in an EELG phase at higher redshift, and the evidence for EELGs being productive ionising photon producers (high $\xi_\mathrm{ion}^\mathrm{HII}$, see Section \ref{sec:beta}), galaxies in an EELG phase therefore have the potential to play an important role in the reionisation of the Universe. 

\section{Conclusions}

Through a JWST/NIRSpec spectroscopic sample of SFGs over a broad redshift range ($3<z<9.5$), we examine what galaxy properties cause a galaxy to enter an EELG phase. 
Out of our sample of 85 SFGs, 42\% (36/85) exhibit [OIII]$\lambda5007$ rest-frame equivalent widths consistent with being $>750$\,\AA .
This sub-sample of EELGs shows high EWs in both the [OIII]$\lambda5007$ and H$\alpha$ emission lines. 
They are characterised by a hard ionising radiation field, observed through the ionisation state of the ISM (with O32 $\gtrsim10$); a sub-solar, although not extremely metal poor, oxygen abundance ($7<\log_{10}(\mathrm{O/H})<8.5$); a high ionising photon production efficiency (with mean $\log_{10}(\xi_\mathrm{ion}^\mathrm{HII}/{\rm erg^{-1}Hz})=25.5\pm0.2$); and low levels of dust attenuation (consistent with minimal or no dust, with a mean $A_{1600}=0.4\pm0.1$). These properties boost the [OIII]$\lambda5007$ line luminosity, and hence the EW. 
The high ionisation state of the ISM and the high emission line EWs observed in EELGs is driven by the hot O-type stars in young stellar populations, which we also observe in the characteristically blue UV spectral slopes ($\beta$) of our EELG sample (with mean observed $\beta=-2.2\pm0.1$, where we quote the standard error on the mean). 
Through SED modelling of the full prism spectrum using \texttt{BEAGLE}, we constrain the SFHs of our sample, determining that EELGs typically exhibit a
young luminosity-weighted ages ($<5$Myr) and a recent SFR (either instantaneous, or averaged over a 3 or 5Myr lookback time) that exceeds the long term average  (measured out to a lookback time of 100Myr preceding the recent-measurement time frame) by a factor of 2, with a median excess of 4.3. 

We note that these properties that EELGs exhibit are all necessary, but not sufficient conditions for a galaxy to enter an EELG phase, with several lower EW ([OIII]$\lambda5007$ EW < 500\AA) systems in our full sample also exhibiting a sub-set of these characteristics (e.g., blue $\beta$ slopes or high short-term/long-term SFR ratios). In these cases where low EWs are found in galaxies matching part of our EELG phase criteria, we must consider the roles of dust, metallicity and the escape fraction of ionising photons in modulating the equivalent width (as it is the UV-ionising photons that do not leak from the galaxy that drive the nebular line emission).

Stacking the SFHs in bins of [OIII]$\lambda5007$ EW we determine that EELGs produce most of their mass within a lookback of 10Myr, while lower EW stacks favour histories where the peak of star formation occurs at greater lookback times with a drop in the SFR over the last 10Myr (the time period responsible for the ionising photon that drive the nebula line emission).

The \texttt{BEAGLE} SED-fitting suggests that SFGs only enter an EELG phase during short-lived periods of rapid star formation. 
We determine from \texttt{BEAGLE} single-stellar population models\footnote{Single-stellar denotes that these population models do not consider stars in binary pairs.} that the highest probability star forming duty cycle for an EELG is 10-20$\%$ (for a metallicity and ionisation parameter: 0.1Z$_\odot$, log$_{10}$(U)=-2.0), adopting a top hat duty cycle paramaterisation. This optimum duty cycle is consistent with the fraction of the last 100Myr when galaxies in our sample exhibit a short-term/long-term SFR excess of 3 (a threshold which defines an on-phase from our EELG sub-sample). 
This suggests that our full spectroscopic sample contains both EELGs currently in an on-phase as well as galaxies with the same duty cycle but captured in an off-phase which are not EELGs but are sufficiently bright in the rest-UV to have entered our spectroscopic selection. 

These off-phases make up a greater proportion of the total SFH and are accompanied by a dimming of the UV luminosity.
Therefore, SFGs may be boosted into our UV-selected sample during an on-phase as an EELG, confirming why we observe that lower mass galaxies in our sample typically exhibit higher EWs (as a similar mass galaxy in an off-phase may not satisfy our UV-selection). 

The bursty SFHs in our EELGs may also be responsible for the high detection rate of Lyman-$\alpha$ emission, with $53\%$ (9/17) of high redshift ($z>5.7$) EELGs confirmed to be Lyman-$\alpha$ emitters, compared to only 18\% (2/11) of high-redshift non-EELGs. The rate of LAE detection highlights that the conditions required to enter an EELG phase also promote the escape of Lyman-$\alpha$ photons, through the ionisation or removal of neutral gas via outflows along our sight line. 

Across our full sample ($3<z<9.5$), 42\% of our sample satisfy our [OIII]$\lambda5007$ EELG threshold. Splitting our full sample at $z=5.7$, we find that the fraction of galaxies in an EELG phase increases with redshift, with $61\pm4$\% of our full sample at $z>5.7$ in an EELG phase. 
At these high redshifts, a greater fraction of galaxies are exhibiting high equivalent width line emission through the combination of the redshift evolution in the metallicity and the duty cycle, along with the relative youth of the Universe meaning that older stellar populations have had less chance to form and contribute to the rest-frame optical continuum.

This combination of the high EELG fraction during the EoR and the high ionisation efficiency in EELGs makes these galaxies productive ionising photon producers. 
EELGs during the EoR may therefore contribute significantly to the ionising radiation field responsible for reionising the Universe.

\section*{Acknowledgements}
This work is based in part on observations made with the NASA/ESA/CSA James Webb Space Telescope.  The data were obtained from the Mikulski Archive for Space Telescopes at the Space Telescope Science Institute, which is operated
by the Association of Universities for Research in Astronomy, Inc., under NASA contract NAS 5-03127 for JWST. 
KB is supported by the Australian Research Council Centre of Excellence for All Sky Astrophysics in 3 Dimensions (ASTRO 3D), through project number CE170100013.
AJB, AJC, JC, IEBW, AS, KB $\&$ GCJ acknowledge funding from the "FirstGalaxies" Advanced Grant from the European Research Council (ERC) under the European Union’s Horizon 2020 research and innovation programme (Grant agreement No. 789056).
ECL acknowledges support of an STFC Webb Fellowship (ST/W001438/1).
MC acknowledge support by the Science and Technology Facilities Council (STFC),ERC Advanced Grant 695671 "QUENCH". 
S.A. acknowledges support from Grant PID2021-127718NB-I00 funded by the Spanish Ministry of Science and Innovation/State Agency of Research (MICIN/AEI/ 10.13039/501100011033). 
S.Ca acknowledges support by European Union’s HE ERC Starting Grant No. 101040227 - WINGS.
S. Alberts acknowledges support from the JWST Mid-Infrared Instrument (MIRI) Science Team Lead, grant 80NSSC18K0555, from NASA Goddard Space Flight Center to the University of Arizona.
DJE is supported as a Simons Investigator and by JWST/NIRCam contract to the University of Arizona, NAS5-02015.
RH acknowledges funding by the Johns Hopkins University, Institute for Data Intensive Engineering and Science (IDIES).
KH, BJ, EE, DPS, RE and MR acknowledge the JWST/NIRCam contract to the University of Arizona, NAS5-02015.
DPS additionallu acknowledges support from the National Science Foundation through the grant AST-2109066.
BER acknowledges support from the NIRCam Science Team contract to the University of Arizona, NAS5-02015. 
The research of CCW is supported by NOIRLab, which is managed by the Association of Universities for Research in Astronomy (AURA) under a cooperative agreement with the National Science Foundation.
RS acknowledges support from a STFC Ernest Rutherford Fellowship (ST/S004831/1).
L.S., T.J.L. and J.S acknowledges support by the Science and Technology Facilities Council (STFC) and ERC Advanced Grant 695671 "QUENCH".
C.Si, JW acknowledges support from the Science and Technology Facilities Council (STFC), by the ERC through Advanced Grant 695671 “QUENCH”, by the UKRI Frontier Research grant RISEandFALL.
M.V.M. acknowledges support from the National Science Foundation via grant NSF AAG 2205519.
and the Wisconsin Alumni Research Foundation via grant MSN251397.
H{\"U} gratefully acknowledges support by the Isaac Newton Trust and by the Kavli Foundation through a Newton-Kavli Junior Fellowship.

\section*{Data Availability}
The raw observational JWST NIRSpec and NIRCam data, and the corresponding reduced data products, are publicly available on MAST and via the JADES team website \url{https://jades-survey.github.io}. The line fluxes measured for our prism spectroscopy are given in \citet{Bunker23b}, when detected above a $5\sigma$ significance. The individual EELG and BEAGLE galaxy properties (Tables \ref{tab:EELG} and \ref{tab:beagle_results_update_1}) are made available in a machine readable format.



\bibliographystyle{mnras}
\bibliography{ref} 




\newpage
\appendix

\section{Alternative EW measurement estimates} \label{sec:App:ew}

When computing the EW we use the broadband photometry to measure the flux density of the continuum. When possible our choice of filter is the one contaminated by the line in question. In this Appendix we check the consistency of our EW with three related methods: using the measured continuum flux density from the spectroscopy; taking the photometry from an adjacent filter - uncontaminated by the emission line in question; and using the line flux from the higher resolution R1000 grating. 

\subsection{Uncontaminated broadband filter comparison}
\begin{figure}
    \centering
    \includegraphics[width=\columnwidth]{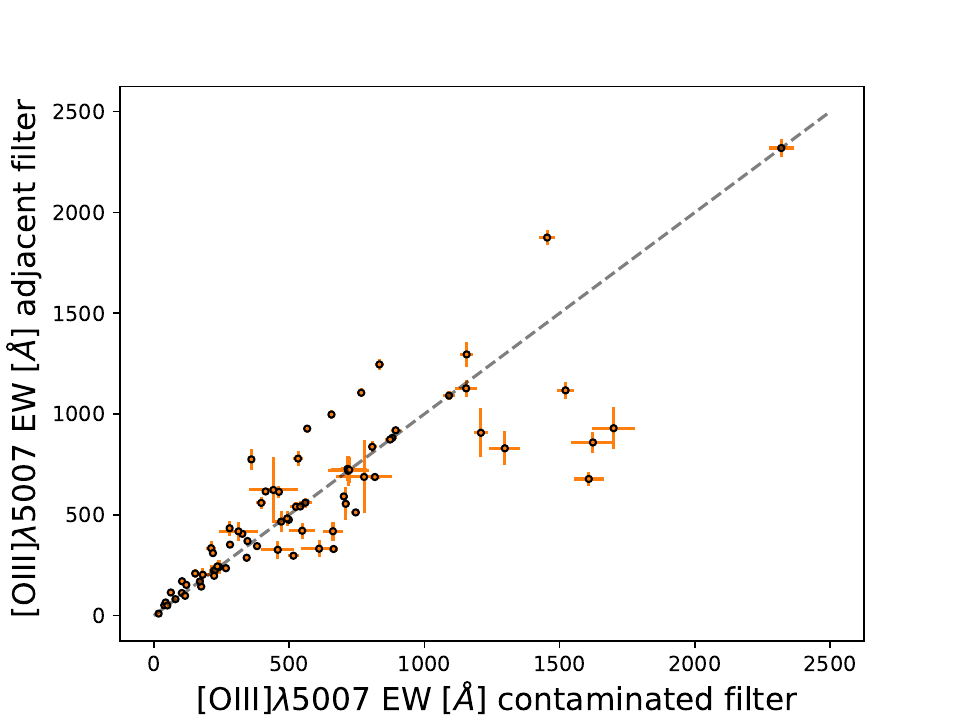}
    \caption{Comparison of [OIII]$\lambda5007$ EW derived using the line contaminated filter and the chosen adjacent filter for the continuum measurement. 
    Good agreement is found albeit with large scatter due to the variation in measured $\beta$-slope compared to our assumed $\beta=-2$.}
    \label{fig:EW_comparison_adj}
\end{figure}

When we do not have NIRCam imaging of the galaxy in the broadband filter covering the observed wavelength of the emission line,  we determine an estimate of the flux density at the wavelength of the line using the nearest (and where possible contamination free) available filter. Again, we remove any contribution from spectroscopically detected line emission to the selected broadband filter, accounting for the wavelength-dependent filter transmission. In figure \ref{fig:EW_comparison_adj} we present the comparison of the [OIII]$\lambda5007$ EW determined using the continuum flux density as measured in the contaminated filter and the adjacent filter. We find that using the nearest available filter provides good agreement with the measurement made using the line contaminated filter - with a scatter consistent with the variation of $\beta$ slopes in our sample.  

For 10/85 galaxies in our full sample we do not have NIRCam photometry that covers the observed wavelength of the [OIII]$\lambda5007$ emission line. This is typically due to the NIRSpec and NIRCam footprints not being in perfect alignment and due to subtly different footprints in the different filter bands (e.g., F444W covers a larger area than the SW filters, the Medium bands from JEMS cover a subtly different footprint to the JADES broadband filters).  

When measuring the EW using an adjacent filter, the quoted errors present the statistical uncertainty on the line flux measurement and signal to noise of the photometry. The quoted uncertainty does not reflect the systematic uncertainty in the $\beta$ assumption.

\subsection{Spectroscopic comparison}

\begin{figure}
    \centering
    \includegraphics[width=\columnwidth]{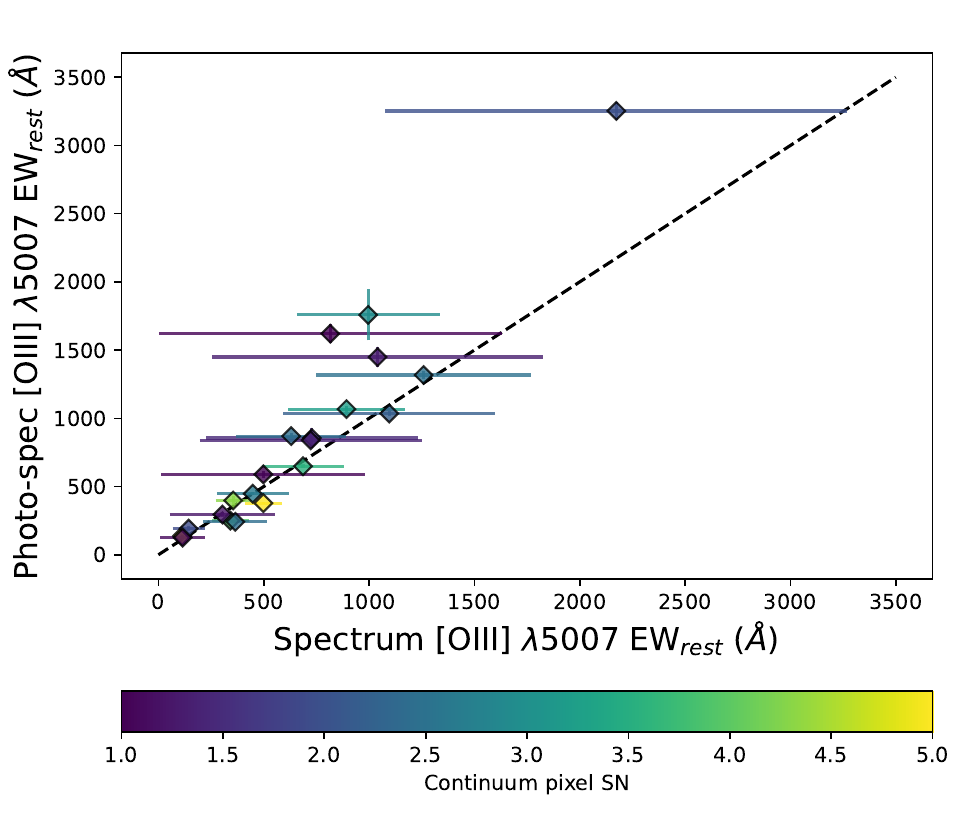}
    \caption{Comparison of [OIII]$\lambda5007$ EW derived using the continuum flux density measured from the line contaminated filter and the prism spectrum. Here we have restricted the full sample to those with continuum S/N per pixel above 1. Good agreement is found when a significant continuum signal to noise (per pixel) is obtained in the spectroscopy. }
    \label{fig:EW_comparison_spec}
\end{figure}

We additionally compare the EWs derived from the broadband photometry with the EWs measured directly from the prism spectrum itself. Here we present the EWs measured using the continuum in the 1D spectra in figure \ref{fig:EW_comparison_spec}, where we show all the spectral measurements where the signal to noise per pixel in the continuum is greater than 1 around the emission line. 

When the spectroscopy does not have a significant detection of the continuum, the inferred EW from the spectroscopy is highly uncertain (consistent with infinite EW in the limiting case of zero continuum). We therefore adopt the continuum estimate based on the broadband photometry, corrected for line contamination using the spectroscopic line fluxes, as this provides better constraints on the EW.

\subsection{prism and R1000 resolution comparison }

\begin{figure}
    \centering
    \includegraphics[width=\columnwidth]{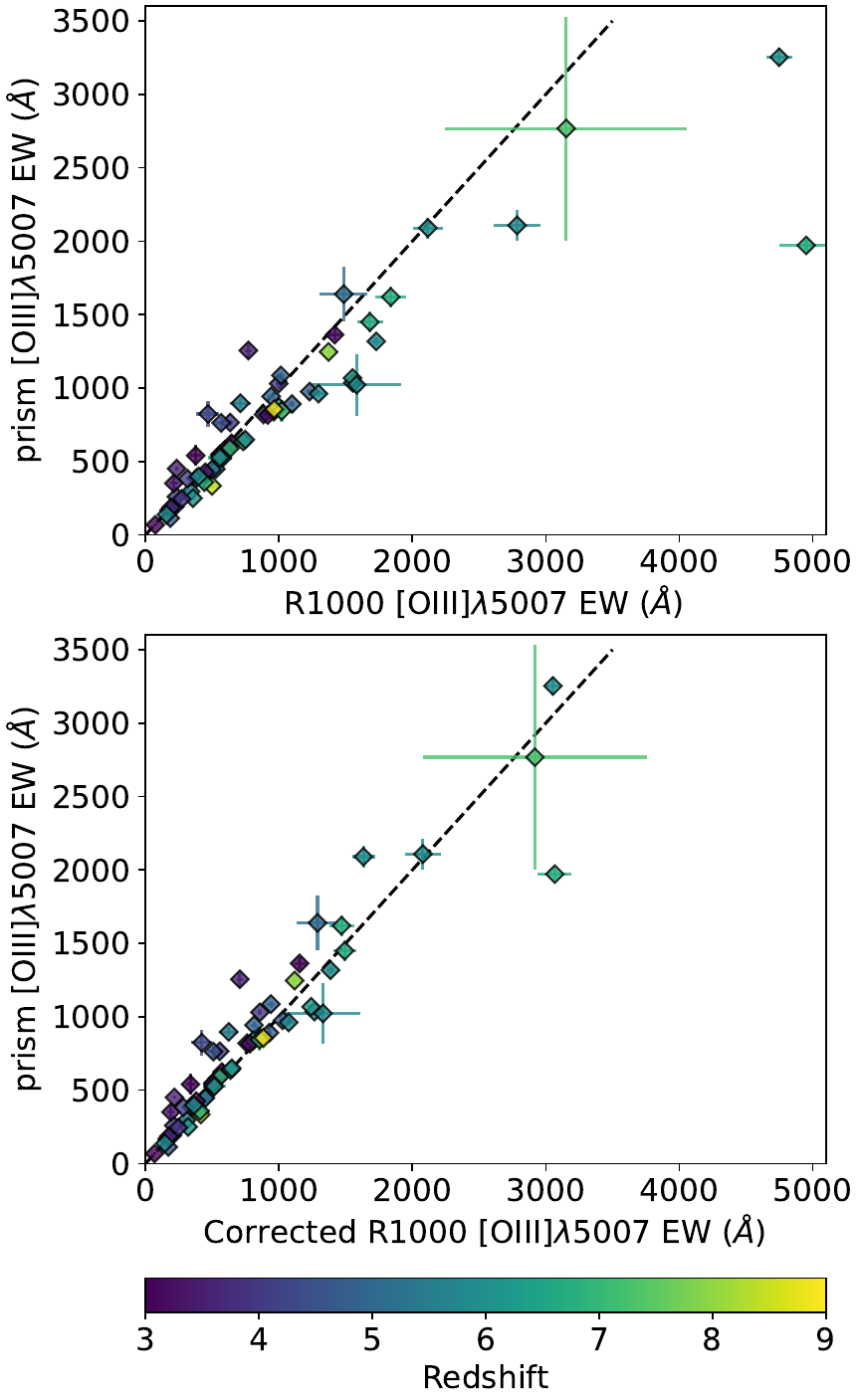}
    \caption{Comparison of [OIII]$\lambda5007$ EWs measured using the line fluxes measured from the prism and the R1000 spectroscopy. Top panel: direct comparison reveals a systematic offset with the R1000 spectroscopy overestimating the EW relative to the prism, seen most prominently at large EW. This offset is due to a systematic overestimate of the measured line flux between the prism and the R1000 spectroscopy, by a factor 1.08, originating from differences in flux calibration during the data reduction (see \citealt{Bunker23} for details). Bottom panel: EW comparison after the R1000 measured line fluxes have been corrected for the 1.08 factor. In both panels the dashed line represents the 1:1 line.}
    \label{fig:EW_comparison}
\end{figure}

We note that the measured emission line fluxes are a factor of $\sim1.08$ larger in the R1000 resolution spectroscopy compared to the prism spectroscopy \citep{Bunker23}. We see this effect in the measured EW in Figure \ref{fig:EW_comparison}, with the EW determined using the R1000 spectroscopy larger than those from the prism. We find that the integrated prism spectra best recovers the reported broadband flux densities, and so we use these measurements as part of our study. 

\section{M$\rm_{UV}$ measurement from photometry and spectroscopy} \label{sec:App:muv}
We want to ensure that the slitloss correction applied to the NIRSpec prism spectroscopy, which uses the position of the target in the MSA shutter and assumes a point-like source (see \citealt{Bunker23b} for details), performs a sufficient job at recovering the total flux of our targets. This is crucial for our estimate of M$\rm_{UV}$. To check the M$\rm_{UV}$ determined from the NIRSpec spectra in Section \ref{sec:beta}, we additionally measure the rest-UV magnitude from direct imaging in the filter closest to rest-frame 1500\AA\, (without including Lyman-break). We use the HST/WFC3 F775W filter for galaxies below $z<4$, the JWST/NIRCam F090W filter for galaxies $4<z<5.6$, the JWST/NIRCam F115W filter for galaxies $5.6<z<7.2$ and the JWST/NIRCam F150W filter for any galaxies at $z>7.2$. The photometry for each galaxy in our sample in each filter is publicly available as part of the JADES NIRCam data release\footnote{\url{https://archive.stsci.edu/hlsp/jades}} \citep{Rieke23b}. 

\begin{figure}
    \centering
    \includegraphics[width=\columnwidth]{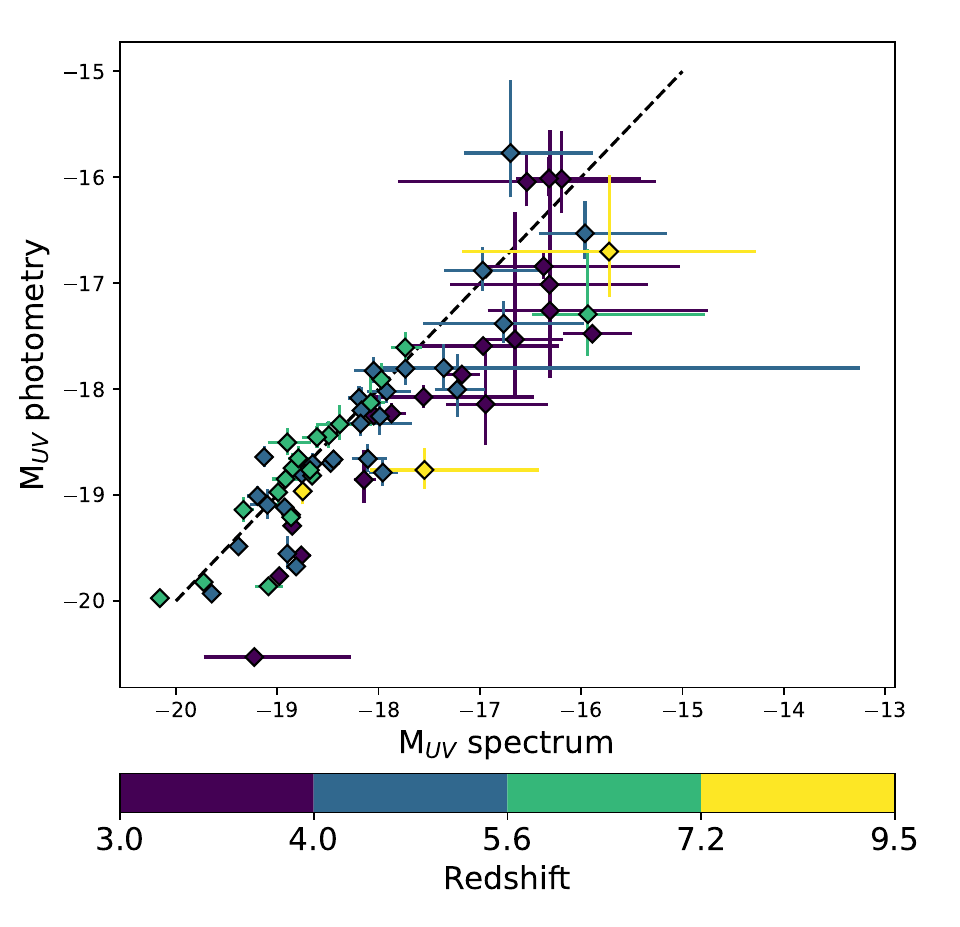}
    \caption{Comparison of the measured M$\rm_{UV}$ from the direct imaging photometry (using a Kron aperture) and from the NIRSpec spectroscopy. We split our sample into four redshift bins based on which imaging filter we used to measure the rest-frame 1500\AA, these utilise the HST/WFC3 F775W and JWST/NIRCam F090W, F115W, F150W filters.}
    \label{fig:Muv_comp}
\end{figure}

A variety of aperture photometry measurements are available for each target. These include a range of fixed circular aperture measurements (up to 1" in diameter), an 80$\%$ flux radius aperture, as well as Kron aperture measurements (see \citep{Rieke23b} for more details). We find consistent results for the measured photometry from the Kron and the largest circular apertures (suggesting many of our targets are resolved and are extended). In Figure \ref{fig:Muv_comp} we present the comparison of the M$\rm_{UV}$ measurements from the photometry and spectroscopy. We find good agreement for the majority of bright galaxies (M$\rm_{UV}$ < -18). Fainter that this, we find a systematic offset with the UV magnitude being underestimated in the spectroscopy by $\sim0.5$ magnitude compared to the photometric measurement, although we note that these measurements are still consistent within the uncertainties. 

\section{UV luminosity-weighted spectral stack} \label{sec:App:weighted_stack}
In Section \ref{sec:stack} we combined the spectra of multiple galaxies in stacks to study the population properties of EELGs and non-EELGs at high and low redshift. To create the stacks we normalised each spectrum to the rest-UV (1500\AA) and then average-combined the galaxies spectra. 
This approach provides equal weight to all galaxies, whereas we could
alternatively combine our sample by weighting each spectrum by the luminosity,
effectively summing the luminosities of the galaxies in the sub-samples to obtain an indication of the total spectrum produced by EELGs.
This alternative approach means UV luminous galaxies are weighted more strongly in the stacks.

\begin{figure*}
    \centering
    \includegraphics[width=\textwidth]{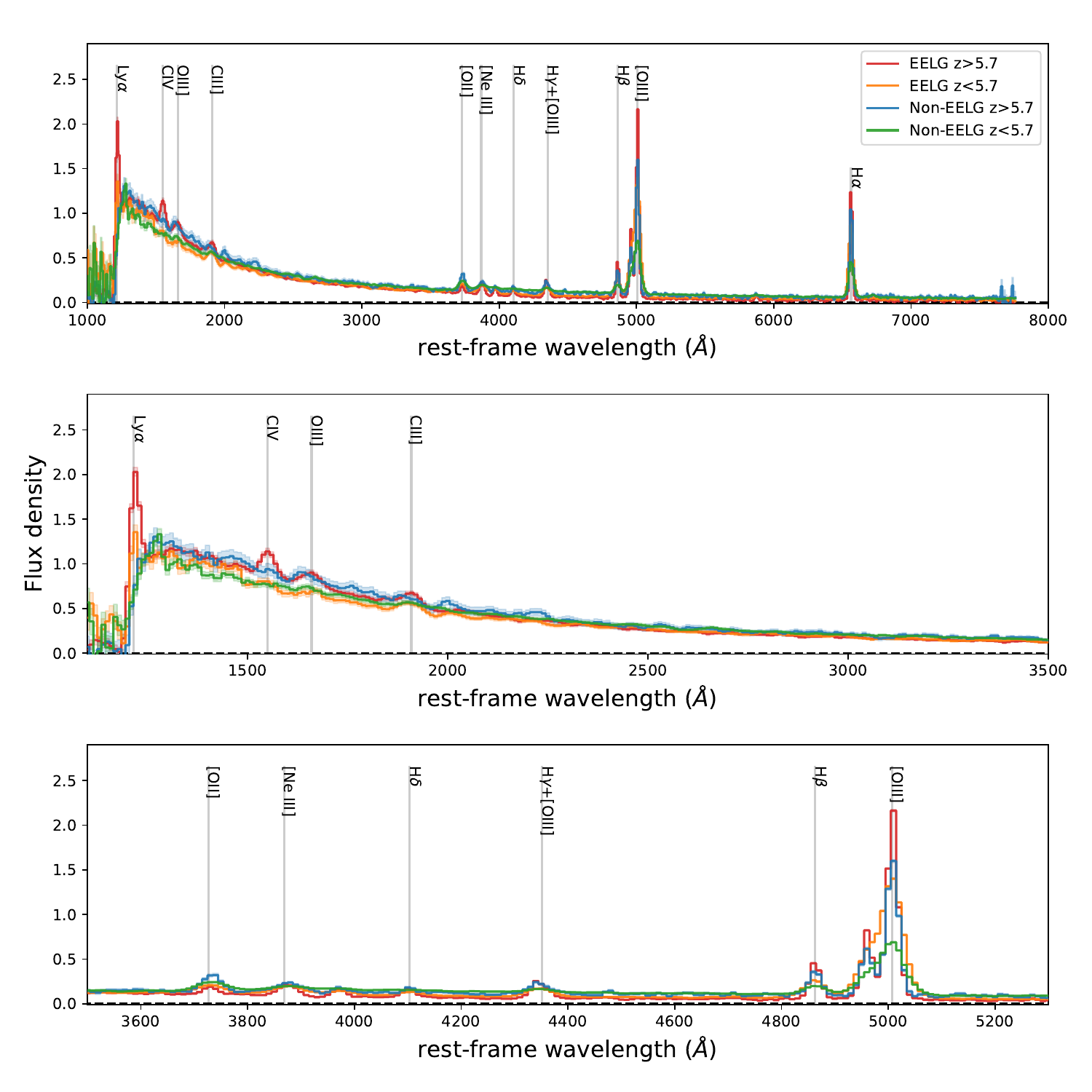}
    \caption{Stacks of the galaxy sub-samples weighted by UV luminosity and split by redshift (at $z=5.7$) and equivalent width (at [OIII]$\lambda5007$ EW $=750$\AA). See Figure \ref{fig:stacks} for the non-weighted stacks. Top panel: The full spectral range. Middle panel: The rest-UV. Bottom panel: The rest-optical.}
    \label{fig:stacks_alternative}
\end{figure*}

We repeat the non-weighted stacking in Section \ref{sec:stack}, now weighting the 1500\AA-normalised spectrum by the UV luminosity. We present the resultant stacks in Figure \ref{fig:stacks_alternative}, which replicates Figure \ref{fig:stacks} with the UV-weighted stacks. As expected these stacks show a higher S/N since we prioritise the UV-bright galaxies in the stacks. We find that the measured emission line and UV-spectral properties are in line with those from the non-weighted stacks. In Table \ref{tab:stack_alternative} we provide the spectral properties for this alternative stacking method. We find consistent stack properties between the two weighting schemes.  

\begin{table*}
\centering
\resizebox{.75\textwidth}{!}{%
\begin{tabular}{ccccc}
Stack             & high-z EELG & low-z EELG & high-z non-EELG & low-z non-EELG \\ \hline
N galaxies   &  17  &  19  &  11  &  37  \\ 
$[$OIII$]\lambda5007$ EW  &  1406$\pm$71  &  1331$\pm$34  &  490$\pm$25  &  337$\pm$5  \\ 
H$\alpha$ EW  &  1102$\pm$102  &  971$\pm$31  &  434$\pm$36  &  273$\pm$5  \\ 
Ly$\alpha$ EW  &  32$\pm$2  &  11$\pm$3  &  $<7$  &  $<25$  \\ 
H$\alpha$/H$\beta$  &  2.88$\pm$0.06  &  3.21$\pm$0.05  &  2.98$\pm$0.15  &  3.74$\pm$0.10  \\ 
A1600  &  0.03$\pm$0.07  &  0.44$\pm$0.06  &  0.15$\pm$0.19  &  1.01$\pm$0.10  \\ 
O32$^*$  &  34.6$\pm$3.4  &  15.3$\pm$1.1  &  6.7$\pm$0.9  &  5.4$\pm$0.7  \\ 
R23$^*$  &  6.9$\pm$0.3  &  8.6$\pm$0.5  &  7.9$\pm$0.9  &  10.2$\pm$1.2  \\ 
Ne3O2$^*$  &  3.1$\pm$0.3  &  1.0$\pm$0.1  &  1.1$\pm$0.2  &  0.4$\pm$0.1  \\ 
12+$\log_{10}\mathrm{(O/H)}$ &$7.48^{+0.06}_{-0.05}$ &$7.72\pm0.07$ & $7.74^{+0.13}_{-0.12}$ & $8.03^{+0.09}_{-0.15}$ \\
$\log_{10}(\xi_\mathrm{ion}^\mathrm{HII})$ &  $25.48\pm0.02$  &  $25.68\pm0.02$  &  $25.36\pm0.04$  &  $25.20\pm0.04$   \\ 
$\beta_{obs}$  &  -2.41$\pm$0.01  &  -2.21$\pm$0.02  &  -2.16$\pm$0.02  &  -1.89$\pm$0.01  \\ 
$\beta_{int}^*$  &  -2.42$\pm$0.05  &  -2.45$\pm$0.05  &  -2.24$\pm$0.12  &  -2.44$\pm$0.06  \\ 
\hline
\end{tabular}
}
\caption{Galaxy properties derived for our UV-weighted stacked spectra, as in Table \ref{tab:stack}. Properties marked with ($^*$) have been corrected for dust. Upper limits are given at $3\sigma$.}
\label{tab:stack_alternative}
\end{table*}

\section{Dust correction to UV $\beta$ slopes} \label{sec:App:dust_beta}
We correct our measurements of the observed rest-frame UV slope ($\beta_{obs}$, from $f_\lambda \propto \lambda^{\beta}$) for the dust attenuation determined from the measured Balmer decrement, adopting a \citet{Calzetti00} dust law and R$_v=4.05$. 
To aid us with this correction we first determine a simple relation between dust attenuation and the change in $\beta$ slope. 

\begin{figure}
    \centering
    \includegraphics[width=\columnwidth]{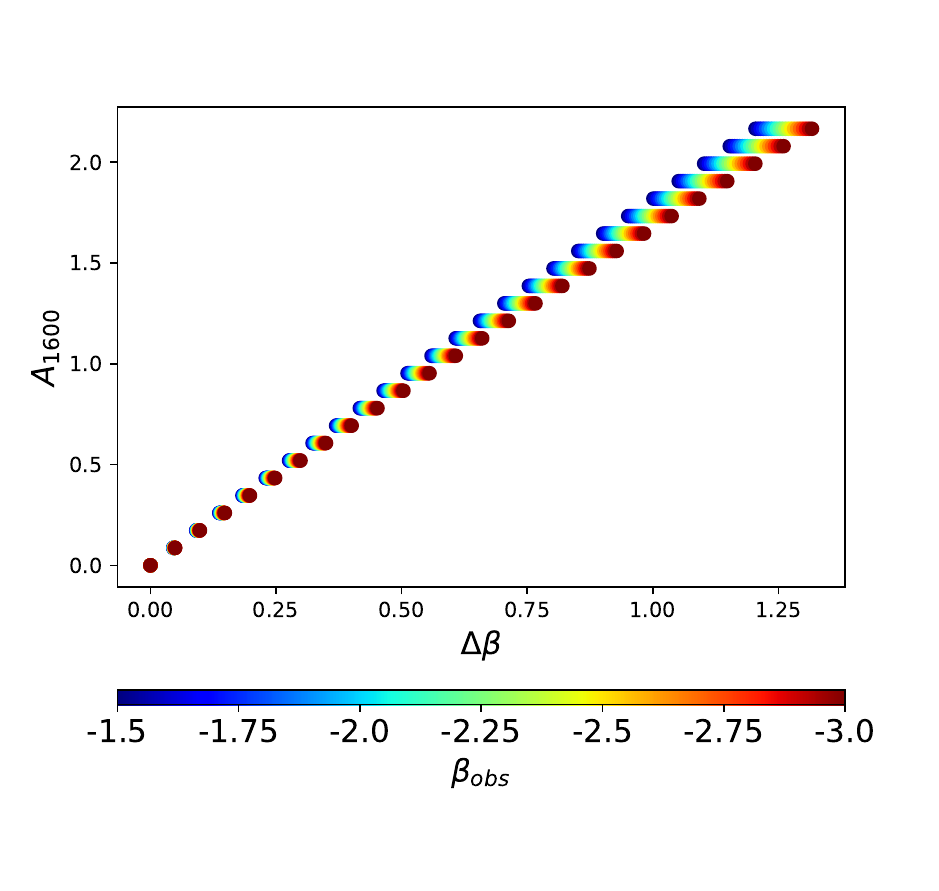}
    \caption{The theoretical relation between the dust attenuation (A$_{1600}$) for a given change in $\beta$ slope, adopting the \citet{Calzetti00} dust law. }
    \label{fig:A1600}
\end{figure}

We create a grid of stellar\footnote{The stellar (continuum) and gas-phase (line emission) dust extinction is related via E$_s$(B-V)=0.44 E(B-V).} extinction ($0<A_{1600}<1$) and observed $\beta$ slopes ($-1.5>\beta_{obs}>-3.5$), and for each grid point we generate an observed UV slope and correct each data point obeying the \citet{Calzetti00} law for the given attenuation. We then fit a new $\beta$ slope ($f_\lambda \propto \lambda^{\beta}$) to the resultant dust corrected data and we compare the observed and dust corrected $\beta$ values. The dust attenuation and change in $\beta$ slope is presented in Figure \ref{fig:A1600}. We fit a relation to $A_{1600}$ as a function of $\Delta \beta$ (given as $\beta_{obs} - \beta_{corr}$) and $\beta_{obs}$

\begin{equation}
    A_{1600} = (2.0178 + 0.0983*\beta_{obs})*\Delta\beta
\end{equation}

which we employ in our data analysis in Section \ref{sec:beta}.

\section{Spectral energy distribution results} \label{sec:App:BEAGLE} 
In Section \ref{sec:beagle} we modelled the NIRSpec spectroscopy and NIRCam photometry using \texttt{BEAGLE} to constrain the SFH of each galaxy. In Tables \ref{tab:beagle_results_update_1} and \ref{tab:beagle_results_update_2} we present the resultant SFH properties. 
\begin{table*}
\centering
\begin{tabular}{cccccccccc}
ID & $z_{\rm spec}$ & [OIII]$\lambda5007$ EW & Stellar mass & Mass-w-age & Lumin-w-age & SFR$_{\rm burst}$ & SFR$_{100}$ & SFR$_3$ / SFR$_{3-100}$ & SFR$_5$ / SFR$_{5-100}$ \\ 
- & - & \AA\ & $\log_{10}($M/M$_\odot)$ & $\log_{10}(t/yr)$ & $\log_{10}(t/yr)$ & M$_\odot$yr$^{-1}$ & M$_\odot$yr$^{-1}$ & - & - \\ 
\hline

10058975 & 9.44 & $856\pm75$ & $8.17^{+0.03}_{-0.03}$ & $7.05^{+0.02}_{-0.02}$ & $6.42^{+0.05}_{-0.03}$ & $6.68\pm1.46$ & $1.47\pm1.46$ & $3.25\pm0.28$ & $1.97\pm0.28$\\ 

8013 & 8.48 & $334\pm17$ & $7.04^{+0.37}_{-0.33}$ & $7.22^{+0.67}_{-0.92}$ & $6.45^{+0.65}_{-0.22}$ & $2.56\pm1.28$ & $0.11\pm1.28$ & $21.17\pm0.55$ & $15.08\pm0.55$\\ 

21842 & 7.98 & $1246\pm37$ & $7.45^{+0.08}_{-0.06}$ & $6.93^{+0.12}_{-0.08}$ & $6.38^{+0.25}_{-0.07}$ & $4.15\pm0.99$ & $0.28\pm0.99$ & $7.78\pm0.56$ & $4.83\pm0.56$\\ 

10013682 & 7.28 & $2767\pm763$ & $6.23^{+0.14}_{-0.08}$ & $6.32^{+0.08}_{-0.06}$ & $6.32^{+0.08}_{-0.06}$ & $0.01\pm0.06$ & $0.02\pm0.06$ & $435.79\pm11.36$ & -\\ 

10013905 & 7.21 & $844\pm77$ & $7.27^{+0.06}_{-0.58}$ & $6.93^{+0.07}_{-0.62}$ & $6.32^{+0.13}_{-0.05}$ & $2.45\pm1.43$ & $0.19\pm1.43$ & $16.24\pm0.69$ & $21.77\pm0.69$\\ 

20961 & 7.05 & $589\pm23$ & $7.76^{+0.05}_{-0.06}$ & $7.12^{+0.07}_{-0.07}$ & $6.90^{+0.11}_{-0.16}$ & $0.56\pm0.06$ & $0.58\pm0.06$ & $0.97\pm0.32$ & $0.97\pm0.32$\\ 

10013609 & 6.93 & $637\pm28$ & $6.94^{+0.05}_{-0.05}$ & $6.35^{+0.03}_{-0.01}$ & $6.36^{+0.03}_{-0.02}$ & $0.01\pm0.06$ & $0.09\pm0.06$ & $275.59\pm9.23$ & -\\ 

4297 & 6.72 & $1971\pm39$ & $6.68^{+0.03}_{-0.03}$ & $6.28^{+0.00}_{-0.01}$ & $6.28^{+0.00}_{-0.01}$ & $0.01\pm0.05$ & $0.05\pm0.05$ & - & -\\ 

3334 & 6.71 & $1620\pm72$ & $7.31^{+0.06}_{-0.06}$ & $7.06^{+0.07}_{-0.06}$ & $6.82^{+0.09}_{-0.09}$ & $0.49\pm0.03$ & $0.21\pm0.03$ & $2.49\pm0.13$ & $2.57\pm0.13$\\ 

16625 & 6.63 & $1449\pm71$ & $7.51^{+0.08}_{-0.08}$ & $7.06^{+0.10}_{-0.12}$ & $6.48^{+0.23}_{-0.19}$ & $3.81\pm1.20$ & $0.33\pm1.20$ & $6.78\pm0.47$ & $4.13\pm0.47$\\ 

10005447 & 6.63 & $359\pm46$ & $6.97^{+0.13}_{-0.14}$ & $7.00^{+0.23}_{-0.18}$ & $6.59^{+0.20}_{-0.08}$ & $0.28\pm0.05$ & $0.09\pm0.05$ & $3.24\pm0.48$ & $3.40\pm0.48$\\ 

18846 & 6.34 & $1068\pm21$ & $7.82^{+0.01}_{-0.01}$ & $6.75^{+0.00}_{-0.00}$ & $6.71^{+0.01}_{-0.01}$ & $0.37\pm0.06$ & $0.66\pm0.06$ & $0.56\pm0.17$ & $0.55\pm0.17$\\ 

18179 & 6.33 & $241\pm6$ & $8.53^{+0.08}_{-0.10}$ & $8.18^{+0.13}_{-0.18}$ & $7.91^{+0.12}_{-0.16}$ & $0.93\pm0.28$ & $1.38\pm0.28$ & $0.67\pm0.33$ & $0.66\pm0.33$\\ 

18976 & 6.33 & $839\pm30$ & $7.30^{+0.05}_{-0.04}$ & $6.75^{+0.02}_{-0.02}$ & $6.69^{+0.03}_{-0.02}$ & $0.17\pm0.11$ & $0.20\pm0.11$ & $0.82\pm0.65$ & $0.82\pm0.65$\\ 

10009693 & 6.30 & $525\pm39$ & $6.32^{+0.31}_{-0.22}$ & $6.42^{+1.03}_{-0.15}$ & $6.41^{+0.51}_{-0.14}$ & $0.12\pm0.23$ & $0.02\pm0.23$ & $13.65\pm1.89$ & $26.34\pm1.89$\\ 

17566 & 6.11 & $250\pm5$ & $8.05^{+0.13}_{-0.12}$ & $7.24^{+0.18}_{-0.16}$ & $7.06^{+0.12}_{-0.13}$ & $0.02\pm0.25$ & $1.12\pm0.25$ & $0.40\pm10.64$ & $1.95\pm10.64$\\ 

19342 & 5.98 & $2090\pm72$ & $7.61^{+0.05}_{-0.09}$ & $7.18^{+0.06}_{-0.07}$ & $6.86^{+0.11}_{-0.19}$ & $2.08\pm0.70$ & $0.41\pm0.70$ & $2.65\pm0.49$ & $1.61\pm0.49$\\ 

10013618 & 5.95 & $138\pm6$ & $8.42^{+0.10}_{-0.05}$ & $7.31^{+0.13}_{-0.08}$ & $7.24^{+0.10}_{-0.07}$ & $0.56\pm0.08$ & $2.64\pm0.08$ & $0.21\pm0.29$ & $0.20\pm0.29$\\ 

9422 & 5.94 & $3253\pm65$ & $7.94^{+0.01}_{-0.01}$ & $6.89^{+0.01}_{-0.01}$ & $6.29^{+0.02}_{-0.04}$ & $16.77\pm1.85$ & $0.88\pm1.85$ & $8.94\pm0.13$ & $5.42\pm0.13$\\ 

6002 & 5.94 & $870\pm20$ & $7.43^{+0.02}_{-0.01}$ & $6.97^{+0.01}_{-0.00}$ & $6.59^{+0.02}_{-0.01}$ & $0.54\pm0.02$ & $0.27\pm0.02$ & $2.08\pm0.05$ & $2.13\pm0.05$\\ 

10013704 & 5.93 & $649\pm13$ & $7.13^{+0.01}_{-0.01}$ & $6.46^{+0.01}_{-0.01}$ & $6.49^{+0.00}_{-0.00}$ & $0.95\pm0.09$ & $0.13\pm0.09$ & $8.64\pm0.10$ & -\\ 

10013620 & 5.92 & $398\pm30$ & $7.95^{+0.03}_{-0.01}$ & $7.00^{+0.04}_{-0.00}$ & $6.81^{+0.06}_{-0.02}$ & $0.92\pm0.03$ & $0.89\pm0.03$ & $1.03\pm0.05$ & $1.03\pm0.05$\\ 

19606 & 5.89 & $962\pm21$ & $7.51^{+0.06}_{-0.05}$ & $7.04^{+0.06}_{-0.05}$ & $6.61^{+0.16}_{-0.13}$ & $2.55\pm0.68$ & $0.32\pm0.68$ & $3.93\pm0.46$ & $2.36\pm0.46$\\ 

10056849 & 5.82 & $1022\pm209$ & $7.26^{+0.09}_{-0.07}$ & $6.95^{+0.17}_{-0.06}$ & $6.37^{+0.21}_{-0.07}$ & $1.86\pm0.62$ & $0.18\pm0.62$ & $5.79\pm0.59$ & $3.68\pm0.59$\\ 

10005113 & 5.82 & $2107\pm103$ & $7.22^{+0.09}_{-0.10}$ & $6.97^{+0.14}_{-0.07}$ & $6.38^{+0.22}_{-0.08}$ & $1.59\pm0.62$ & $0.17\pm0.62$ & $5.78\pm0.65$ & $3.67\pm0.65$\\ 

22251 & 5.80 & $1318\pm26$ & $7.70^{+0.15}_{-0.08}$ & $7.17^{+0.31}_{-0.24}$ & $6.78^{+0.23}_{-0.16}$ & $2.50\pm0.45$ & $0.50\pm0.45$ & $5.74\pm0.20$ & $6.38\pm0.20$\\ 

4404 & 5.78 & $1035\pm21$ & $7.58^{+0.02}_{-0.02}$ & $6.84^{+0.01}_{-0.01}$ & $6.52^{+0.01}_{-0.01}$ & $1.25\pm0.03$ & $0.38\pm0.03$ & $3.55\pm0.05$ & $3.75\pm0.05$\\ 

3968 & 5.77 & $295\pm19$ & $7.67^{+0.07}_{-0.09}$ & $7.20^{+0.17}_{-0.13}$ & $7.12^{+0.14}_{-0.16}$ & $0.31\pm0.05$ & $0.47\pm0.05$ & $0.64\pm0.70$ & $0.64\pm0.70$\\ 

6384 & 5.61 & $192\pm16$ & $8.10^{+0.06}_{-0.06}$ & $7.41^{+0.15}_{-0.13}$ & $7.35^{+0.11}_{-0.12}$ & $0.26\pm0.08$ & $1.27\pm0.08$ & $0.20\pm1.07$ & $0.20\pm1.07$\\ 

16745 & 5.57 & $379\pm8$ & $8.07^{+0.01}_{-0.01}$ & $6.98^{+0.01}_{-0.01}$ & $6.94^{+0.00}_{-0.00}$ & $1.59\pm0.05$ & $1.17\pm0.05$ & $1.37\pm0.11$ & $1.38\pm0.11$\\ 

6246 & 5.57 & $448\pm12$ & $7.57^{+0.02}_{-0.02}$ & $6.98^{+0.02}_{-0.02}$ & $6.90^{+0.02}_{-0.02}$ & $0.83\pm0.09$ & $0.37\pm0.09$ & $2.34\pm0.17$ & $2.40\pm0.17$\\ 

10016374 & 5.51 & $1758\pm186$ & $7.45^{+0.08}_{-0.05}$ & $7.00^{+0.18}_{-0.08}$ & $6.73^{+0.18}_{-0.08}$ & $1.11\pm0.05$ & $0.28\pm0.05$ & $4.26\pm0.16$ & $4.57\pm0.16$\\ 

9743 & 5.45 & $897\pm35$ & $7.29^{+0.37}_{-0.52}$ & $7.15^{+0.80}_{-0.89}$ & $6.47^{+0.72}_{-0.20}$ & $2.86\pm2.49$ & $0.19\pm2.49$ & $14.53\pm0.91$ & $12.89\pm0.91$\\ 

9452 & 5.13 & $446\pm20$ & $7.94^{+0.28}_{-0.24}$ & $7.28^{+0.61}_{-0.27}$ & $6.96^{+0.56}_{-0.25}$ & $0.59\pm1.00$ & $0.86\pm1.00$ & $0.68\pm1.72$ & $1.45\pm1.72$\\ 

4902 & 5.12 & $113\pm4$ & $8.61^{+0.05}_{-0.11}$ & $7.65^{+0.10}_{-0.09}$ & $7.50^{+0.06}_{-0.07}$ & $0.44\pm0.27$ & $4.04\pm0.27$ & $0.11\pm2.03$ & $0.10\pm2.03$\\ 

10015338 & 5.07 & $1639\pm187$ & $7.58^{+0.03}_{-0.03}$ & $6.75^{+0.01}_{-0.01}$ & $6.65^{+0.02}_{-0.02}$ & $0.58\pm0.09$ & $0.38\pm0.09$ & $1.54\pm0.18$ & $1.55\pm0.18$\\ 

10009320 & 5.06 & $220\pm27$ & $6.86^{+0.17}_{-0.17}$ & $7.09^{+0.27}_{-0.19}$ & $6.83^{+0.25}_{-0.20}$ & $0.12\pm0.07$ & $0.07\pm0.07$ & $1.62\pm0.93$ & $2.06\pm0.93$\\ 

5759 & 5.05 & $523\pm13$ & $7.29^{+0.07}_{-0.07}$ & $7.01^{+0.15}_{-0.11}$ & $6.67^{+0.19}_{-0.11}$ & $0.56\pm0.04$ & $0.20\pm0.04$ & $3.00\pm0.44$ & $3.13\pm0.44$\\ 

8113 & 4.90 & $943\pm19$ & $7.55^{+0.03}_{-0.02}$ & $6.87^{+0.03}_{-0.02}$ & $6.74^{+0.03}_{-0.03}$ & $0.91\pm0.08$ & $0.36\pm0.08$ & $2.69\pm0.11$ & $2.79\pm0.11$\\ 

10005217 & 4.89 & $1086\pm24$ & $7.13^{+0.42}_{-0.33}$ & $7.88^{+0.78}_{-1.62}$ & $6.89^{+0.80}_{-0.86}$ & $4.18\pm0.91$ & $0.12\pm0.91$ & $27.10\pm0.22$ & $16.86\pm0.22$\\ 

17260 & 4.89 & $131\pm13$ & $7.74^{+0.08}_{-0.13}$ & $7.13^{+0.09}_{-0.07}$ & $6.95^{+0.13}_{-0.10}$ & $0.16\pm0.04$ & $0.55\pm0.04$ & $0.28\pm0.66$ & $0.28\pm0.66$\\ 

5457 & 4.87 & $387\pm13$ & $7.47^{+0.06}_{-0.05}$ & $6.91^{+0.05}_{-0.03}$ & $6.80^{+0.04}_{-0.04}$ & $0.35\pm0.04$ & $0.30\pm0.04$ & $1.17\pm0.21$ & $1.17\pm0.21$\\ 

4009 & 4.86 & $313\pm9$ & $7.95^{+0.06}_{-0.05}$ & $7.10^{+0.07}_{-0.06}$ & $6.96^{+0.09}_{-0.09}$ & $0.97\pm0.11$ & $0.89\pm0.11$ & $1.10\pm0.71$ & $1.10\pm0.71$\\ 

7938 & 4.82 & $891\pm18$ & $7.75^{+0.14}_{-0.02}$ & $6.96^{+0.15}_{-0.00}$ & $6.65^{+0.24}_{-0.07}$ & $1.34\pm0.03$ & $0.56\pm0.03$ & $2.51\pm0.03$ & $2.59\pm0.03$\\ 

18090 & 4.79 & $976\pm20$ & $7.82^{+0.06}_{-0.06}$ & $7.09^{+0.06}_{-0.06}$ & $6.86^{+0.09}_{-0.08}$ & $5.32\pm0.73$ & $0.66\pm0.73$ & $4.31\pm0.19$ & $3.18\pm0.19$\\ 

10001892 & 4.77 & $663\pm176$ & $6.98^{+0.18}_{-0.16}$ & $6.98^{+0.11}_{-0.08}$ & $6.87^{+0.11}_{-0.07}$ & $0.01\pm0.02$ & $0.10\pm0.02$ & $0.06\pm3.05$ & $0.06\pm3.05$\\ 

17072 & 4.71 & $114\pm6$ & $7.67^{+0.05}_{-0.13}$ & $7.33^{+0.12}_{-0.22}$ & $7.28^{+0.10}_{-0.22}$ & $0.12\pm0.02$ & $0.47\pm0.02$ & $0.26\pm0.41$ & $0.25\pm0.41$\\ 

8083 & 4.67 & $1808\pm36$ & $8.21^{+0.07}_{-0.10}$ & $7.70^{+0.14}_{-0.20}$ & $7.23^{+0.15}_{-0.22}$ & $13.41\pm1.60$ & $1.41\pm1.60$ & $4.82\pm0.14$ & $3.49\pm0.14$\\ 

5076 & 4.50 & $42\pm7$ & $7.99^{+0.04}_{-0.04}$ & $7.65^{+0.10}_{-0.13}$ & $7.56^{+0.10}_{-0.10}$ & $0.06\pm0.01$ & $0.98\pm0.01$ & $0.06\pm1.61$ & $0.06\pm1.61$\\ 

7304 & 4.49 & $233\pm21$ & $7.50^{+0.06}_{-0.06}$ & $7.06^{+0.07}_{-0.04}$ & $6.93^{+0.10}_{-0.09}$ & $0.27\pm0.05$ & $0.31\pm0.05$ & $0.84\pm0.97$ & $0.84\pm0.97$\\ 

10000626 & 4.47 & $763\pm42$ & $6.88^{+0.16}_{-0.13}$ & $7.33^{+0.58}_{-0.27}$ & $6.97^{+0.39}_{-0.24}$ & $0.19\pm0.01$ & $0.08\pm0.01$ & $2.66\pm0.13$ & $2.76\pm0.13$\\ 

8073 & 4.39 & $614\pm13$ & $7.75^{+0.06}_{-0.14}$ & $7.39^{+0.21}_{-0.11}$ & $7.23^{+0.13}_{-0.11}$ & $0.47\pm0.27$ & $0.56\pm0.27$ & $0.84\pm0.63$ & $0.83\pm0.63$\\ 

10001916 & 4.28 & $823\pm87$ & $6.24^{+0.29}_{-0.17}$ & $7.90^{+0.54}_{-0.60}$ & $6.90^{+0.56}_{-0.44}$ & $0.43\pm0.15$ & $0.01\pm0.15$ & $39.99\pm0.36$ & $24.47\pm0.36$\\ 

7892 & 4.25 & $820\pm16$ & $7.32^{+0.02}_{-0.02}$ & $6.86^{+0.03}_{-0.02}$ & $6.52^{+0.02}_{-0.01}$ & $0.71\pm0.02$ & $0.21\pm0.02$ & $3.64\pm0.11$ & $3.85\pm0.11$\\ 

6519 & 4.24 & $1461\pm31$ & $7.66^{+0.06}_{-0.03}$ & $6.96^{+0.08}_{-0.04}$ & $6.64^{+0.10}_{-0.06}$ & $1.62\pm0.04$ & $0.46\pm0.04$ & $3.83\pm0.04$ & $4.07\pm0.04$\\ 

7762 & 4.15 & $383\pm8$ & $8.08^{+0.06}_{-0.05}$ & $7.44^{+0.10}_{-0.09}$ & $7.17^{+0.09}_{-0.08}$ & $0.01\pm0.06$ & $1.21\pm0.06$ & $0.62\pm9.35$ & $1.39\pm9.35$\\ 
 
\end{tabular}
\caption{Reported \texttt{BEAGLE} spectral energy distribution modelled star formation history properties. Galaxies are ordered in descending spectroscopic redshift. The final two columns relate to the short-term/long-term time averaged star formation rates.}
\label{tab:beagle_results_update_1}
\end{table*}

\begin{table*}
\begin{tabular}{cccccccccc}
ID & $z_{\rm spec}$ & [OIII]$\lambda5007$ EW & Stellar mass & Mass-w-age & Lumin-w-age & SFR$_{\rm burst}$ & SFR$_{100}$ & SFR$_3$ / SFR$_{3-100}$ & SFR$_5$ / SFR$_{5-100}$ \\ 
- & - & \AA\ & $\log_{10}($M/M$_\odot)$ & $\log_{10}(t/yr)$ & $\log_{10}(t/yr)$ & M$_\odot$yr$^{-1}$ & M$_\odot$yr$^{-1}$ & - & - \\ 
\hline
17777 & 4.14 & $259\pm10$ & $8.00^{+0.08}_{-0.07}$ & $7.77^{+0.21}_{-0.16}$ & $7.58^{+0.20}_{-0.20}$ & $0.33\pm0.08$ & $0.82\pm0.08$ & $0.39\pm0.39$ & $0.39\pm0.39$\\ 

7507 & 4.04 & $765\pm20$ & $7.41^{+0.04}_{-0.03}$ & $6.98^{+0.04}_{-0.03}$ & $6.61^{+0.07}_{-0.05}$ & $0.57\pm0.06$ & $0.26\pm0.06$ & $2.29\pm0.37$ & $2.36\pm0.37$\\ 

10015344 & 4.03 & $94\pm6$ & $8.73^{+0.02}_{-0.02}$ & $7.30^{+0.03}_{-0.02}$ & $7.12^{+0.06}_{-0.05}$ & $2.48\pm0.39$ & $5.33\pm0.39$ & $0.46\pm0.22$ & $0.45\pm0.22$\\ 

10013578 & 4.02 & $507\pm44$ & $7.86^{+0.02}_{-0.02}$ & $7.29^{+0.07}_{-0.05}$ & $7.13^{+0.05}_{-0.06}$ & $0.88\pm0.04$ & $0.73\pm0.04$ & $1.21\pm0.07$ & $1.22\pm0.07$\\ 

4270 & 4.02 & $1030\pm21$ & $7.83^{+0.05}_{-0.05}$ & $7.11^{+0.05}_{-0.06}$ & $6.65^{+0.06}_{-0.06}$ & $4.68\pm0.24$ & $0.68\pm0.24$ & $8.41\pm0.07$ & $7.09\pm0.07$\\ 

10015193 & 3.96 & $285\pm13$ & $8.06^{+0.02}_{-0.02}$ & $6.92^{+0.02}_{-0.02}$ & $6.85^{+0.01}_{-0.01}$ & $1.14\pm0.08$ & $1.14\pm0.08$ & $0.99\pm0.09$ & $0.99\pm0.09$\\ 

10016186 & 3.94 & $245\pm5$ & $8.32^{+0.01}_{-0.01}$ & $7.30^{+0.01}_{-0.01}$ & $7.26^{+0.01}_{-0.01}$ & $0.78\pm0.02$ & $2.10\pm0.02$ & $0.36\pm0.04$ & $0.36\pm0.04$\\ 

18970 & 3.73 & $186\pm4$ & $9.24^{+0.03}_{-0.03}$ & $8.07^{+0.06}_{-0.06}$ & $7.84^{+0.06}_{-0.06}$ & $0.21\pm0.77$ & $8.18\pm0.77$ & $0.48\pm3.69$ & $0.75\pm3.69$\\ 

22924 & 3.71 & $54\pm10$ & $7.95^{+0.09}_{-0.07}$ & $7.78^{+0.27}_{-0.20}$ & $7.70^{+0.25}_{-0.19}$ & $0.08\pm0.04$ & $0.75\pm0.04$ & $0.10\pm1.08$ & $0.10\pm1.08$\\ 

19519 & 3.61 & $550\pm12$ & $7.87^{+0.02}_{-0.02}$ & $7.06^{+0.03}_{-0.03}$ & $6.90^{+0.04}_{-0.04}$ & $1.06\pm0.08$ & $0.75\pm0.08$ & $1.44\pm0.10$ & $1.45\pm0.10$\\ 

10009506 & 3.60 & $542\pm11$ & $7.89^{+0.01}_{-0.02}$ & $7.21^{+0.04}_{-0.03}$ & $7.01^{+0.04}_{-0.04}$ & $1.24\pm0.08$ & $0.77\pm0.08$ & $1.64\pm0.09$ & $1.66\pm0.09$\\ 

7809 & 3.60 & $451\pm9$ & $8.27^{+0.03}_{-0.03}$ & $7.26^{+0.07}_{-0.08}$ & $7.04^{+0.11}_{-0.15}$ & $1.84\pm0.43$ & $1.88\pm0.43$ & $0.98\pm0.38$ & $0.98\pm0.38$\\ 

4282 & 3.60 & $260\pm37$ & $8.16^{+0.08}_{-0.08}$ & $8.26^{+0.19}_{-0.21}$ & $8.09^{+0.16}_{-0.16}$ & $0.09\pm0.06$ & $0.48\pm0.06$ & $0.19\pm0.79$ & $0.18\pm0.79$\\ 

4493 & 3.59 & $19\pm2$ & $10.10^{+0.02}_{-0.02}$ & $8.68^{+0.03}_{-0.05}$ & $7.26^{+0.03}_{-0.03}$ & $416.0\pm26.5$ & $29.0\pm26.5$ & $24.4\pm0.1$ & $30.1\pm0.1$\\ 

10035295 & 3.58 & $1255\pm43$ & $7.53^{+0.11}_{-0.10}$ & $7.24^{+0.24}_{-0.17}$ & $6.67^{+0.19}_{-0.13}$ & $3.29\pm0.80$ & $0.34\pm0.80$ & $7.24\pm0.26$ & $6.05\pm0.26$\\ 

3184 & 3.47 & $579\pm22$ & $8.11^{+0.05}_{-0.04}$ & $7.29^{+0.09}_{-0.08}$ & $7.04^{+0.09}_{-0.07}$ & $0.01\pm0.11$ & $1.30\pm0.11$ & $0.45\pm11.33$ & $1.85\pm11.33$\\ 

7629 & 3.46 & $242\pm19$ & $7.96^{+0.12}_{-0.15}$ & $8.00^{+0.20}_{-0.29}$ & $7.76^{+0.19}_{-0.29}$ & $0.00\pm0.02$ & $0.52\pm0.02$ & $0.49\pm5.92$ & $0.78\pm5.92$\\ 

10001587 & 3.44 & $350\pm80$ & $7.19^{+0.26}_{-0.27}$ & $8.45^{+0.38}_{-0.56}$ & $8.17^{+0.46}_{-0.56}$ & $0.04\pm0.01$ & $0.05\pm0.01$ & $0.82\pm0.60$ & $0.82\pm0.60$\\ 

3322 & 3.40 & $198\pm30$ & $7.74^{+0.12}_{-0.14}$ & $7.85^{+0.23}_{-0.29}$ & $7.63^{+0.24}_{-0.31}$ & $0.00\pm0.04$ & $0.41\pm0.04$ & $0.17\pm7.96$ & $0.63\pm7.96$\\ 

19431 & 3.32 & $425\pm12$ & $6.97^{+0.15}_{-0.10}$ & $7.21^{+0.83}_{-0.32}$ & $6.76^{+0.42}_{-0.19}$ & $0.52\pm0.03$ & $0.09\pm0.03$ & $6.62\pm0.11$ & $7.51\pm0.11$\\ 

10013597 & 3.32 & $351\pm46$ & $7.61^{+0.04}_{-0.04}$ & $7.20^{+0.14}_{-0.10}$ & $7.06^{+0.14}_{-0.16}$ & $0.34\pm0.09$ & $0.41\pm0.09$ & $0.82\pm0.40$ & $0.82\pm0.40$\\ 

16478 & 3.25 & $48\pm7$ & $7.97^{+0.06}_{-0.05}$ & $7.69^{+0.10}_{-0.08}$ & $7.61^{+0.12}_{-0.09}$ & $0.09\pm0.02$ & $0.92\pm0.02$ & $0.09\pm0.39$ & $0.09\pm0.39$\\ 

18322 & 3.16 & $1363\pm41$ & $6.96^{+0.09}_{-0.12}$ & $6.92^{+0.35}_{-0.22}$ & $6.30^{+0.25}_{-0.16}$ & $2.75\pm0.82$ & $0.09\pm0.82$ & $28.37\pm0.39$ & $19.72\pm0.39$\\ 

10040 & 3.15 & $167\pm12$ & $7.64^{+0.14}_{-0.10}$ & $7.38^{+0.56}_{-0.21}$ & $6.89^{+0.30}_{-0.16}$ & $1.13\pm0.48$ & $0.44\pm0.48$ & $2.69\pm0.46$ & $2.79\pm0.46$\\ 

21150 & 3.09 & $816\pm16$ & $8.28^{+0.00}_{-0.00}$ & $7.25^{+0.01}_{-0.01}$ & $7.20^{+0.01}_{-0.01}$ & $0.98\pm0.03$ & $1.90\pm0.03$ & $0.51\pm0.04$ & $0.50\pm0.04$\\ 

10004721 & 3.07 & $68\pm6$ & $6.81^{+0.04}_{-0.05}$ & $7.22^{+0.04}_{-0.03}$ & $7.14^{+0.03}_{-0.04}$ & $0.09\pm0.02$ & $0.07\pm0.02$ & $1.41\pm0.25$ & $1.42\pm0.25$\\ 

8245 & 3.07 & $624\pm55$ & $7.18^{+0.10}_{-0.10}$ & $7.27^{+0.29}_{-0.24}$ & $7.04^{+0.27}_{-0.23}$ & $0.17\pm0.12$ & $0.15\pm0.12$ & $1.15\pm0.85$ & $1.15\pm0.85$\\ 
2923 & 3.01 & $540\pm71$ & $7.39^{+0.06}_{-0.07}$ & $7.15^{+0.11}_{-0.08}$ & $6.95^{+0.13}_{-0.13}$ & $0.15\pm0.09$ & $0.24\pm0.09$ & $0.59\pm0.64$ & $0.59\pm0.64$\\ 
\end{tabular}
\caption{Table \ref{tab:beagle_results_update_1} continued.}
\label{tab:beagle_results_update_2}
\end{table*}


\bsp	
\label{lastpage}
\end{document}